\documentclass[11pt]{article}
\PassOptionsToPackage{table}{xcolor}
\usepackage{acl}
\usepackage{newtxtext}
\usepackage[utf8]{inputenc}
\usepackage[T1]{fontenc}
\usepackage{amsmath,amssymb,amsthm}
\usepackage{graphicx}
\usepackage{booktabs}
\usepackage{multirow}
\usepackage{url}
\usepackage{xcolor}
\usepackage{tabularx}
\definecolor{famshade}{RGB}{240,244,250}
\definecolor{famname}{HTML}{0F4D92}
\definecolor{famform}{HTML}{106B54}
\newcommand{\famstack}[3]{\parbox[t]{\linewidth}{\raggedright
  \textcolor{#1}{\texttt{\textbf{#2}}}\\[-3pt]
  {\scriptsize\textcolor{black!45}{#3}}}}
\newcommand{\famcell}[2]{\famstack{famname}{#1}{#2}}
\newcommand{\famcellform}[2]{\famstack{famform}{#1}{#2}}
\definecolor{Best}{HTML}{F5B7BF}
\definecolor{Secondbest}{HTML}{D5DEFF}
\definecolor{Ours}{HTML}{FCE4D6}
\usepackage{microtype}
\usepackage{titletoc}
\titlecontents{section}[0em]{\addvspace{2pt}\bfseries}{\contentslabel{1.5em}}{}{\hfill\contentspage}
\usepackage{wrapfig}
\usepackage{placeins}
\usepackage{enumitem}
\usepackage[font=small,skip=4pt]{caption}
\setlist[itemize]{leftmargin=1.4em,itemsep=1pt,topsep=2pt,parsep=0pt}
\setlist[enumerate]{leftmargin=1.8em,itemsep=1pt,topsep=2pt,parsep=0pt}
\usepackage{float}
\usepackage[ruled,vlined]{algorithm2e}
\SetAlFnt{\small}\SetAlCapFnt{\small}
\hypersetup{colorlinks=true,linkcolor=blue,citecolor=blue,urlcolor=blue}
\usepackage{fvextra}
\usepackage{adjustbox}
\extrafloats{100}
\usepackage{eso-pic}
\AddToShipoutPictureBG{\color{white}\rule{\paperwidth}{\paperheight}}

\newcommand{\method}{\textsc{Music\-RLVR}}
\newcommand{\bench}{\textsc{MusicConstraintBench}}
\newcommand{\BaseEightkilltestsuite}{0.123}

\newcommand{\BaseThirtytwokilltestsuite}{0.363}

\newcommand{\Basekilltestsuite}{0.160}

\newcommand{\Chatmusicianbenchstructure}{0.000}

\newcommand{\Chatmusiciankilltestsuite}{0.160}

\newcommand{\KlBetaHi}{0.010}
\newcommand{\KlBetaLo}{0.047}
\newcommand{\KlBetaMid}{0.024}
\newcommand{\KlLrA}{0.006}
\newcommand{\KlLrB}{0.024}
\newcommand{\KlLrC}{0.049}
\newcommand{\KlLrD}{0.157}
\newcommand{\LlamaParse}{0.970}

\newcommand{\Llamakilltestsuite}{0.380}

\newcommand{\Minimaxkilltestsuite}{0.367}

\newcommand{\Oursbenchstructure}{0.808}

\newcommand{\Ourskilltestsuite}{0.663}

\newcommand{\RwLrA}{0.713}
\newcommand{\RwLrB}{0.832}
\newcommand{\RwLrC}{0.893}
\newcommand{\RwLrD}{0.936}

\newcommand{\Sftbenchstructure}{0.600}

\newcommand{\StructW}{0.675}
\newcommand{\StructWident}{32}
\newcommand{\StructWmix}{0.637}

\newcommand{\QualityMatchedItems}{297}
\newcommand{\QualityRlDistinctBars}{0.426}

\newcommand{\QualitySftDistinctBars}{0.417}

\newcommand{\QualityUniquePrompts}{220}
\newcommand{\StrictAuditEvaluations}{175}
\newcommand{\StrictAuditItems}{43920}
\newcommand{\StrictAuditLost}{201}

\newcommand{\EdgeFreeMinimax}{0.195}
\newcommand{\EdgeFreeN}{329}
\newcommand{\EdgeFreeOurs}{0.210}
\newcommand{\EdgeFreeSft}{0.100}
\newcommand{\PEdgeFreeMinimax}{0.685}
\newcommand{\PEdgeFreeSft}{2.03\times 10^{-6}}
\newcommand{\LamContrastonezero}{45}
\newcommand{\LamContrastzerofive}{55}
\newcommand{\LamContrastzeroseven}{104}
\newcommand{\LamIdentonezero}{74}
\newcommand{\LamIdentzerofive}{65}
\newcommand{\LamIdentzeroseven}{15}

\newcommand{\PLamonezeroMixed}{0.002}
\newcommand{\PLamonezeroStructure}{4.74\times 10^{-15}}
\newcommand{\PLamzerofiveMixed}{0.368}
\newcommand{\PLamzerofiveStructure}{3.54\times 10^{-9}}
\newcommand{\AggAbab}{0.791}
\newcommand{\AggAbabIdent}{0.140}

\newcommand{\AggKFour}{0.475}
\newcommand{\AggKOne}{0.850}
\newcommand{\AggKRatio}{1.8}

\newcommand{\AggMixedStruct}{0.644}

\newcommand{\AggSymContrast}{0.233\pm0.024}
\newcommand{\AggSymIdent}{0.762\pm0.018}
\newcommand{\AggUniqueStruct}{0.637}

\newcommand{\MainLlamaKRatio}{14.4}
\newcommand{\MainLlamaMixedKFour}{0.050}
\newcommand{\MatchedWarmMixed}{0.577}

\newcommand{\MatchedDirectMixed}{0.797}
\newcommand{\MatchedDirectSingle}{0.985}

\newcommand{\MatchedDirectUnseen}{0.595}
\newcommand{\MatchedDirectHighk}{0.575}
\newcommand{\MatchedDirectStructure}{0.967}
\newcommand{\MatchedDirectDense}{0.362}
\newcommand{\MatchedDirectEdge}{0.300}

\newcommand{\PoolN}{2{,}180}
\newcommand{\PoolGraded}{0.577}
\newcommand{\PoolBinary}{0.549}

\newcommand{\PoolGradedWins}{221}
\newcommand{\PoolBinaryWins}{160}
\newcommand{\PoolGradedBinaryP}{0.0021}
\newcommand{\PoolGradedWarmWins}{549}
\newcommand{\PoolWarmWins}{216}
\newcommand{\PoolGradedWarmP}{<10^{-4}}
\newcommand{\PoolBinaryWarmP}{<10^{-4}}
\newcommand{\PoolGradedCount}{1257}

\newcommand{\ParaDropDirect}{0.122}
\newcommand{\ParaDropBase}{0.050}
\newcommand{\ParaDropAug}{-0.018}
\newcommand{\ParaPAug}{0.311}
\newcommand{\AugMixedCost}{0.057}
\newcommand{\PoolBinaryCount}{1196}
\newcommand{\PoolWarmCount}{924}

\newcommand{\FamLeapBinary}{0.833}
\newcommand{\FamLeapGraded}{0.839}
\newcommand{\FamLengthBinary}{0.558}
\newcommand{\FamLengthGraded}{0.663}

\newcommand{\DeadBinaryEarly}{0.550}
\newcommand{\DeadBinaryLate}{0.823}
\newcommand{\DeadBinaryAll}{0.717}

\newcommand{\DeadGradedEarly}{0.307}
\newcommand{\DeadGradedLate}{0.823}
\newcommand{\DeadGradedAll}{0.660}

\newcommand{\DeadWarmEarly}{0.120}
\newcommand{\DeadWarmLate}{0.343}
\newcommand{\DeadWarmAll}{0.245}

\title{Do Language Models Need Music Supervision?\\
Verifiable Rewards for Multi-Constraint Symbolic Music Generation}

\author{
Haoyue Liu\textsuperscript{1,2}\quad
Xiaoying Tang\textsuperscript{1,2,$\dagger$}
\\[5pt]
{\normalfont\small
\textsuperscript{1}School of Science and Engineering, The Chinese University of Hong Kong, Shenzhen 518172, China}\\
{\normalfont\small
\textsuperscript{2}Shenzhen Future Network of Intelligence Institute (FNii-Shenzhen)}
}

\begin{document}
\maketitle

\begin{abstract}
Language models can now generate symbolic music from natural-language instructions. However, existing models still struggle to satisfy multiple explicit musical constraints simultaneously. To systematically evaluate this capability, we introduce \bench{}, a benchmark of 2,180 items spanning eight families of programmatically verifiable constraints. On this benchmark, Llama-3.1-70B satisfies 0.630 of single-constraint items but only 0.044 of four-constraint ones. To address this challenge, we propose \method{}, which trains
a language model with group relative policy optimisation (GRPO) on verifier rewards alone, needing
no human annotation, reward model or music-domain supervised fine-tuning. \method{} incorporates
(1) a hard validation gate that rejects malformed scores, (2) graded per-family credit that,
unlike a binary reward, separates partially correct outputs, and (3) an all-satisfied bonus
for meeting every constraint at once. Extensive experiments show that, after 300 GRPO updates, \method{} raises Qwen3-4B-Instruct-2507 from $\Basekilltestsuite$ to $\MatchedDirectMixed$ on mixed constraints, outperforming Llama-3.1-70B, and generalises to unseen property combinations, out-of-range parameters and more constraints than any training prompt.
The recipe transfers to Qwen3-8B, and neither trained model loses significant accuracy on
general benchmarks.
\end{abstract}

\section{Introduction}

\begin{figure}[t]
\centering
\includegraphics[width=\columnwidth]{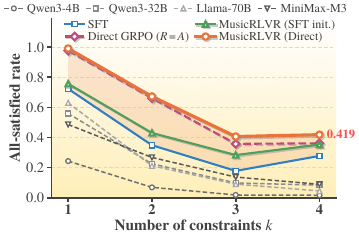}
\caption{All-satisfied rate against $k$ over all $2{,}180$ \bench{} items, for the matched runs
and the baselines of Table~\ref{tab:main}. Each slice pools every suite with that $k$ ($354$,
$614$, $512$ and $700$ items), so slices differ in suite and family mixture
(Table~\ref{tab:bykpooled}). The right-hand label gives the Direct \method{} rate at $k=4$.}
\label{fig:degrade}
\end{figure}

Language models can now generate machine-readable symbolic scores from
natural-language instructions
\citep{yuan2024chatmusician,qu2025mupt,wang2025notagen}.
In applications requiring explicit control, users may specify multiple
musical properties, including key, meter, length, pitch range, final note,
rhythmic vocabulary, melodic motion, and form.
These properties can be evaluated programmatically from parsed scores,
making constraint-following music generation a
\emph{property-verifiable generation} problem.

However, existing models still struggle to satisfy such constraints jointly. A request may simultaneously require a tune to be \emph{in D minor, in $6/8$,
sixteen bars long, ending on the tonic, with limited melodic leaps and an AABA form}; because these
requirements must hold jointly, violating any one of them makes the output non-compliant. No
dedicated benchmark isolates this capability, so we construct \bench{}, a benchmark of $2{,}180$
ABC-generation items spanning eight constraint families. As illustrated in
Figure~\ref{fig:degrade}, the all-satisfied rate drops sharply as the number of simultaneous
constraints increases: Llama-3.1-70B satisfies $0.630$ of single-constraint items but only
$0.044$ of four-constraint ones. Notably, the natural solution of using the deterministic
verifiers directly as reinforcement-learning rewards inherits the same conjunction problem when
the reward is binary: once any constraint fails, outputs with very different degrees of partial
compliance receive the same reward. In group relative policy optimisation (GRPO), a group whose
rollouts all receive the same reward provides no learning signal, and under this binary reward
$\DeadBinaryEarly$ of rollout groups in our matched run have zero reward variance over the first
$50$ updates.

To tackle the above problems, we address the following core question:
\begin{center}
\emph{Can deterministic verifiers alone teach a model\\
to jointly follow multiple musical constraints?}
\end{center}
Hence, we introduce \method{}, a verifier-driven reinforcement learning method that addresses the
aforementioned question through three components: (1) a hard validation gate that rejects
malformed or unparseable scores before any credit is assigned, since partial credit alone would
reward a fragment that satisfies local properties without forming a tune; (2) graded per-family
credit that, unlike the binary reward, separates partially correct outputs; and (3) an
all-satisfied bonus for meeting every requested constraint at once. We summarise the
contributions of this paper as follows:
\begin{itemize}[leftmargin=1.2em,itemsep=2pt,topsep=2pt,parsep=0pt]

\item \textbf{The first benchmark for joint score-constraint compliance.}
We introduce \bench{}, the first benchmark of its kind: $2{,}180$ items spanning eight verifier-defined constraint families, plus $400$ paraphrases, testing constraint density, unseen family combinations, parameter extrapolation, and wording robustness.

\item \textbf{Turning property verifiers into fine-grained training rewards.}
We introduce \method{}, which combines a hard validation gate, graded per-family credit, and an
all-satisfied bonus without human annotation or a learned reward model, giving credit
across requested families and, for form, within one family. Removing or reweighting any of these
components lowers the all-satisfied rate over all $2{,}180$ items from $0.580$ to
$0.502$--$0.553$ (\S\ref{sec:a4}).

\item \textbf{Training constraint following without music-domain supervised fine-tuning (SFT).}
Experiments show that, after 300 GRPO updates, Direct \method{} raises the Mixed all-satisfied rate of Qwen3-4B-Instruct-2507 from $0.160$ to
$\MatchedDirectMixed$, outperforming every zero-shot baseline of Table~\ref{tab:main}, including
Llama-3.1-70B, and generalises to property combinations never observed together during training,
to parameter values outside the training range and to free-form requests. The recipe transfers to
Qwen3-8B, which rises from $0.123$ to $0.657$ on Mixed, and neither policy loses significant
accuracy on IFBench, MMLU-Redux or GSM8K.

\end{itemize}

\begin{figure*}[t]
\centering
\includegraphics[width=0.9\textwidth]{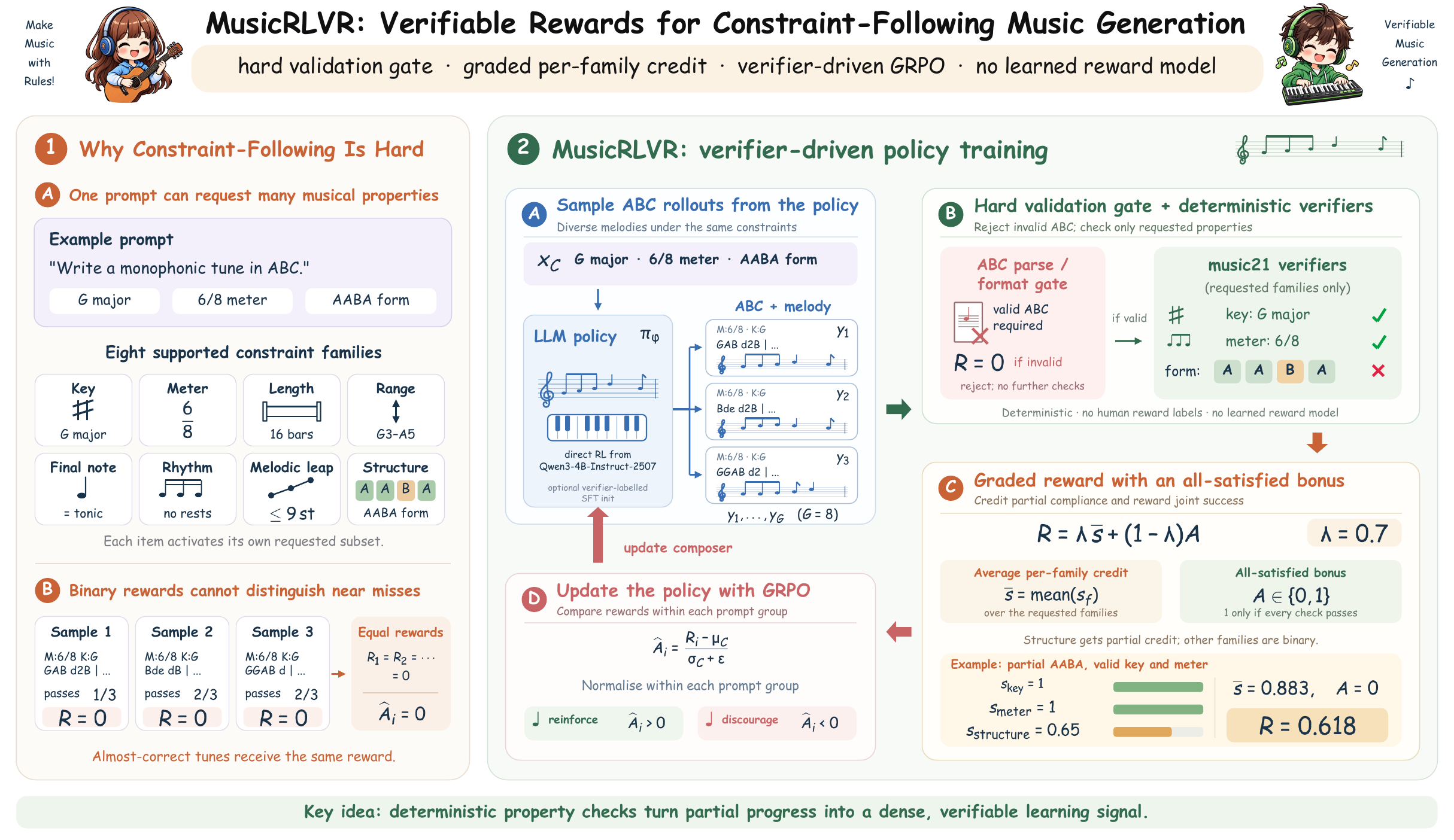}
\caption{\textbf{Overview of \method{}.}
\textbf{Left:} Binary all-satisfied rewards cannot distinguish near misses.
\textbf{Right:} \method{} samples $G$ ABC rollouts, uses a hard validation gate,
and verifies requested properties deterministically. Average per-family credit
(binary except for graded structure) and an all-satisfied bonus form the reward,
normalised per prompt group for GRPO.}
\label{fig:overview}
\end{figure*}

\section{Related Work}

Symbolic music models \citep{sturm2016music,huang2018music,yuan2024chatmusician,qu2025mupt,wang2025notagen,von2023figaro,thickstun2023anticipatory,agostinelli2023musiclm,copet2023simple}
are trained by likelihood or against a learned music--text encoder \citep{wu2025clamp}, with hard
rules applied, if at all, at decoding time
\citep{lu2021neurologic,huang2024symbolic,kaliakatsos2025incorporating}. Reinforcement learning
for music optimises human or learned quality signals
\citep{cideron2024musicrl,jonason2025smart,ziv2026mr,he2026symphonygen} or music-theory
heuristics \citep{jaques2017sequence}. However, these approaches do not directly target joint satisfaction of
prompt-specified, programmatically verifiable score constraints.
In contrast, we use deterministic verifiers as rewards, with constraint
families and parameters varying across prompts.

Reinforcement learning with verifiable rewards (RLVR) builds on PPO-based RLHF
\citep{schulman2017proximal,ouyang2022training} and group-relative baselines
\citep{ahmadian2024back,shao2024deepseekmath}, underpins reasoning models
\citep{guo2025deepseek,lambert2024tulu} and extends beyond mathematics
\citep{ma2026general,gunjal2026rubrics}; DAPO \citep{yu2026dapo} and Dr.~GRPO
\citep{liu2025understanding} patch dead groups and length bias, whereas we grade the reward
itself. FollowBench \citep{jiang2024followbench}, \mbox{IFEval} \citep{zhou2023instruction},
and IFBench \citep{pyatkin2026generalizing} study text constraint following;
the IFBench work also explores RLVR training.
\citet{wang2025towards} build verifiable sheet-music reasoning questions.
Our setting instead evaluates generated scores against prompt-specific constraints,
without a predefined target composition. Because musical fragments may satisfy
individual properties without constituting valid tunes, we apply a hard validation
gate before assigning partial credit.

\section{Method}
\label{sec:method}

In this section, we introduce \method{}, a reinforcement learning method that trains a
general-purpose model to follow requested constraints, with deterministic verifiers as its only
task-specific supervision. As shown in Figure~\ref{fig:overview}, this is
achieved in four steps: (A) the policy samples $G$ ABC rollouts; (B) a hard validation gate rejects
invalid ABC and verifiers check the requested properties (\S\ref{sec:families}); (C) a graded
reward with an all-satisfied bonus scores the verdicts (\S\ref{sec:reward}); and (D) GRPO updates
the policy with group-relative advantages (\S\ref{sec:training}; Algorithm~\ref{alg:musicrlvr}).

\subsection{Problem Formulation}

In this section, we formalise the task and its success criterion. Let $\mathcal{F}$ be a set of
\emph{constraint families}; we use the eight families
in Table~\ref{tab:families}. A \emph{constraint} $c = (f, \theta)$ pairs a family
$f \in \mathcal{F}$ with parameters $\theta$ (e.g.\ $f = \texttt{key}$,
$\theta = (\text{G}, \text{major})$). Each family has a deterministic \emph{verifier}
$v_f : \mathcal{S} \times \Theta_f \rightarrow \{0, 1\}$, where $\mathcal{S}$ is the set of
parseable scores and $\Theta_f$ the parameter space of family $f$. A \emph{problem} is a set of constraints $C = \{c_1, \dots, c_k\}$ with distinct
families, rendered into a prompt $x_C$ by a fixed template. The policy $\pi_\phi$, with trainable
parameters $\phi$, emits a string $y$ in ABC text notation (Appendices~\ref{app:abc}--\ref{app:rendering}); $\rho$
extracts the ABC and applies the validation gate of Appendix~\ref{app:verifiers}, returning
$\rho(y) \in \mathcal{S}$ or $\bot$ on rejection. We define the binary all-satisfied indicator
\begin{equation}
A(y, C) =
\begin{cases}
\prod_{(f,\theta) \in C} v_f\!\left(\rho(y), \theta\right) & \rho(y) \neq \bot,\\[2pt]
0 & \rho(y) = \bot.
\end{cases}
\label{eq:allsat}
\end{equation}
At evaluation, all-satisfied rate is the fraction
of all test items whose outputs pass the validation gate and satisfy every requested musical constraint, i.e.\ the mean of $A$.

\subsection{Constraint Families and Verifiers}
\label{sec:families}

In this section, we present the verifiers that supply every reward signal, how they are
validated and how the key criterion is calibrated.

% <<< tables/families
% Compact main-text form of Table~\ref{tab:families}: one line per family. The prompt
% sentences the template emits live in the full table, Table~\ref{tab:families_full}.
% Hand-maintained from src/constraints.py (KEYS, METERS, BAR_COUNTS, RANGES,
% RHYTHM_VOCABS, LEAP_LIMITS, STRUCTURES).
\begin{table*}[t]
\centering
\caption{The eight constraint families, each decided by a deterministic \texttt{music21} routine on the
parsed score, spanning global, bar, note, transition and cross-section properties. The third
column is the default sampler's domain; the verifier domain $\Theta_f$ is wider, which the Edge
split exploits. Table~\ref{tab:families_full} adds each prompt sentence.}
\label{tab:families}
\small
\setlength{\tabcolsep}{5pt}
\renewcommand{\arraystretch}{1.15}
\rowcolors{2}{famshade}{white}
% No @{} at the outer edges: colortbl paints each cell's background with a
% \tabcolsep overhang on both sides, so with @{} the shading spills 5pt past
% the rules on each side. Keeping the outer separators makes the band land
% exactly on the rules.
\begin{tabularx}{\textwidth}{@{\hspace{\tabcolsep}}>{\raggedright\arraybackslash}p{1.72cm} >{\raggedright\arraybackslash}p{2.1cm} >{\raggedright\arraybackslash}X}
\toprule
\rowcolor{white}
\textbf{Family} & \textbf{Checked over} & \textbf{Default sampler values} \\
\midrule
\textcolor{famname}{\texttt{\textbf{key}}} & global & 13 keys: 8 major, 5 minor \\
\textcolor{famname}{\texttt{\textbf{meter}}} & every bar & $4/4$, $3/4$, $6/8$, $2/4$; each bar must sum to the signature \\
\textcolor{famname}{\texttt{\textbf{length}}} & global & $8$, $12$ or $16$ bars, no pickup \\
\textcolor{famname}{\texttt{\textbf{range}}} & every note & 4 inclusive pitch spans, e.g.\ G3--A5 \\
\textcolor{famname}{\texttt{\textbf{final}}} & last note & a pitch class, any octave (a scale degree 1/3/5 when a key is present) \\
\textcolor{famname}{\texttt{\textbf{rhythm}}} & every note & 4 admissible duration sets; no dotted values, no rests \\
\textcolor{famname}{\texttt{\textbf{leap}}} & adjacent notes & largest melodic interval: $7$, $9$ or $12$ semitones \\
\textcolor{famname}{\texttt{\textbf{structure}}} & 4-bar units & \textsc{aaba}, \textsc{aabb} or \textsc{abab} over $16$ bars; equal letters identical, different letters distinct \\
\bottomrule
\end{tabularx}
\end{table*}

% >>> tables/families

\paragraph{Verifying properties at multiple musical scales.}
Table~\ref{tab:families} lists the eight constraint families. Seven constrain one attribute
each, such as key, meter or pitch range; the eighth, \texttt{structure}, is a cross-section form
constraint: a sixteen-bar tune in a stated form over
four-bar units (\textsc{AABA}, \textsc{AABB}, \textsc{ABAB}), units sharing a letter identical
note-for-note and units with different letters differing. A unit $u_a$ ($a \in \{1,\dots,4\}$) is
its sequence of $M = 4$ bar signatures (pitch and duration per event); a tune without exactly
$4M$ bars scores $0$ on every structure term. Writing $E$ for the index pairs required equal,
$D$ for those required different (minimal sets per form in Appendix~\ref{app:verifiers}) and
$\mathbb{I}[\cdot]$ for an indicator,
\begin{equation}
\begin{aligned}
v_{\texttt{structure}} = {}&
\mathbb{I}\!\left[\textstyle\bigwedge_{(a,b) \in E} u_a = u_b\right]\\
&\cdot\,\mathbb{I}\!\left[\textstyle\bigwedge_{(a,b) \in D} u_a \neq u_b\right].
\end{aligned}
\label{eq:structure}
\end{equation}
We write $v_{\text{rep}}$ and $v_{\text{contrast}}$ for the two factors, the \emph{repetition}
and \emph{contrast} halves. Their
conjunction is zero whenever either fails, however close the output is. Writing $u_a[m]$ for the
$m$-th bar signature of unit $u_a$, the graded repetition credit
\begin{equation}
\bar{e} = \frac{1}{|E| M}\sum_{(a,b) \in E} \sum_{m=1}^{M}
\mathbb{I}\bigl[u_a[m] = u_b[m]\bigr]
\label{eq:gradedstruct}
\end{equation}
counts how much of the required repetition is already in place rather than whether all of it is.
The verifiers are validated without annotators: all $32$ constructed scores with known property
values, covering the eight families, receive the correct verdict; six single-property mutations
of a fully compliant tune each flip exactly the targeted family; and the meter check agrees with
the declared meter of all $400$ IrishMAN tunes tested (Appendix~\ref{app:verifiers}).

\paragraph{Calibrating the key criterion.}
The \texttt{key} verifier is the one hybrid among the eight, because a literal check fails in
both directions: the \texttt{K:} header is gameable, and key-profile analysis
\citep{krumhansl2001cognitive,temperley1999s} (Aarden--Essen profiles) agrees with the corpus
header on only $0.750$ of $400$ corpus tunes. We accept a tune iff the header matches \emph{and}
the analysed key is the requested one or its relative (Appendix~\ref{app:keycal}).

\subsection{Gated Graded Reward}
\label{sec:reward}

In this section, we present the reward that turns verifier verdicts into a learning signal.

\paragraph{Binary all-satisfied reward and its limitations.}
For a prompt $x_C$, GRPO samples a group of $G$ rollouts $\{y_1, \dots, y_G\}$ and centres their
rewards within that group,
\begin{equation}
\begin{aligned}
\hat{A}_i &= \frac{R(y_i, C) - \mu_C}{\sigma_C + \epsilon},
\quad \mu_C = \frac{1}{G}\sum_{j=1}^{G} R(y_j, C),\\
\sigma_C^2 &= \frac{1}{G-1}\sum_{j=1}^{G} \bigl(R(y_j, C) - \mu_C\bigr)^2 ,
\end{aligned}
\label{eq:grpoadv}
\end{equation}
with $\epsilon = 10^{-4}$ a small constant. Despite matching the evaluation criterion exactly,
taking $R = A$ (Equation~\ref{eq:allsat}) has two limitations: (1) it separates no partial
ordering, as a rollout that misses one constraint scores the same as one that misses all; and
(2) such ties remove the learning signal, since a group whose rollouts all score alike
contributes only the Kullback--Leibler (KL) term (Appendix~\ref{app:zerostd}).

\paragraph{Addressing the limitations by graded per-family credit.}
To address the aforementioned limitations, we grade the reward per requested family and keep an
all-satisfied bonus. In detail, for a completion $y$ on problem $C$ with $k = |C|$,
\begin{equation}
R(y, C) =
\begin{cases}
0, & \rho(y) = \bot,\\[4pt]
\lambda\,\dfrac{1}{k}\displaystyle\sum_{(f,\theta) \in C} s_f(\rho(y), \theta) & \\
\quad +\,(1-\lambda)\, A(y, C), & \text{else},
\end{cases}
\label{eq:reward}
\end{equation}
with $\lambda = 0.7$, which leaves the binary term $0.3$ of the reward even at $k=1$
(\S\ref{sec:a4} and Appendix~\ref{app:lambda} vary it).
Here $s_f: \mathcal{S} \times \Theta_f \rightarrow [0,1]$ equals $v_f$ for the seven non-structure
families; for \texttt{structure} it is graded,
$s_{\texttt{structure}} = 0.7\,\bar{e} + 0.3\,v_{\text{contrast}}$ (a fixed split, separate from
$\lambda$). The $\lambda$-weighted average separates rollouts satisfying different \emph{numbers
of families}; holding the contrast verdict fixed, the graded $s_{\texttt{structure}}$ also
separates rollouts that fail the form check but match different fractions of bar positions,
which is credit \emph{within} a family. We grade repetition and keep contrast binary because exact
bar matches measure partial repetition directly (a symmetric variant is in
Appendix~\ref{app:structvariants}).

\paragraph{Guarding against fragments by a hard validation gate.}
However, paying partial credit to every output would let a fragment satisfying \texttt{key} and
\texttt{range}, but forming no tune, collect $\lambda \cdot 2/k$
\citep[cf.][]{skalse2022defining}. As a remedy, the gate scores every rejected output $0$ before
any credit is assigned. Training and main evaluation share the gate; under the stricter format checks
of Appendix~\ref{app:strictaudit}, the matched checkpoints of Table~\ref{tab:main} keep every
success.

\subsection{Training}
\label{sec:training}

In this section, we describe GRPO training with the reward of
Equation~\ref{eq:reward}, optionally preceded by music-domain SFT. \emph{Direct} applies (1) to Qwen3-4B-Instruct-2507, whereas \emph{SFT init.} applies (2)
before (1); both use the same reward, prompt pool and GRPO settings, with the initial policy as
the KL reference.

\paragraph{Optimising with the verifier as the only source of task supervision.}
In detail, both routes use GRPO \citep{shao2024deepseekmath}: every rollout of a group is first
scored by Equation~\ref{eq:reward}, its group-relative advantage (Equation~\ref{eq:grpoadv}) is
then assigned to every token, and the policy finally maximises the PPO clipped surrogate
\citep{schulman2017proximal} with a per-token KL penalty $\beta$ towards the route's initial
checkpoint (Equation~\ref{eq:grpo} in Appendix~\ref{app:algorithm}). No value network is
trained; the group mean $\mu_C$ is the baseline. We run
$G = 8$, $\beta = 0.04$, clipping range $\varepsilon_{\mathrm{clip}} = 0.2$, learning rate $1{\times}10^{-6}$, $300$ optimiser steps on $6{,}000$
synthetic prompts, with Qwen3-4B-Instruct-2507 \citep{yang2025qwen3} trained full-parameter (\S\ref{app:infra}). The prompt pool upsamples
structure-containing problems from $21\%$ to $33\%$ (\S\ref{sec:a4} compares both), and
Table~\ref{tab:hyperparams} gives all settings.

\paragraph{Warm-starting on targets that the verifiers label themselves.}
The optional SFT uses $12{,}890$ instruction--tune pairs self-labelled from the IrishMAN
corpus \citep{wu2023tunesformer}: for each corpus tune the verifiers read candidate constraint
parameters off the parsed score, sample $k \sim \mathcal{U}\{1,2,3\}$ families, and admit the
pair only if $A(y, C(y)) = 1$. Held-out family pairs (\S\ref{sec:bench}) are excluded from both
training pools. Full construction details are in Appendices~\ref{app:sftdata}--\ref{app:prompt}.

\section{\bench{}}
\label{sec:bench}

\begin{figure}[t]
\centering
\includegraphics[width=\columnwidth]{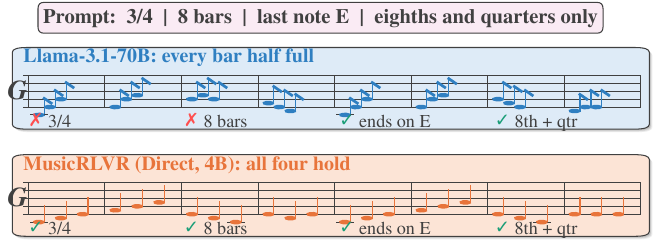}
\caption{One High-$k$ item, the first that Direct \method{} solves and Qwen3-32B,
Llama-3.1-70B and MiniMax-M3 all fail. Llama-3.1-70B fills half of every $3/4$ bar.}
\label{fig:staff}
\end{figure}

In this section, we introduce \bench{}, which comprises five complementary tests: \emph{constraint density} (Single, Mixed,
High-$k$), \emph{compositional generalisation} (Seen, Unseen, Dense), \emph{parameter
extrapolation} (Edge), \emph{multi-section control} (Structure), and \emph{linguistic variation}
(paraphrases). Single ($n=200$) contains one constraint, High-$k$ ($n=160$) four, and Mixed
($n=300$) spans $k\in\{1,2,3,4\}$ and serves as the per-family diagnostic. Seen ($n=200$)
contains two or three families that co-occur in training, whereas Unseen ($n=200$) requires one
of four held-out pairs; Structure ($n=120$) always includes the cross-section form constraint.
Dense ($n=500$; four constraints with a held-out pair) and Edge ($n=500$; parameters outside the
RL prompt domain, \S\ref{sec:a3}) complete the $\PoolN$ items; both were
added after the reward design was final. The held-out pairs are (\texttt{key}, \texttt{rhythm}), (\texttt{meter},
\texttt{leap}), (\texttt{final}, \texttt{structure}), and (\texttt{length}, \texttt{range});
none occurs together in either training stage \citep{lake2018generalization,kim2020cogs}.
They form a perfect matching over the eight families, so each family is held out once and any
held-out-pair-free set has at most $k=4$ (Appendix~\ref{app:heldout}). Finally, re-rendering all
Single and Seen items with alternative wording yields a $400$-item paraphrase suite with unchanged
constraint values (Appendix~\ref{app:benchdetails}). Figure~\ref{fig:staff} shows one High-$k$
item that Llama-3.1-70B fails and Direct \method{} solves.

\begin{table*}[t]
\centering
\captionsetup{skip=2pt}
\caption{All-satisfied rate under greedy decoding on the eight \bench{} suites and over all $2{,}180$
items. Qwen3-4B is Qwen3-4B-Instruct-2507; the 4B RL rows are matched runs at equal budget, and RFT (rejection-sampling fine-tuning)
fine-tunes Qwen3-4B on its own verifier-accepted samples at that budget (Appendix~\ref{app:rft}). Each
column marks its {\fboxsep=1.2pt\colorbox{Best}{best}} and {\fboxsep=1.2pt\colorbox{Secondbest}{runner-up}};
the {\fboxsep=1.2pt\colorbox{Ours}{shaded row}} is our main configuration. Counts and paired tests:
Appendix~\ref{app:matched}.}
\label{tab:main}
\footnotesize
\renewcommand{\arraystretch}{1.00}
\setlength{\aboverulesep}{1.5pt}\setlength{\belowrulesep}{1.5pt}
% Full text width, so the table edges line up with both columns; equal data columns.
\setlength{\tabcolsep}{3pt}
\begin{tabularx}{\textwidth}{l*{9}{>{\centering\arraybackslash}X}}
\toprule
& \multicolumn{6}{c}{Six original splits} & \multicolumn{2}{c}{Later splits} & Overall \\
\cmidrule(lr){2-7}\cmidrule(lr){8-9}\cmidrule(lr){10-10}
Model & Mixed & Single & Seen & Unseen & High-$k$ & Structure & Dense & Edge & All \\
\midrule
\rowcolor{famname!8}\multicolumn{10}{l}{\textcolor{famname}{\textit{Zero-shot}}} \\
Qwen3-4B & 0.160 & 0.225 & 0.100 & 0.040 & 0.037 & 0.100 & 0.008 & 0.010 & 0.068 \\
Qwen3-8B & 0.123 & 0.210 & 0.115 & 0.050 & 0.006 & 0.000 & 0.020 & 0.032 & 0.064 \\
Qwen3-32B & 0.363 & 0.535 & 0.250 & 0.190 & 0.094 & 0.267 & 0.074 & 0.112 & 0.204 \\
Llama-3.1-70B & 0.380 & 0.695 & 0.300 & 0.195 & 0.056 & 0.167 & 0.040 & 0.058 & 0.197 \\
MiniMax-M3 & 0.367 & 0.580 & 0.315 & 0.190 & 0.100 & 0.008 & 0.088 & 0.160 & 0.215 \\
ChatMusician-7B & 0.160 & 0.290 & 0.130 & 0.090 & 0.006 & 0.000 & 0.020 & 0.046 & 0.084 \\
\midrule
\rowcolor{famname!8}\multicolumn{10}{l}{\textcolor{famname}{\textit{Trained (Qwen3-4B-Instruct-2507)}}} \\
SFT & 0.477 & 0.745 & 0.415 & 0.440 & 0.231 & 0.600 & 0.290 & 0.076 & 0.346 \\
RFT & 0.543 & \cellcolor{Secondbest}\underline{0.960} & 0.445 & 0.355 & 0.163 & 0.758 & 0.098 & 0.128 & 0.342 \\
\rowcolor{black!6}
Direct GRPO ($R=A$) & \cellcolor{Secondbest}\underline{0.797} & \cellcolor{Secondbest}\underline{0.960} & \cellcolor{Secondbest}\underline{0.745} & 0.565 & \cellcolor{Secondbest}\underline{0.469} & \cellcolor{Secondbest}\underline{0.908} & 0.322 & \cellcolor{Secondbest}\underline{0.300} & \cellcolor{Secondbest}\underline{0.545} \\
\rowcolor{black!6}
\method{} (SFT init.) & 0.577 & 0.880 & 0.585 & 0.530 & 0.350 & 0.383 & \cellcolor{Secondbest}\underline{0.366} & 0.134 & 0.424 \\
\rowcolor{Ours}
\method{} (Direct) & \cellcolor{Best}\textbf{0.807} & \cellcolor{Best}\textbf{0.985} & \cellcolor{Best}\textbf{0.815} & \cellcolor{Secondbest}\underline{0.595} & \cellcolor{Best}\textbf{0.575} & \cellcolor{Best}\textbf{0.967} & 0.362 & \cellcolor{Best}\textbf{0.310} & \cellcolor{Best}\textbf{0.580} \\
\midrule
\rowcolor{famname!8}\multicolumn{10}{l}{\textcolor{famname}{\textit{Trained (Qwen3-8B)}}} \\
\rowcolor{black!6}
\method{} (Direct) & 0.657 & 0.915 & 0.705 & \cellcolor{Best}\textbf{0.645} & 0.431 & 0.900 & \cellcolor{Best}\textbf{0.404} & 0.252 & 0.530 \\
\bottomrule
\end{tabularx}
\end{table*}

\section{Experiments}
\label{sec:results}

In this section, we carry out experiments to address the following questions:
\begin{itemize}[leftmargin=1.2em,itemsep=0pt,topsep=2pt,parsep=0pt]
\item \textbf{Q1}: Do existing models comply? See \S\ref{sec:a1}.
\item \textbf{Q2}: Is music-domain SFT needed? See \S\ref{sec:a2}.
\item \textbf{Q3}: Does the gain generalise? See \S\ref{sec:a3}.
\item \textbf{Q4}: Which reward parts matter? See \S\ref{sec:a4}.
\item \textbf{Q5}: Where are the gains largest? See \S\ref{sec:a5}.
\end{itemize}
Additional results can be found in Appendix~\ref{app:addexp}.

\subsection{Experiment Setup}

To answer the above five questions, we score greedy completions with the all-satisfied
criterion of \S\ref{sec:method} (temperature $1.0$ for the sampling-budget study).
\textbf{Baselines.} Qwen3-4B-Instruct-2507, Qwen3-8B and Qwen3-32B \citep{yang2025qwen3},
Llama-3.1-70B \citep{grattafiori2024llama}, MiniMax-M3, and ChatMusician-7B
\citep{yuan2024chatmusician} are evaluated using the same task prompts;
the music-domain SFT checkpoint serves as a supervised baseline
(Appendix~\ref{app:sftexamples}).
\textbf{Training runs.} The three \emph{matched runs}---Direct
GRPO ($R=A$, i.e.\ $\lambda=0$), Direct \method{}, and SFT-initialised \method{}---share seed
$43$, $300$ updates, and $48$ completions per update, and are evaluated once per suite
(paired McNemar tests \citep{mcnemar1947note}: Appendix~\ref{app:matched}).
Reward ablations (\S\ref{sec:a4}) start from the Direct
configuration; sensitivity plots use Direct runs; SFT-initialised diagnostics
remain in Appendix~\ref{app:sweeps}.

\subsection{A1: Existing Models Struggle with Multi-Constraint Music Instructions}
\label{sec:a1}

We assess the zero-shot models' constraint compliance on \bench{}. As shown in Table~\ref{tab:main}
and Figure~\ref{fig:degrade}, we observe that (1) high parseability does not imply joint compliance: Llama-3.1-70B writes a parseable score for $\LlamaParse$ of Mixed items
yet satisfies every requested constraint on only $\Llamakilltestsuite$, with Qwen3-32B and
MiniMax-M3 alongside it at $\BaseThirtytwokilltestsuite$ and $\Minimaxkilltestsuite$; (2) neither
size nor music specialisation sets the ranking: Qwen3-8B reaches $\BaseEightkilltestsuite$ on
Mixed, below the $\Basekilltestsuite$ of the smaller Qwen3-4B, and the music-specialised
ChatMusician-7B reaches only $\Chatmusiciankilltestsuite$; (3) compliance is lost as constraints
accumulate: Llama-3.1-70B falls $14.2{\times}$ from one to four constraints, from $0.630$ to $0.044$, whereas Direct \method{} falls $2.4{\times}$, from $0.992$
to $0.419$ (Appendix~\ref{app:byk}).

\begin{figure*}[t]
\centering
\includegraphics[width=\textwidth]{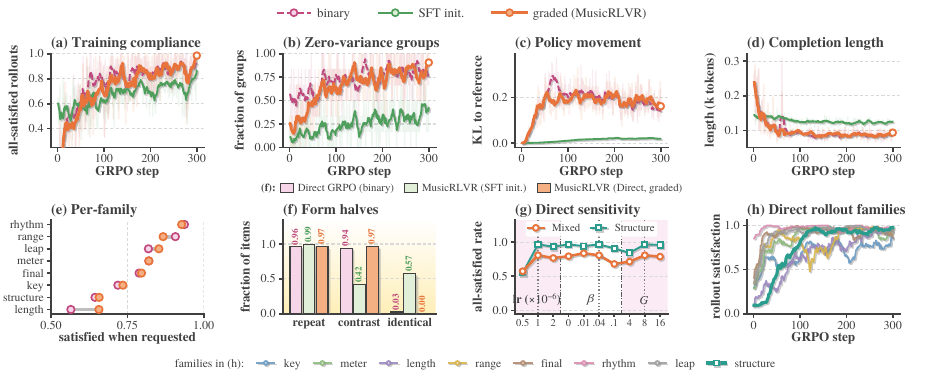}
\caption{\textbf{Training dynamics.} (a--f) Matched runs of Table~\ref{tab:main}: ``binary'' and
``graded'' are the two Direct arms ($\lambda=0$ and $0.7$), ``SFT init.'' starts from the SFT
checkpoint. \textbf{(a)} All-satisfied rollouts. \textbf{(b)} Zero-variance groups. \textbf{(c)}
KL. \textbf{(d)} Completion length. \textbf{(e)} Per-family satisfaction. \textbf{(f)}
Form-verifier components. (g--h) Direct analyses: \textbf{(g)} sensitivity to learning rate,
$\beta$ and $G$ (dotted: reference); \textbf{(h)} per-family rollout satisfaction.}
\label{fig:matched}
\end{figure*}

\subsection{A2: \method{} Teaches Constraint Following Without Music-Domain SFT}
\label{sec:a2}

To test whether music-domain SFT is necessary, we train
Qwen3-4B-Instruct-2507 directly with \method{}. As shown in Table~\ref{tab:main}, we observe that (1) Direct \method{} raises
the Mixed all-satisfied rate from $\Basekilltestsuite$ to
$\MatchedDirectMixed$ and surpasses every zero-shot baseline of Table~\ref{tab:main} on all eight suites, reaching
$\MatchedDirectSingle$ on Single, $\MatchedDirectStructure$ on Structure and
$\MatchedDirectHighk$ on High-$k$ against a best zero-shot of merely $0.100$, with every
completion clearing the validation gate on the six original suites. Item by item, it satisfies $609$ of the $2{,}180$ items that Qwen3-32B, Llama-3.1-70B
and MiniMax-M3 all fail, against $5$ the other way round (Appendix~\ref{app:qualitative}).
(2) Music-domain SFT is not a prerequisite: under the same reward, prompt pool and budget,
the SFT-initialised route reaches $\MatchedWarmMixed$ on Mixed against
$\MatchedDirectMixed$ for Direct and trails on seven of eight suites;
% the discordant total is derived, not a separate macro: \numexpr keeps it in step with the
% two win counts when inline/matched_extra_stats.tex is regenerated.
pooled over the $\PoolN$ items they disagree on
$\the\numexpr\PoolGradedWarmWins+\PoolWarmWins\relax$,
$\PoolGradedWarmWins$ in Direct's favour ($p\PoolGradedWarmP$), and Direct stays significantly
ahead on six of the eight suites after Holm correction (Appendix~\ref{app:matched}). (3) Compared with binary-reward Direct GRPO, graded \method{}
further improves performance on all eight suites, notably
High-$k$ ($0.469\rightarrow0.575$), Seen ($0.745\rightarrow0.815$)
and Structure ($0.908\rightarrow0.967$), while raising the
overall all-satisfied rate from $0.545$ to $0.580$
(Table~\ref{tab:main}; Appendix~\ref{app:matched}).

\begin{table}[t]
\centering
\caption{One greedy decode against the best of eight samples reranked by the verifier
(temperature $1.0$; Appendix~\ref{app:passk}). Each row marks its
{\fboxsep=1.2pt\colorbox{Best}{best}} and {\fboxsep=1.2pt\colorbox{Secondbest}{runner-up}}.}
\label{tab:rerank8}
\footnotesize
\setlength{\tabcolsep}{4pt}
% <<< tables/rerank8_table (make_rerank8_table.py)
\begin{tabular}{lcccc}
\toprule
& \multicolumn{2}{c}{Qwen3-4B (zero-shot)} & \multicolumn{2}{c}{\method{} (Direct)} \\
\cmidrule(lr){2-3}\cmidrule(lr){4-5}
Suite & Greedy & Best of 8 & Greedy & Best of 8 \\
\midrule
Mixed & 0.160 & 0.393 & \cellcolor{Secondbest}\underline{0.807} & \cellcolor{Best}\textbf{0.897} \\
Single & 0.225 & 0.685 & \cellcolor{Secondbest}\underline{0.985} & \cellcolor{Best}\textbf{1.000} \\
Seen & 0.100 & 0.330 & \cellcolor{Secondbest}\underline{0.815} & \cellcolor{Best}\textbf{0.875} \\
Unseen & 0.040 & 0.180 & \cellcolor{Secondbest}\underline{0.595} & \cellcolor{Best}\textbf{0.750} \\
High-$k$ & 0.037 & 0.050 & \cellcolor{Secondbest}\underline{0.575} & \cellcolor{Best}\textbf{0.756} \\
Structure & 0.100 & 0.367 & \cellcolor{Secondbest}\underline{0.967} & \cellcolor{Best}\textbf{0.975} \\
Dense & 0.008 & 0.052 & \cellcolor{Secondbest}\underline{0.362} & \cellcolor{Best}\textbf{0.624} \\
Edge & 0.010 & 0.090 & \cellcolor{Secondbest}\underline{0.310} & \cellcolor{Best}\textbf{0.448} \\
\bottomrule
\end{tabular}
% >>> tables/rerank8_table
\end{table}

\begin{figure}[t]
\centering
\includegraphics[width=\columnwidth]{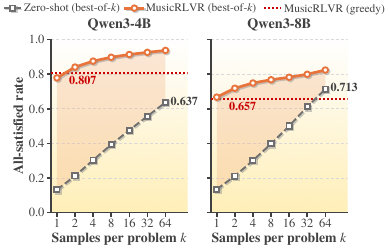}
\caption{Best-of-$k$ verifier reranking on Mixed ($64$ samples at temperature $1.0$). The dotted
line is one greedy decode of the trained policy.}
\label{fig:budget}
\end{figure}

Neither self-distillation nor inference-time selection substitutes for training. Additional comparisons show that (1) fine-tuning Qwen3-4B on its own verifier-accepted samples, drawn at the
sampling budget of GRPO, reaches $0.543$ on Mixed and trails Direct on all eight suites
(Table~\ref{tab:main}; Appendix~\ref{app:rft}); (2) reranking $64$ samples of the untrained
Qwen3-4B with the verifier reaches $0.637$, below Direct's single greedy decode, and the
untrained Qwen3-8B needs $32$ to $64$ reranked samples to match its trained policy
(Figure~\ref{fig:budget}); (3) reranking eight samples of the trained policy raises every suite
further (Table~\ref{tab:rerank8}), to $0.897$ on Mixed against $0.393$ for the untrained model,
so training and selection are complementary.

\begin{table}[t]
\centering
\caption{Accuracy (\%) before and after training on IFBench \citep{pyatkin2026generalizing},
MMLU-Redux \citep{gema2025we} and GSM8K \citep{cobbe2021training}. No difference is significant
(paired exact McNemar, $p \ge 0.096$). Shading as in Table~\ref{tab:main}.}
\label{tab:generalmain}
\footnotesize
\setlength{\tabcolsep}{3pt}
\begin{adjustbox}{max width=\columnwidth}
% <<< tables/general_main (make_robustness_tables.py)
\begin{tabular}{lcccc}
\toprule
& \multicolumn{2}{c}{IFBench} & & \\
\cmidrule(lr){2-3}
Model & strict & loose & MMLU-Redux & GSM8K \\
\midrule
Qwen3-4B (zero-shot) & 32.0 & 33.3 & 83.9 & 93.9 \\
\rowcolor{Ours}
\method{} (Direct) & 29.3 & 31.7 & 84.4 & 94.1 \\
\midrule
Qwen3-8B (zero-shot) & 25.7 & 30.3 & 81.5 & 93.0 \\
\rowcolor{black!6}
\method{} (Direct, Qwen3-8B) & 25.3 & 30.0 & 82.6 & 93.3 \\
\bottomrule
\end{tabular}
% >>> tables/general_main
\end{adjustbox}
\end{table}

The recipe also carries over to Qwen3-8B (Appendix~\ref{app:qwen8b}), which rises from $0.123$
to $0.657$ on Mixed and posts the highest Unseen and Dense rates in Table~\ref{tab:main}. As
shown in Table~\ref{tab:generalmain}, neither policy trades away general capability
(Appendix~\ref{app:general}).

\subsection{A3: \method{} Generalises Beyond the Training Distribution}
\label{sec:a3}

\begin{table}[t]
\centering
\caption{All-satisfied rate beyond the training range of $k$ (at most $3$). $k=4$ is Dense;
$k=5$--$7$ hold $300$ items each, every one with a held-out pair. Colours as in
Table~\ref{tab:main}.}
\label{tab:kladder}
\footnotesize
\setlength{\tabcolsep}{4pt}
\begin{adjustbox}{max width=\columnwidth}
% <<< tables/kladder_main (make_kladder_tables.py)
\begin{tabular}{lcccc}
\toprule
Model & $k=4$ & $k=5$ & $k=6$ & $k=7$ \\
\midrule
Qwen3-4B (zero-shot) & 0.008 & 0.003 & 0.000 & 0.000 \\
SFT & 0.290 & 0.140 & 0.117 & 0.073 \\
RFT & 0.098 & 0.053 & 0.007 & 0.003 \\
Direct GRPO ($R=A$) & 0.322 & 0.170 & 0.060 & 0.063 \\
\method{} (SFT init.) & \cellcolor{Secondbest}\underline{0.366} & 0.187 & \cellcolor{Secondbest}\underline{0.200} & \cellcolor{Secondbest}\underline{0.117} \\
\method{} (Direct) & 0.362 & \cellcolor{Secondbest}\underline{0.230} & 0.160 & 0.087 \\
\method{} (Direct, Qwen3-8B) & \cellcolor{Best}\textbf{0.404} & \cellcolor{Best}\textbf{0.273} & \cellcolor{Best}\textbf{0.250} & \cellcolor{Best}\textbf{0.153} \\
\bottomrule
\end{tabular}
% >>> tables/kladder_main
\end{adjustbox}
\end{table}

\begin{table*}[t]
\centering
\captionsetup{skip=2pt}
\caption{Reward ablations in the Direct configuration. Each row below the default changes one
part of Equation~\ref{eq:reward}, keeping the seed, budget, prompts (except the last row) and
evaluation of Table~\ref{tab:main}; $\lambda=0$ is Direct GRPO ($R=A$). Details:
Appendix~\ref{app:directabl}. Colours as in Table~\ref{tab:main}.}
\label{tab:ablation}
\footnotesize
\renewcommand{\arraystretch}{1.00}
\setlength{\aboverulesep}{1.5pt}\setlength{\belowrulesep}{1.5pt}
\setlength{\tabcolsep}{3pt}
% <<< tables/ablation_direct_main (make_ablation_direct_tables.py)
\begin{tabularx}{\textwidth}{l*{9}{>{\centering\arraybackslash}X}}
\toprule
Variant & Mixed & Single & Seen & Unseen & High-$k$ & Structure & Dense & Edge & All \\
\midrule
\rowcolor{Ours}
\method{} (Direct) & \cellcolor{Best}\textbf{0.807} & \cellcolor{Best}\textbf{0.985} & \cellcolor{Best}\textbf{0.815} & \cellcolor{Secondbest}\underline{0.595} & \cellcolor{Best}\textbf{0.575} & \cellcolor{Best}\textbf{0.967} & \cellcolor{Best}\textbf{0.362} & \cellcolor{Best}\textbf{0.310} & \cellcolor{Best}\textbf{0.580} \\
\midrule
Gate on $A$ only & 0.763 & 0.975 & 0.770 & 0.575 & 0.469 & 0.867 & 0.346 & 0.238 & 0.534 \\
\midrule
$\lambda=0$ ($R=A$) & \cellcolor{Secondbest}\underline{0.797} & 0.960 & 0.745 & 0.565 & 0.469 & 0.908 & 0.322 & 0.300 & 0.545 \\
$\lambda=0.5$ & 0.703 & \cellcolor{Best}\textbf{0.985} & 0.665 & 0.545 & 0.406 & 0.883 & 0.246 & \cellcolor{Secondbest}\underline{0.302} & 0.502 \\
$\lambda=1$ (no bonus) & 0.753 & 0.950 & 0.735 & \cellcolor{Best}\textbf{0.635} & \cellcolor{Secondbest}\underline{0.494} & 0.908 & 0.358 & 0.298 & \cellcolor{Secondbest}\underline{0.553} \\
\midrule
Binary form credit & 0.777 & \cellcolor{Secondbest}\underline{0.980} & \cellcolor{Secondbest}\underline{0.785} & 0.590 & 0.481 & 0.900 & \cellcolor{Secondbest}\underline{0.360} & 0.242 & 0.546 \\
Symmetric form credit & 0.757 & 0.950 & 0.760 & 0.500 & 0.394 & \cellcolor{Secondbest}\underline{0.942} & 0.282 & 0.272 & 0.515 \\
Equal form weights & 0.753 & 0.940 & 0.740 & 0.570 & 0.475 & \cellcolor{Secondbest}\underline{0.942} & 0.352 & 0.290 & 0.544 \\
\midrule
Form share $21\%$ & 0.753 & \cellcolor{Secondbest}\underline{0.980} & 0.740 & 0.550 & 0.381 & \cellcolor{Best}\textbf{0.967} & 0.270 & 0.276 & 0.518 \\
\bottomrule
\end{tabularx}
% >>> tables/ablation_direct_main
\end{table*}

We next assess whether the trained policy generalises beyond its training distribution. As shown in
Tables~\ref{tab:main} and~\ref{tab:kladder}, we observe that (1) the gain transfers to held-out
pairs. Every Unseen item requires one of the four family pairs
excluded from both training pools; Direct reaches $\MatchedDirectUnseen$ there, three times the
best zero-shot rate of $0.195$ (Appendix~\ref{app:coverage}).

(2) It extends to new parameter values. Edge moves meters to $5/4$, $2/2$, $9/8$, $12/8$ and lengths to
$4$, $24$, $32$ bars and extends leap, range and rhythm (Table~\ref{tab:edgespec}). Direct scores $\MatchedDirectEdge$ against the
strongest zero-shot $0.160$. On items whose meter or length no RL
prompt requests, Direct reaches $0.287$ and $0.181$, against at most $0.138$ and $0.142$
for any zero-shot model (Table~\ref{tab:edgefactors}).

(3) It extends to more constraints than trained. Training prompts request at most three constraints
(Appendix~\ref{app:datastats}), so every High-$k$ and Dense item already asks for one more;
Direct reaches $\MatchedDirectHighk$ and $\MatchedDirectDense$ there. Suites at $k=5$, $6$ and
$7$ go further (Appendix~\ref{app:kladder}): the untrained Qwen3-4B satisfies $1$ of their $900$
items, whereas Direct \method{} keeps $0.230$, $0.160$ and $0.087$, and the Qwen3-8B policy
$0.273$, $0.250$ and $0.153$, the highest at every rung (Table~\ref{tab:kladder}).

(4) It extends to free-form wording. On $60$ free-form requests that phrase the constraints as
a user might (Appendix~\ref{app:freeform}), Direct reaches $0.583$, the highest of the 4B models
(Qwen3-4B: $0.067$), and the Qwen3-8B policy reaches $0.717$.

\subsection{A4: Every Reward Component Helps}
\label{sec:a4}

To assess reward components, we ablate Equation~\ref{eq:reward} in the Direct
configuration, one at a time. As shown in Table~\ref{tab:ablation}, every change lowers the rate over all $2{,}180$ items from $0.580$ to $0.502$--$0.553$
(paired exact McNemar, Holm-adjusted $p \le 0.002$). The default is higher on $61$ of the $64$
variant--suite pairs, and every variant loses most on High-$k$, $0.381$--$0.494$ against $0.575$,
where at least four families must hold at once. More specifically, (1) crediting outputs
that the gate rejects but music21 parses lowers all eight suites, Structure by $0.100$ and
High-$k$ by $0.106$; (2) dropping per-family credit ($\lambda=0$) or the all-satisfied bonus ($\lambda=1$)
lowers High-$k$ to $0.469$ and $0.494$;
(3) graded form credit lifts Structure to $0.967$ from $0.900$--$0.942$ under binary, symmetric
and equal-weight form rewards, and training with $21\%$ form prompts gives the lowest High-$k$
rate, $0.381$, even though that pool holds more three-constraint prompts.

Beyond these ablations, we find that (1) graded credit cuts groups without a reward gradient from $\DeadBinaryEarly$ to $\DeadGradedEarly$ over the first $50$ updates and
from $\DeadBinaryAll$ to $\DeadGradedAll$ over all $300$ (Figure~\ref{fig:matched}(b)); (2) in the
matched runs, with each family scored only on the items requesting it, the largest positive gains occur for
\texttt{length} ($\FamLengthBinary\rightarrow\FamLengthGraded$) and
\texttt{leap} ($\FamLeapBinary\rightarrow\FamLeapGraded$),
highlighting improvements in bar-count and melodic-motion control (Figure~\ref{fig:matched}(e); Appendix~\ref{app:familybreak}); (3) Direct sensitivity experiments show that the default learning
rate supports strong performance on Mixed and Structure, with
similarly high scores across tested $\beta$ and $G$ values
(Figure~\ref{fig:matched}(g)); rollout satisfaction improves
across all eight families (Figure~\ref{fig:matched}(h)).

\subsection{A5: The Largest Gains Are on Counting and Structure}
\label{sec:a5}

Examining per-family satisfaction across all $2{,}180$ items
(Appendix~\ref{app:familybreak}), we find that (1) larger zero-shot models already handle the
tonal families: on \texttt{key} and \texttt{final}, Qwen3-32B and Llama-3.1-70B reach
$0.719$--$0.868$ against $0.242$ and $0.278$ for Qwen3-4B, and Direct \method{} stays close with
$0.735$ and $0.796$; (2) scale falls short on counting and structure: no zero-shot model exceeds
$0.378$ on \texttt{meter}, $0.281$ on \texttt{length} or $0.150$ on \texttt{structure}, whereas
Direct reaches $0.820$, $0.657$ and $0.658$; (3) Direct also far exceeds every zero-shot model on
\texttt{rhythm}, \texttt{range} and \texttt{leap}, $0.853$--$0.928$ against at most $0.671$.

\section{Conclusion and Future Work}

We frame constraint-following music generation as a \emph{property-verifiable generation}
problem and introduce \bench{} and \method{}, using only verifier rewards.
After 300 GRPO updates, Direct \method{} raises Qwen3-4B's Mixed all-satisfied rate from
$\Basekilltestsuite$ to $\MatchedDirectMixed$, surpassing all zero-shot baselines,
including Llama-3.1-70B. Gains generalise beyond training, with no statistically
significant changes on general benchmarks, demonstrating constraint following
without music-domain SFT. For future work, we will extend the verifiers to
polyphony and richer forms.

\section*{Limitations}

Our study is limited to eight constraint families for monophonic ABC in Western tonal
idioms and does not address polyphony or richer musical forms. Within this scope,
we test generalisation to held-out family pairs, out-of-range parameters,
more constraints than any training prompt, and free-form requests
(\S\ref{sec:a3}).

\section*{Ethics Statement}

The study involves no human subjects, no personal data and no user-facing deployment. Supervised
targets derive from the publicly available IrishMAN corpus of traditional Irish music
\citep{wu2023tunesformer}; the reinforcement prompts are produced by deterministic samplers
rather than drawn from a corpus. The intended uses are settings that need a
score to meet a stated specification, such as music education, game audio and accessibility
tools. The misuse we consider most likely is reading a compliance score as a quality score; the
verifiers certify the requested properties. Appendix~\ref{app:impact} gives the longer discussion.

\bibliography{refs}
\appendix
% Appendix only: wide tables are full-width floats here, and the default limit of two
% per page top leaves most appendix pages half empty. The main text keeps the defaults.
\raggedbottom
\setcounter{dbltopnumber}{4}
\renewcommand{\dbltopfraction}{0.92}
\renewcommand{\dblfloatpagefraction}{0.75}
\setcounter{topnumber}{4}
\renewcommand{\topfraction}{0.92}
\renewcommand{\textfraction}{0.06}
\renewcommand{\floatpagefraction}{0.75}
% Float pages start at the top instead of centring their floats, so the tall example
% figures do not sit in the middle of an otherwise empty page.
\makeatletter
\setlength{\@fptop}{0pt}\setlength{\@dblfptop}{0pt}
\makeatother
\clearpage
\begin{center}{\Large\textsc{Contents of Appendix}}\end{center}
\startcontents[appendix]
{\hypersetup{linkcolor=black}\small\printcontents[appendix]{}{1}{}}
\newpage

\section{Algorithm}
\label{app:algorithm}

In this section, we present the detailed algorithm of \method{}. Algorithm~\ref{alg:musicrlvr}
runs verifier-driven GRPO directly or with an optional supervised warm start built from verified corpus-derived labels
($\theta_f(y) := p_f(\rho(y))$, \S\ref{sec:training}). Unless marked Direct, the
diagnostics below describe the SFT-initialised route, retaining a common initialisation
for the reward-design comparisons.

\paragraph{Objective.} For a prompt $x_C$ from the reinforcement pool $\mathcal{D}_{\mathrm{RL}}$,
the behaviour policy $\pi_{\phi_{\text{old}}}$ samples a group $\{y_1, \dots, y_G\}$ and the
advantage $\hat{A}_i$ of Equation~\ref{eq:grpoadv} is assigned to every token of $y_i$. Writing
$y_{i,t}$ for the $t$-th token of $y_i$ and $r_{i,t}(\phi)$ for the token-level likelihood ratio
$\pi_\phi(y_{i,t} \mid x_C, y_{i,<t}) / \pi_{\phi_{\text{old}}}(y_{i,t} \mid x_C, y_{i,<t})$,
GRPO maximises
\begin{equation}
\begin{aligned}
\mathcal{J}(\phi) = \mathbb{E}\Bigl[&\tfrac{1}{G}\textstyle\sum_{i=1}^{G}
\tfrac{1}{|y_i|}\textstyle\sum_{t=1}^{|y_i|}
\Bigl\{\min\bigl(r_{i,t}\hat{A}_i,\\
&\mathrm{clip}(r_{i,t}, 1{-}\varepsilon_{\mathrm{clip}}, 1{+}\varepsilon_{\mathrm{clip}})\hat{A}_i\bigr)\\
&- \beta\, \widehat{K}_{i,t}\Bigr\}\Bigr],
\end{aligned}
\label{eq:grpo}
\end{equation}
where the expectation is over problems $C$ and their sampled groups, $|y_i|$ counts completion
tokens, $\varepsilon_{\mathrm{clip}} = 0.2$ is the clipping range,
$\widehat{K}_{i,t} = e^{d}-d-1$ with $d = \log\pi_{\text{ref}}(y_{i,t}\mid\cdot) - \log\pi_\phi(y_{i,t}\mid\cdot)$
is the per-token KL estimator of \citet{shao2024deepseekmath}, $\pi_{\phi_{\text{old}}}$ is a
detached copy of $\pi_\phi$ (one update per rollout batch), and $\pi_{\text{ref}}$ is the route's
initial checkpoint (hyperparameters in Appendix~\ref{app:hyperparams}).

\begin{algorithm}[t]
\SetAlgoLined
\DontPrintSemicolon
\caption{\method{}: verifier-only GRPO with optional self-labelled SFT}
\label{alg:musicrlvr}
\KwIn{initial policy $\pi_{\phi_0}$; optional SFT flag $u$ and corpus tunes $\mathcal{Y}$; families $\mathcal{F}$ with verifiers $\{v_f\}$ and
parser $\rho$; held-out pairs $H$; held-out retries $N$; reinforcement pool $\mathcal{D}_{\mathrm{RL}}$
(no prompt contains a pair in $H$; structure share $33\%$); reward weight $\lambda$; group size $G$;
$\beta$, $\varepsilon_{\mathrm{clip}}$, $\epsilon$; steps $T$; learning-rate schedule $\eta_s$}
\KwOut{trained policy $\pi_\phi$}
$\pi_{\text{ref}} \leftarrow \pi_{\phi_0}$\;
\If{$u$ is enabled}{
\tcp{Optional self-labelled supervised fine-tuning (\S\ref{sec:training})}
$\mathcal{D} \leftarrow \emptyset$\;
\ForEach{$y \in \mathcal{Y}$}{
  $C(y) \leftarrow$ \textsc{null}; draw $k \sim \mathcal{U}\{1,2,3\}$\;
  \For{attempt $= 1, \dots, N$}{
    sample $F \subseteq \mathcal{F}_y$ (families measurable on $y$), $|F| = \min(k, |\mathcal{F}_y|)$\;
    $C \leftarrow \{(f, \theta_f(y)) : f \in F\}$ with $\theta_f(y) = p_f(\rho(y))$ read off
      $\rho(y)$\;
    \lIf{$C$ contains no pair in $H$}{$C(y) \leftarrow C$; \textbf{break}}
  }
  \If{$C(y) = $ \textsc{null}}{
    $C(y) \leftarrow$ one constraint drawn from $y$'s properties\tcp*{singleton: trivially held-out-free}
  }
  \If{$A(y, C(y)) = 1$}{
    $\mathcal{D} \leftarrow \mathcal{D} \cup \{(x_{C(y)}, y)\}$\tcp*{else this draw of $y$ is discarded}
  }
  \tcp{with prob.\ $0.2$: a structure-only pair from $y$'s first two 4-bar units, re-verified the same way}
}
$\pi_{\text{ref}} \leftarrow$ fine-tune $\pi_{\phi_0}$ on $\mathcal{D}$\;
}
\end{algorithm}

\begin{algorithm}[t]
\SetAlgoLined
\DontPrintSemicolon
\addtocounter{algocf}{-1}
\caption{\method{} (continued): verifier-driven GRPO}
$\pi_\phi \leftarrow \pi_{\text{ref}}$\;
\tcp{Verifier-driven GRPO (\S\ref{sec:training})}
\For{step $s = 1, \dots, T$}{
  $\phi_{\text{old}} \leftarrow \phi$ (detached)\;
  sample a batch of prompts $x_C$, $C \sim \mathcal{D}_{\mathrm{RL}}$\;
  \ForEach{$x_C$ in the batch}{
    sample a group $\{y_1, \dots, y_G\} \sim \pi_{\phi_{\text{old}}}(\cdot \mid x_C)$\;
    score every $y_i$ by $R(y_i, C)$ (Equation~\ref{eq:reward})\;
    $\hat{A}_i \leftarrow \bigl(R(y_i, C) - \mu_C\bigr) / (\sigma_C + \epsilon)$ for every $i$
      (Equation~\ref{eq:grpoadv})\tcp*{$\hat A_i = 0$ for every $i$ if $\sigma_C = 0$}
  }
  accumulate $\nabla_\phi(-\mathcal{J}(\phi))$ over the batch, clip its norm to $1.0$, AdamW step at $\eta_s$\tcp*{Equation~\ref{eq:grpo}}
}
\end{algorithm}

\section{Experimental Settings}
\label{app:settings}

In this section, we present the detailed experimental settings, including the infrastructure,
hyper-parameters, SFT data, prompt template and dataset statistics.

\subsection{Infrastructure}
\label{app:infra}
In this section, we describe the training stack, which uses ms-swift's GRPO implementation with a
custom reward plugin that shells out to the evaluation verifiers, so training and benchmarking
share one \texttt{music21} build. On these accelerators, co-locating the vLLM rollout engine
with the training state is infeasible for full-parameter $4$B training: the sleeping
engine's memory pool still counts against the training process and the backward pass
fails at a fixed allocation point regardless of batch size. Serving rollouts from a
dedicated device solves it. Two further settings are required on this stack:
disabling the NZ tensor layout (it corrupts weight synchronisation during RL) and
capping the rollout engine's context length, which otherwise reserves key-value cache
for the model's full $262{,}144$-token window.

\subsection{Training Hyperparameters}
\label{app:hyperparams}

In Table~\ref{tab:hyperparams}, we list the settings of the optional SFT, the matched GRPO runs of
Table~\ref{tab:main} and the seed-$42$ diagnostic GRPO run; both GRPO configurations keep $48$
completions per update. The saved trainer configurations carry the remaining optimiser defaults.

\begin{table*}[t]
\centering
\caption{Hyperparameters of the optional SFT, the matched GRPO runs (seed $43$) and the seed-$42$
diagnostic GRPO run. TP: tensor-parallel degree; mem.\ frac.: the rollout engine's memory
fraction.}
\label{tab:hyperparams}
\small
\begin{tabular}{lccc}
\toprule
& Optional SFT & GRPO (matched) & GRPO (diagnostic) \\
\midrule
Base model & \multicolumn{3}{c}{Qwen3-4B-Instruct-2507} \\
Parameters trained & Full & Full & Full \\
Training data & 12,890 pairs & 6,000 prompts & 6,000 prompts \\
Optimiser & AdamW & AdamW & AdamW \\
Learning rate & $1 \times 10^{-5}$ & $1 \times 10^{-6}$ & $1 \times 10^{-6}$ \\
LR schedule & Cosine & Cosine & Cosine \\
Warmup (fraction of steps) & 3\% & 2\% & 2\% \\
Batch size (per device $\times$ devices) & $4 \times 3$ pairs & $2 \times 2$ completions & $2 \times 3$ completions \\
Gradient accumulation & 4 & 12 & 8 \\
Effective batch per optimiser step & 48 pairs & 48 completions & 48 completions \\
Epochs / steps & 2 epochs (538 steps) & 300 steps & 300 steps \\
Max sequence / completion length & 1536 / -- & 1536 / 768 & 1536 / 768 \\
Group size ($G$) & -- & 8 & 8 \\
KL coefficient ($\beta$) & -- & 0.04 & 0.04 \\
Clipping range ($\varepsilon_{\mathrm{clip}}$) & -- & 0.2 & 0.2 \\
Rollout temperature & -- & 1.0 & 1.0 \\
$\lambda$ (Equation~\ref{eq:reward}) & -- & 0.7 & 0.7 \\
Structure share in data & 16.5\% & 33.1\% (upsampled from $21\%$) & 33.1\% \\
Precision & bf16 & bf16 & bf16 \\
Devices & 3 & 2 train + 1 rollout & 3 train + 1 rollout \\
Rollout engine (vLLM) & -- & TP $=1$, mem.\ frac.\ $0.92$ & TP $=1$, mem.\ frac.\ $0.9$ \\
Framework & ms-swift 4.2.3 & ms-swift 4.2.3 & ms-swift 4.2.3 \\
Wall time & $\sim$45 min & $\sim$3.7 hours & $\sim$3 hours \\
\bottomrule
\end{tabular}
\end{table*}

\begin{figure*}[t]
\centering
\includegraphics[width=\textwidth]{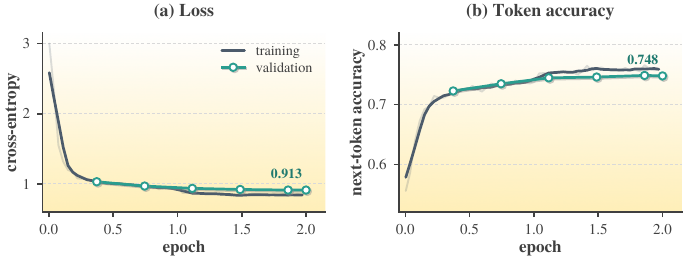}
\caption{\textbf{The supervised stage.} Cross-entropy (a) and next-token accuracy (b)
over the two SFT epochs, for the $12{,}890$ training pairs and the $125$-pair validation
split. Validation loss decreases to $0.913$ and validation accuracy
plateaus at $0.748$.}
\label{fig:sft}
\end{figure*}

\noindent Figure~\ref{fig:sft} shows the supervised stage, run for a fixed two-epoch budget. Two diagnostics from the GRPO logs are worth
recording alongside it. The clipped fraction of Equation~\ref{eq:grpo} is identically zero
at all $300$ steps, as expected when one optimiser update follows each rollout batch, so the
clipping range never binds. In the default diagnostic GRPO run no completion reaches the $768$-token cap: the longest
sampled completion over training is $492$ tokens, and the trainer's truncation counter stays
at zero.

\subsection{Second Base Model}
\label{app:qwen8b}

In this section, we describe the Qwen3-8B run, which applies the Direct recipe unchanged: $\lambda = 0.7$, seed $43$, $300$ updates of
$48$ completions with $G = 8$, learning rate $10^{-6}$, $\beta = 0.04$ and a $768$-token completion
cap. Qwen3-8B is a hybrid reasoning model; training and evaluation use its chat template with
thinking disabled, as does the zero-shot Qwen3-8B row of Table~\ref{tab:main}. The larger model
trains on two policy devices with ZeRO-3 and CPU offload of parameters and optimiser state
(micro-batch $1$, accumulation $24$), which changes memory layout but not the update; a further
device serves rollouts. The run takes $15.7$ hours.

\paragraph{Where the two base models differ.} The 4B base is the instruction-tuned
Qwen3-4B-Instruct-2507 release, whereas Qwen3-8B is the original hybrid release used with
thinking disabled; zero-shot, the smaller model is the stronger of the two ($0.160$ against
$0.123$ on Mixed). After training, the 4B and 8B policies reach $0.807$ and $0.657$ on Mixed.
On the Structure suite the 8B policy reaches $0.900$ ($108/120$), despite its base parsing only
$19$ of the $120$ prompts, showing that the same Direct recipe can recover strong form control
from a weak zero-shot starting point. The 4B policy remains stronger on Structure at $0.967$,
while the 8B policy posts the highest Unseen and Dense rates in Table~\ref{tab:main}, supporting transfer of the same recipe across model scales.

\subsection{SFT Data Construction}
\label{app:sftdata}

In this section, we describe how the supervised training data is built from the IrishMAN corpus
\citep{wu2023tunesformer}, a collection of $\sim$45{,}000 traditional Irish tunes in
ABC notation. The construction pipeline has four steps:

\begin{enumerate}
\item \textbf{Filter and parse.} Tunes whose body contains the builder's list of unsupported tokens, or that have fewer than $4$ or more than $32$ written bars, are discarded; the
survivors are parsed with \texttt{music21} and kept only if the parsed bar count agrees with
the written one.
\item \textbf{Property extraction.} Property-extraction routines associated with the evaluation verifiers
derive candidate constraint parameters from each parsed tune: its key (hybrid policy), meter, bar count, pitch
range, final note, the set of distinct note durations, and the largest melodic
interval.
\item \textbf{Instruction rendering.} A random subset of the extracted properties is
selected (with $k$ sampled uniformly from $\{1, 2, 3\}$, capped by the number of measurable
properties) and rendered into a
constraint instruction using the same template as evaluation. The tune itself is the
target output.
\item \textbf{Verification gate.} Every (instruction, tune) pair is re-verified: the
tune is checked against the rendered constraints using the evaluation verifiers. Any
pair that fails verification is discarded. This ensures that no training example
teaches the model to produce output that violates its own constraints.
\end{enumerate}

Form examples (\texttt{structure} family) are constructed separately: for a tune with at
least eight bars whose first two four-bar units differ, the two units are arranged into a
sampled \textsc{AABA}/\textsc{AABB}/\textsc{ABAB} shape (attempted with probability $0.2$ per
tune; the resulting pair carries the structure constraint only). The resulting
$12{,}890$ pairs form the training set, with a further $125$ pairs held out for validation. Importantly, the
four held-out family pairs (\S\ref{sec:bench}) are enforced during construction: any
pair whose constraints contain a held-out combination is regenerated with a different
family subset.

\subsection{SFT Training Examples}
\label{app:sftexamples}

In this section, we present three representative (instruction, output) pairs from the SFT training
set to illustrate the data quality and the range of constraint combinations.

\begin{figure*}[t]
\small\raggedright
\textbf{SFT example 1: single constraint (\texttt{meter}).}\par
\smallskip\noindent\textbf{Instruction:} Write a single monophonic tune in ABC
notation \ldots\ The tune must satisfy ALL of the following constraints:
-- the meter must be 6/8, and every bar must contain exactly the right total duration.

\smallskip\noindent\textbf{Target (from IrishMAN):}
\begin{Verbatim}[fontsize=\small]
X:89124
T:Tune
L:1/8
M:6/8
K:G
GBd gdB | GBd cBA | GBd gdB | cBA edc | GBd gdB | GBd cBA | cBA dcB | cAF G3 |
DFA cBA | GAB d2 d | DFA cBA | BAG AGE | DFA cBA | GAB d2 d | cBc dcB | cAF G3 |
\end{Verbatim}
\noindent Every one of the $16$ bars contains exactly six eighth-note units, and the tune
carries no repeat sign, so it passes the gate and the meter verifier.
\end{figure*}

\begin{figure*}[t]
\small\raggedright
\textbf{SFT example 2: two constraints (\texttt{length} + \texttt{key}).}\par
\smallskip\noindent\textbf{Instruction:} \ldots\ -- the tune must be exactly 32 bars long
(no pickup bar). -- the tune must be in D major.

\smallskip\noindent\textbf{Target:}
\begin{Verbatim}[fontsize=\small]
X:106003
T:Tune
L:1/8
Q:1/8=180
M:6/8
K:D
DFA cdc | AFA BGE | DFA dcd | AFA gfe | DFA cdc | AFA BGE | afd bge | AFD E2 D |
DFA cdc | AFA BGE | DFA dcd | AFA gfe | DFA cdc | AFA BGE | afd bge | AFD E2 D |
fef gfg | afd AFD | fef gfg | DFA B2 A | fdf gfg | afd bge | fga Bcd | AFD E2 D |
fef gfg | afd AFD | fef gfg | DFA B2 A | fdf gfg | afd bge | fga Bcd | AFD E2 D |
\end{Verbatim}
\noindent The header reads \texttt{K:D} and content analysis returns D major; exactly
$32$ bars. The constraint parameters were read off this tune's own properties, and the
pair was re-verified before admission.
\end{figure*}

\begin{figure*}[t]
\small\raggedright
\textbf{SFT example 3: structure constraint (\texttt{structure}, AABA).}\par
\smallskip\noindent\textbf{Instruction:} \ldots\ -- the tune must be exactly 16 bars with
an AABA form in 4-bar units: labeled units with the same letter must be note-for-note
identical, and units with different letters must differ.

\smallskip\noindent\textbf{Target (assembled from the source tune's first two 4-bar units):}
\begin{Verbatim}[fontsize=\small]
X:141183
T:Tune
L:1/4
M:3/4
K:Bmin
B2 c | d3 | d2 a | f e d | B2 c | d3 | d2 a | f e d |
c3 | e3 | f2 B | B3 | B2 c | d3 | d2 a | f e d |
\end{Verbatim}
\noindent Units 1, 2, and 4 are identical (A sections, bars 1--4, 5--8, 13--16);
unit 3 (bars 9--12) differs, satisfying the AABA form. The two units are the source
tune's first eight bars; every assembled pair is verified before entering the dataset.
\end{figure*}

\subsection{Prompt Template}
\label{app:prompt}
In this section, we present the fixed template that renders every benchmark item into a prompt.
The shared instruction prefix (the first paragraph) is part of the user message of every item; the
constraint block is populated from the item's constraint list. No
chain-of-thought, few-shot examples, or reasoning instructions are given.

\begin{quote}
\small\ttfamily
Write a single monophonic tune in ABC notation.\\
Rules: write every bar out explicitly (no repeat signs like |: or :|), one voice
only, no chords, no lyrics, no grace notes. Start with an X: header, include T:,
M:, L:, K: headers, then the music. Output ONLY the ABC notation and nothing
else.\\[4pt]
The tune must satisfy ALL of the following constraints:\\
-- \textit{$\langle$constraint 1 in natural language$\rangle$}\\
-- \textit{$\langle$constraint 2 in natural language$\rangle$}\\
\hspace*{1em}$\vdots$
\end{quote}

The constraint text for each family is generated by a deterministic renderer that
maps parameters to a fixed English sentence. For example, the \texttt{key} family
with $\theta = (\text{D}, \text{major})$ renders as ``\textit{the tune must be in
D major}''; the \texttt{structure} family with $\theta = (\text{AABA}, 4)$ renders
as ``\textit{the tune must be exactly 16 bars with an AABA form in 4-bar units:
labeled units with the same letter must be note-for-note identical, and units with
different letters must differ}''. The paraphrase suite
(\S\ref{sec:bench}) replaces these renderers with alternative phrasings while
preserving the constraint values.

\subsection{Dataset Statistics}
\label{app:datastats}

In this section, we report the composition of the benchmark splits and the training data.

\paragraph{Benchmark splits.}
Table~\ref{tab:benchstats} summarises the composition of the six original benchmark splits.
The mixed suite spans $k \in \{1,2,3,4\}$ and covers all eight families; the
structure split is the only one where every item includes the cross-section form
\texttt{structure} family. Family frequencies depend on the sampler and exclusion rules (Table~\ref{tab:composition}); the
structure split contains its namesake family in every item and one of five companion
families in $66$ of $120$.

\begin{table*}[t]
\centering
\caption{Composition of \bench{} splits: number of items, distribution over $k$
(number of simultaneous constraints), and family coverage, as ranges where shown.}
\label{tab:benchstats}
\small
\begin{tabular}{lrll}
\toprule
Split & $n$ & $k$ distribution & Families (count) \\
\midrule
Mixed & 300 & 1:100, 2:100, 3:60, 4:40 & All eight \\
Single & 200 & 1:200 & All eight (16--29 each) \\
Seen comb. & 200 & 2:116, 3:84 & All eight (51--72 each) \\
Unseen comb. & 200 & 2:88, 3:112 & All eight (54--70 each) \\
High-$k$ & 160 & 4:160 & All eight (56--104 each) \\
Structure & 120 & 1:54, 2:66 & \texttt{structure}: 120; 5 others \\
\bottomrule
\end{tabular}
\end{table*}

\paragraph{Training data.}
The supervised stage uses $12{,}890$ instruction--tune pairs self-labelled from the
IrishMAN corpus. The reinforcement stage uses $6{,}000$ synthetic prompts with
$k \in \{1, 2, 3\}$ ($2{,}046 / 3{,}060 / 894$ respectively). After the
reward-design iteration of Appendix~\ref{app:diagablation}, the share of prompts
containing the \texttt{structure} family was raised to $33.1\%$ ($1{,}983 / 6{,}000$)
to increase exposure to form constraints; every other
family appears in $1{,}068$--$1{,}347$ prompts.

\subsection{Prompt Coverage, Training Overlap, and Split Composition}
\label{app:coverage}
In this section, we analyse prompt coverage, overlap with the training prompts and split
composition. The weighting and fixed-$k$ comparisons below use the diagnostic SFT-initialised
reference; the suite definitions and prompt counts apply to both studies.

The $2{,}180$ suite items are sampled and contain $1{,}623$ distinct prompt strings across all
eight splits. Table~\ref{tab:coverage}
compares each prompt with the user messages in both training datasets. ``Set
overlap'' ignores constraint order and compares family--parameter sets instead.
Unseen, High-$k$ and Dense share no complete prompt or constraint set with either
training stage, and neither stage requests any of the four held-out pairs. Single and
Structure prompts all occur in training.

\begin{table}[t]
\centering
\caption{Coverage and overlap with either training stage. Overlap columns count
sampled test rows, while ``unique prompts'' counts distinct strings.}
\label{tab:coverage}
\small
% <<< tables/prompt_coverage
\begin{adjustbox}{max width=\columnwidth}
\begin{tabular}{lcccc}
\toprule
Split & Items & Unique prompts & Exact overlap & Set overlap \\
\midrule
Mixed & 300 & 222 & 201 & 203 \\
Single & 200 & 39 & 200 & 200 \\
Seen & 200 & 177 & 147 & 151 \\
Unseen & 200 & 159 & 0 & 0 \\
High-$k$ & 160 & 156 & 0 & 0 \\
Structure & 120 & 50 & 120 & 120 \\
Dense & 500 & 490 & 0 & 0 \\
Edge & 500 & 437 & 32 & 39 \\
\bottomrule
\end{tabular}
\end{adjustbox}

% >>> tables/prompt_coverage
\end{table}

\paragraph{Equal weighting of distinct prompts.}
Table~\ref{tab:uniqueprompts} first averages repeated rows for each prompt within
one evaluation, then averages over distinct prompts. The RL column uses the same
fixed checkpoint and evaluation files as Table~\ref{tab:historicalmain}. The advantage over
SFT persists on all eight splits under this weighting. On Structure, whose $120$
items repeat $50$ prompts, the rates are $\AggUniqueStruct$ and $0.522$, respectively. Item-paired tests in the main analysis retain the original sampler
weights, so they measure the sampled suites rather than $2{,}180$ distinct instructions.
Table~\ref{tab:uniquepromptsmatched} applies the same weighting to the three matched
checkpoints of Table~\ref{tab:main}, read from the same saved per-item files.

\begin{table}[t]
\centering
\caption{Original item weighting and equal weighting of unique prompts, using
the main-table criterion and the diagnostic reference checkpoint.}
\label{tab:uniqueprompts}
\small
% <<< tables/unique_prompt_scores
\begin{adjustbox}{max width=\columnwidth}
\begin{tabular}{lcccc}
\toprule
Split & SFT, items & SFT, unique & RL, items & RL, unique \\
\midrule
Mixed & 0.477 & 0.402 & $0.663$ & $0.615$ \\
Single & 0.745 & 0.705 & $0.835$ & $0.820$ \\
Seen & 0.415 & 0.407 & $0.635$ & $0.618$ \\
Unseen & 0.440 & 0.396 & $0.595$ & $0.547$ \\
High-$k$ & 0.231 & 0.237 & $0.356$ & $0.359$ \\
Structure & 0.600 & 0.522 & $0.808$ & $0.637$ \\
Dense & 0.290 & 0.290 & $0.370$ & $0.371$ \\
Edge & 0.076 & 0.078 & $0.162$ & $0.164$ \\
\bottomrule
\end{tabular}
\end{adjustbox}

% >>> tables/unique_prompt_scores
\end{table}

\begin{table*}[t]
\centering
\caption{The same weighting applied to the three matched checkpoints of
Table~\ref{tab:main}: ``items'' repeats the sampler weights of Table~\ref{tab:main}, ``unique''
averages within each distinct prompt first. The ordering between the three checkpoints is unchanged on every split but Edge,
where the two Direct arms swap by $0.008$.}
\label{tab:uniquepromptsmatched}
\small
\setlength{\tabcolsep}{4pt}
\begin{tabular}{lcccccc}
\toprule
& \multicolumn{2}{c}{Direct GRPO ($R=A$)} & \multicolumn{2}{c}{\method{} (SFT init.)}
& \multicolumn{2}{c}{\method{} (Direct)} \\
\cmidrule(lr){2-3}\cmidrule(lr){4-5}\cmidrule(lr){6-7}
Split & items & unique & items & unique & items & unique \\
\midrule
Mixed & 0.797 & 0.741 & 0.577 & 0.527 & 0.807 & 0.760 \\
Single & 0.960 & 0.949 & 0.880 & 0.875 & 0.985 & 0.981 \\
Seen & 0.745 & 0.729 & 0.585 & 0.567 & 0.815 & 0.814 \\
Unseen & 0.565 & 0.509 & 0.530 & 0.479 & 0.595 & 0.547 \\
High-$k$ & 0.469 & 0.458 & 0.350 & 0.353 & 0.575 & 0.571 \\
Structure & 0.908 & 0.860 & 0.383 & 0.457 & 0.967 & 0.940 \\
Dense & 0.322 & 0.322 & 0.366 & 0.363 & 0.362 & 0.358 \\
Edge & 0.300 & 0.291 & 0.134 & 0.135 & 0.310 & 0.283 \\
\bottomrule
\end{tabular}
\end{table*}

\paragraph{Fixed $k$ does not fix family composition.}
Table~\ref{tab:composition} gives family frequencies in High-$k$, Dense and Edge.
The first two share $k=4$ but are independently sampled and have different family
mixtures. Edge additionally mixes pair-free and held-out items ($171/500$ contain
a held-out pair). Appendix~\ref{app:rescore} scores the pair-free and
held-out-pair halves of Edge separately. Table~\ref{tab:denseuncertainty} bootstraps the
Dense-minus-High-$k$ difference for each model, separating item-sampling variation from this
composition difference.

\begin{table}[t]
\centering
\caption{Number and percentage of items requesting each family. Percentages sum
to more than $100\%$ because items contain multiple families.}
\label{tab:composition}
\small
% <<< tables/split_composition
\begin{adjustbox}{max width=\columnwidth}
\begin{tabular}{lccc}
\toprule
Family & High-$k$ ($n=160$) & Dense ($n=500$) & Edge ($n=500$) \\
\midrule
\texttt{key} & 76 (47.5\%) & 288 (57.6\%) & 168 (33.6\%) \\
\texttt{meter} & 82 (51.2\%) & 282 (56.4\%) & 160 (32.0\%) \\
\texttt{length} & 60 (37.5\%) & 209 (41.8\%) & 127 (25.4\%) \\
\texttt{range} & 100 (62.5\%) & 236 (47.2\%) & 164 (32.8\%) \\
\texttt{final} & 104 (65.0\%) & 234 (46.8\%) & 180 (36.0\%) \\
\texttt{rhythm} & 84 (52.5\%) & 279 (55.8\%) & 152 (30.4\%) \\
\texttt{leap} & 78 (48.8\%) & 270 (54.0\%) & 157 (31.4\%) \\
\texttt{structure} & 56 (35.0\%) & 202 (40.4\%) & 148 (29.6\%) \\
\bottomrule
\end{tabular}
\end{adjustbox}

% >>> tables/split_composition
\end{table}

\begin{table*}[t]
\centering
\caption{Dense minus High-$k$: absolute all-satisfied rate differences with percentile intervals from
$20{,}000$ within-split item bootstrap replicates (seed $20260909$).}
\label{tab:denseuncertainty}
\small
% <<< tables/dense_uncertainty
\begin{tabular}{lcccc}
\toprule
Model & High-$k$ & Dense & $\Delta$ & 95\% interval \\
\midrule
SFT & 0.231 & 0.290 & $+0.059$ & $[-0.019, +0.132]$ \\
\method{} (re-evaluated) & 0.394 & 0.370 & $-0.024$ & $[-0.111, +0.062]$ \\
Llama-3.1-70B & 0.056 & 0.040 & $-0.016$ & $[-0.057, +0.021]$ \\
\bottomrule
\end{tabular}

% >>> tables/dense_uncertainty
\end{table*}

All three intervals include zero: at $k=4$, requiring a held-out pair does not measurably lower
compliance for any of the three models. Table~\ref{tab:composition} gives the family composition the two
splits differ in.

\subsection{Reproducibility}
\label{app:repro}
In this section, we list the scripts that regenerate the paper's tables, macros and figures.
\texttt{make\_assets.py} regenerates the original quantitative tables and numerical
macros; \texttt{make\_audit\_assets.py} adds the overlap, sensitivity
and gate-audit tables; \texttt{src/rescore\_subsets.py} produces the Edge-subset and
stricter-key tables. The historical reporting pass, \texttt{make\_checkpoint\_assets.py --sync},
computes reference-table, by-$k$, Edge-subset and fixed-$k$ comparisons from saved verdicts,
checks ablation equivalence under the format audit, and supplies the fixed-checkpoint decompositions;
it also supplies the corresponding historical figure panels.
The matched tables and macros are generated by
\texttt{make\_matched\_s43\_assets.py} after all three runs finish, and the pooled by-$k$ figure
by \texttt{make\_fig\_degrade\_pooled.py}. This script checks
saved per-item IDs, success counts and provenance hashes for all $24$ evaluations, then
renders the two paired comparisons with eight-suite Holm correction.
Figure scripts read the same saved reports or training logs; each later analysis (reranking,
$k$ ladder, Edge factors, output variety) names its script where it is reported.
The family-definition and rendering tables are maintained descriptions of the
sampler. The six original suites were used during reward development; Dense
and Edge were added afterwards, without that exposure. All training and local evaluation ran on Ascend 910B2 accelerators ($64$\,GB HBM each,
CANN $8.2$.RC1) with vLLM $0.11.0$ through \texttt{vllm-ascend}; the seed-$42$ diagnostic GRPO runs used three
policy accelerators and one rollout accelerator. The matched study used two
policy accelerators and one rollout accelerator. Local models were served at
context $4{,}096$ and generated at most $1{,}200$ tokens per item, greedy. The hosted
baselines were queried at temperature $0$ through the NVIDIA NIM API
(\texttt{integrate.api.nvidia.com}), MiniMax-M3 as \texttt{minimaxai/minimax-m3} in August
$2026$, with per-model completion budgets of $800$ tokens for Llama-3.1-70B, $4{,}000$ for
MiniMax-M3 and $2{,}048$ of context for ChatMusician-7B; budgets and timeouts per model are in
the release scripts.
The release includes per-item completions, original and strict gate decisions,
aggregate reports, audit scripts, and training logs. Historical reports remain
unchanged; \texttt{src/verifiers\_legacy.py} reproduces the training and main evaluation metric,
while \texttt{src/verifiers.py} implements the stricter gate. The reward plugin pins
the historical gate unless strict-gate training is explicitly selected.
\texttt{src/run\_verify.py} defaults to this same $A$ criterion (\texttt{legacy} mode).

\begin{table*}[t]
\centering
\caption{Parameter values of the Edge split against the standard suites. Families are
sampled by the same generator; only the value sets differ. \texttt{key},
\texttt{final} and \texttt{structure} take the same values in both, so they are
omitted.}
\label{tab:edgespec}
\small
\begin{tabular}{lll}
\toprule
Family & Standard suites & Edge \\
\midrule
\texttt{meter} & $2/4$, $3/4$, $4/4$, $6/8$ & $5/4$, $2/2$, $9/8$, $12/8$ \\
\texttt{length} & $8$, $12$, $16$ bars & $4$, $24$, $32$ bars \\
\texttt{leap} & $\leq 7$, $9$, $12$ semitones & $\leq 2$, $3$, $4$ semitones \\
\texttt{rhythm} & three- and two-duration sets & \{16th, 8th\}, \{quarter, half\}, \\
 & & \{quarter\}, \{half\} \\
\texttt{range} & mixed spans & one-octave spans (C4--C5, D4--D5, \\
 & & E4--E5, G3--G4, A3--A4) \\
\bottomrule
\end{tabular}
\end{table*}

\subsection{Evaluation Nondeterminism}
\label{app:evalnoise}
In this section, we quantify evaluation nondeterminism. Compliance evaluation uses temperature
$0$, yet vLLM's batched decoding is not bitwise reproducible: evaluating the diagnostic
checkpoint (\texttt{outputs/grpo\_4b\_v2/checkpoint-300}) twice, on 08-22 and 08-24,
with identical serving flags (Table~\ref{tab:evalnoise}), none of the $160$ high-$k$
completions was byte-identical across the two runs. A plausible mechanism is that the
ordering of floating-point reductions depends on batch composition, and that the first
generated token (the arbitrary \texttt{X:} reference number) sits on a near-tie
between many candidates, so a small numerical difference there can send the whole
continuation down a different path.

\begin{table}[t]
\centering
\caption{The same checkpoint evaluated twice under greedy decoding. ``Flipped'' counts
items whose binary all-satisfied outcome differed between the two. Both have
parse rate $1.000$.}
\label{tab:evalnoise}
\small
\begin{adjustbox}{max width=\columnwidth}
\begin{tabular}{lcccccc}
\toprule
Split & $n$ & First & Second & Flipped & $0{\rightarrow}1$ & $1{\rightarrow}0$ \\
\midrule
Mixed & 300 & 0.663 & 0.703 & 52 (17.3\%) & 32 & 20 \\
High-$k$ & 160 & 0.356 & 0.394 & 32 (20.0\%) & 19 & 13 \\
Structure & 120 & 0.808 & 0.758 & 26 (21.7\%) & 10 & 16 \\
\bottomrule
\end{tabular}
\end{adjustbox}
\end{table}

The two evaluations differ by $0.040$ on Mixed and $0.050$ on Structure.

The first evaluation supplies the historical reference and diagnostic values except Dense and Edge, which
come from the re-evaluation, and Figure~\ref{fig:dyn6}(d), which uses a separate
full-retention run.
The fixed-$k$ comparison uses the re-evaluated High-$k$ score
($0.394\rightarrow0.370$).

\subsection{Data and Code Release}
\label{app:release}

We will release the code and synthetic benchmark suites under an MIT licence,
with corpus-derived data and model artefacts subject to their upstream terms:

\begin{itemize}
\item \textbf{\bench{} benchmark suites:} eight frozen JSONL files containing all
$2{,}180$ benchmark items plus the $400$-item paraphrase suite. Each item includes
the prompt and the constraint specifications; the generation seed is a parameter of the suite
sampler rather than a per-row field.
\item \textbf{Verifier code:} the eight constraint verifiers, the ABC parser, the
reward function, the historical verifier, the corrected strict gate, and both
sets of scoring reports.
\item \textbf{SFT and RL data:} the $12{,}890$ supervised training pairs and
$6{,}000$ RL prompts.
\item \textbf{Model outputs:} per-item completions and per-item verifier decisions
(with all sub-signals) for every checkpoint and baseline evaluated in this paper.
\item \textbf{Evaluation reports:} aggregate reports, significance tests, quality
statistics, prompt-overlap and unique-prompt analyses, strict-gate audits, and the
training logs from which the reward curves are drawn.
\item \textbf{Asset generation scripts:} \texttt{make\_assets.py} for the initial tables,
\texttt{make\_audit\_assets.py} for supplementary analyses, the final reporting pass
\texttt{make\_checkpoint\_assets.py --sync}, and one
\texttt{make\_fig\_*.py} per figure, all reading saved reports,
so every number in the paper is traceable to a specific evaluation output.
\end{itemize}

The IrishMAN corpus \citep{wu2023tunesformer} used for SFT training is publicly
available; source-specific rights and attribution should be preserved for
corpus-derived targets. Model distribution follows the licence of the exact
Qwen3-4B-Instruct-2507 checkpoint \citep{yang2025qwen3}.
No human subjects were involved in this research; all evaluation is automated.

\FloatBarrier
\section{Benchmark and Verifier Details}
\label{app:benchdetails}

In this section, we present the verifier implementation and its audit, the key calibration, the
held-out pairs, ABC notation and examples of the prompt renderings.

\subsection{Verifier Implementation Details}
\label{app:verifiers}

In this section, we describe the validation gate and the family verifiers. The historical gate
$\rho$, which is the validation gate of \S\ref{sec:method} and is also called the parse gate below, is used in training and in all tables labelled $A$, and runs first. It extracts the ABC block from the completion and rejects the
output if extraction fails, if the text contains \texttt{|:} or \texttt{:|}, if \texttt{music21}
raises or exceeds a $20$-second parse timeout, or if the selected first part has fewer than eight
notes, fewer than four non-empty measures, or any chord object. Outputs that clear the gate are
then scored by the family verifiers. Appendix~\ref{app:strictaudit} adds stricter format checks on
the same saved completions, under which the matched checkpoints of Table~\ref{tab:main} keep
every success.

Each of the eight constraint families is implemented as a deterministic function
that takes a \texttt{music21} \texttt{Score} object and the constraint parameters,
returning a binary verdict. We describe the non-trivial implementation choices
below.

\paragraph{\texttt{key}.} As discussed in \S\ref{sec:families}, the verifier uses a
hybrid policy: the \texttt{K:} header must match the request, \emph{and} key-profile
analysis of the note content (\texttt{score.analyze("key")}, which in \texttt{music21}~$10.5.0$
selects the Aarden--Essen profiles) must return the requested
key or its relative major/minor. On a calibration set of $400$ IrishMAN tunes, this hybrid
policy achieves $0.890$ agreement with the corpus key annotation, compared to $0.750$ for content analysis alone and $1.000$
(trivially) for header-only checking.

\paragraph{\texttt{meter}.} The verifier checks two conditions: (1) the time
signature in the \texttt{M:} header matches the request, and (2) every non-empty bar's total
duration (notes and rests, in quarter-note equivalents) equals that signature's bar length
within $10^{-4}$. Incomplete initial or final bars receive no tolerance.

\paragraph{\texttt{length}.} The number of non-empty bars is counted after parsing and compared
with the request. The prompt's ``no pickup bar'' is enforced here as well: the first and last
written bar must each carry a full measure of the tune's own time signature, so an output that
opens with an anacrusis and closes with its complement fails even when no \texttt{meter}
constraint is requested.

\paragraph{\texttt{range}.} Every sounding pitch is compared against the specified
minimum and maximum (as MIDI numbers). Rests are ignored.

\paragraph{\texttt{final}.} The pitch class of the last sounding note is compared
against the request; octave is always ignored.

\paragraph{\texttt{rhythm}.} Every note's duration in quarter-note units, rounded to four
decimals, must belong to the admissible set, and any rest fails the check. A dotted or tied
value therefore passes only if its parsed duration is itself admissible.

\paragraph{\texttt{leap}.} Every pair of consecutive sounding pitches is examined;
the verifier returns $0$ if any interval exceeds the stated maximum (in semitones).

\paragraph{\texttt{structure}.} A tune with other than sixteen non-empty bars fails outright;
otherwise it is split into four units of four bars each. Each bar is reduced to its
signature (event type, MIDI pitch and duration rounded to four decimals, rests included).
The minimal pair sets are \textsc{AABA}: $E=\{(1,2),(1,4)\}$, $D=\{(1,3)\}$;
\textsc{AABB}: $E=\{(1,2),(3,4)\}$, $D=\{(1,3)\}$; \textsc{ABAB}: $E=\{(1,3),(2,4)\}$,
$D=\{(1,2)\}$. For each pair in $E$ (Equation~\ref{eq:structure}) the verifier compares the bar
signatures position by position; for each pair in $D$ it checks that at least one position differs. The graded training signal $\bar{e}$ is the fraction of
bar positions that match across all pairs in $E$; the binary evaluation signal
requires all to match.

\paragraph{Validation without annotators.} Except for the hybrid \texttt{key} check, every
family is a deterministic function of the parsed score, so the question is whether the
implementation computes the property it claims. \texttt{src/validate\_verifiers.py} builds $32$
ABC scores whose property values are fixed by construction (\texttt{key} $2$, \texttt{meter}
$4$, \texttt{length} $3$, \texttt{range} $5$, \texttt{final} $3$, \texttt{rhythm} $5$,
\texttt{leap} $5$, \texttt{structure} $5$, including the inclusive bounds) and all $32$ verdicts
are correct. It then starts from one tune that satisfies six families at once and mutates one
property at a time (\texttt{length}, \texttt{meter}, \texttt{range}, \texttt{final},
\texttt{rhythm}, \texttt{leap}); each mutation flips exactly the targeted verdict and leaves the
other five unchanged. Three conventions are reported rather than asserted: a pickup bar counts
against \texttt{length}, an enharmonic spelling of the final pitch class is accepted, and a tune
whose notes outline the relative key of its header is accepted by \texttt{key}.
\texttt{src/validate\_meter\_corpus.py} runs the meter check on $400$ IrishMAN tunes: the parsed
time signature equals the declared \texttt{M:} field on all $400$.

\subsection{Stricter Fixed-Instruction Gate Audit}
\label{app:strictaudit}
In this section, we audit the diagnostic checkpoints of Table~\ref{tab:historicalmain}
and the three matched RL checkpoints of Table~\ref{tab:main} under stricter format checks.

The historical gate can accept a second voice because most checks inspect only
the first part, whereas key analysis inspects the whole score. It also does not
reject every wrapper, missing header, lyric or grace note. We retain that verifier
as \texttt{verifiers\_legacy.py} to reproduce the saved training runs, and implement
a stricter gate in \texttt{verifiers.py}. It requires a single tune with the
specified headers, no unconsumed wrapper text, no repeat notation, lyrics or grace
notes, and no multiple parts, multiple voices, chord objects or overlapping notes.
Whitespace is normalised when detecting wrappers. The eight family checks remain
unchanged. This hardens the fixed-instruction checks.

The stricter metric is
\[
 A_{\mathrm{strict}}(y,C)=A(y,C)\,\mathbb{I}[g_{\mathrm{strict}}(y)].
\]
\texttt{audit\_strict\_gate.py} evaluates the additional gate on saved completions;
no regeneration or retraining is involved. For the configurations within this manuscript's
scope, the audit covers $\StrictAuditEvaluations$ evaluation files and $\StrictAuditItems$
rows: the three matched checkpoints of Table~\ref{tab:main} on all eight suites, plus
the baseline, training-configuration, ablation, sweep, paraphrase and retained dynamics
checkpoints. All file-level outcomes, rejection reasons and source hashes are saved separately.
Tables~\ref{tab:strictscores}--\ref{tab:strictrejects} show the original six suites;
Table~\ref{tab:strictextra} adds Dense and Edge. Across the audit subset,
$\StrictAuditLost$ previously successful rows fail the stricter gate, and \emph{none of them
belongs to the three matched RL rows of Table~\ref{tab:main}}: those three checkpoints incur no additional
gate rejection and lose no success on any of the $6{,}540$ items they were scored on, so every
rate in those three RL rows is unchanged under $A_{\mathrm{strict}}$. The main comparison and
reward curves use $A$; this audit measures the additional fixed-format requirements, which are
not included in the main score.

\begin{table*}[t]
\centering
\caption{All-satisfied rate under the stricter fixed-instruction gate $A_{\mathrm{strict}}$ on
the original-suite completions. The last three rows are the matched checkpoints of
Table~\ref{tab:main}; every one of their rates equals the corresponding $A$ rate.}
\label{tab:strictscores}
\scriptsize
\setlength{\tabcolsep}{5pt}
% <<< tables/strict_scores
\begin{tabular}{lcccccc}
\toprule
Model & Mixed & Single & Seen & Unseen & High-$k$ & Structure \\
\midrule
Qwen3-4B & 0.160 & 0.225 & 0.100 & 0.040 & 0.037 & 0.100 \\
Qwen3-8B & 0.120 & 0.210 & 0.115 & 0.050 & 0.006 & 0.000 \\
Qwen3-32B & 0.357 & 0.535 & 0.240 & 0.190 & 0.094 & 0.267 \\
Llama-3.1-70B & 0.377 & 0.695 & 0.300 & 0.190 & 0.050 & 0.167 \\
MiniMax-M3 & 0.367 & 0.580 & 0.315 & 0.190 & 0.100 & 0.008 \\
ChatMusician-7B & 0.000 & 0.000 & 0.000 & 0.000 & 0.000 & 0.000 \\
SFT & 0.477 & 0.745 & 0.415 & 0.440 & 0.231 & 0.600 \\
\method{} (diagnostic reference) & $0.663$ & $0.835$ & $0.635$ & $0.595$ & $0.356$ & $0.808$ \\
\midrule
Direct GRPO ($R=A$) & 0.797 & 0.960 & 0.745 & 0.565 & 0.469 & 0.908 \\
\method{} (Direct) & 0.807 & 0.985 & 0.815 & 0.595 & 0.575 & 0.967 \\
\method{} (SFT init.) & 0.577 & 0.880 & 0.585 & 0.530 & 0.350 & 0.383 \\
\bottomrule
\end{tabular}

% >>> tables/strict_scores
\end{table*}

\begin{table*}[t]
\centering
\caption{Additional gate rejections across the six original suites ($1{,}180$ rows per model). Lost
successes previously had $A=1$ and now have $A_{\mathrm{strict}}=0$; an extra rejection need not
change success if a family already failed. Reasons list only the first detected violation per
output.}
\label{tab:strictrejects}
\scriptsize
% <<< tables/strict_rejections
\begin{tabular}{lrrp{0.42\textwidth}}
\toprule
Model & New rejects & Lost successes & Leading reasons \\
\midrule
Qwen3-4B & 7 & 0 & non abc wrapper or unconsumed text: 7 \\
Qwen3-8B & 6 & 1 & non abc wrapper or unconsumed text: 5, grace notes forbidden: 1 \\
Qwen3-32B & 22 & 4 & non abc wrapper or unconsumed text: 19, repeat signs forbidden: 2 \\
Llama-3.1-70B & 26 & 3 & non abc wrapper or unconsumed text: 26 \\
MiniMax-M3 & 1 & 0 & non abc wrapper or unconsumed text: 1 \\
ChatMusician-7B & 749 & 151 & missing required header: 734, non abc wrapper or unconsumed text: 15 \\
SFT & 0 & 0 & -- \\
\method{} (diagnostic reference) & 0 & 0 & -- \\
\midrule
Direct GRPO ($R=A$) & 0 & 0 & -- \\
\method{} (Direct) & 0 & 0 & -- \\
\method{} (SFT init.) & 0 & 0 & -- \\
\bottomrule
\end{tabular}

% >>> tables/strict_rejections
\end{table*}

\begin{table*}[t]
\centering
\caption{Historical and strict all-satisfied rates on the two additional suites,
using the same saved completions.}
\label{tab:strictextra}
\small
% <<< tables/strict_extra
\begin{tabular}{lcccc}
\toprule
Model & Dense $A$ & Dense $A_{\rm strict}$ & Edge $A$ & Edge $A_{\rm strict}$ \\
\midrule
Qwen3-4B & 0.008 & 0.008 & 0.010 & 0.010 \\
Qwen3-8B & 0.020 & 0.020 & 0.032 & 0.032 \\
Qwen3-32B & 0.074 & 0.074 & 0.112 & 0.102 \\
Llama-3.1-70B & 0.040 & 0.038 & 0.058 & 0.056 \\
MiniMax-M3 & 0.088 & 0.088 & 0.160 & 0.160 \\
ChatMusician-7B & 0.020 & 0.000 & 0.046 & 0.000 \\
SFT & 0.290 & 0.290 & 0.076 & 0.076 \\
\method{} (diagnostic, re-evaluated) & 0.370 & 0.370 & 0.162 & 0.162 \\
\midrule
Direct GRPO ($R=A$) & 0.322 & 0.322 & 0.300 & 0.300 \\
\method{} (Direct) & 0.362 & 0.362 & 0.310 & 0.310 \\
\method{} (SFT init.) & 0.366 & 0.366 & 0.134 & 0.134 \\
\bottomrule
\end{tabular}

% >>> tables/strict_extra
\end{table*}

The SFT and fixed SFT-to-GRPO checkpoints incur no additional gate rejections
on any of the eight suites. ChatMusician's strict
success rates are zero because its otherwise accepted outputs omit required
headers or include wrappers. This audit therefore changes some baseline scores; a strict-format
failure is a formatting verdict.

The corrected gate has regression checks for valid monophonic output and for
multiple voices, chords, grace notes, lyrics, missing title, prose wrappers and
repeat signs. Historical training uses the frozen gate.

\subsection{Key Verifier Calibration}
\label{app:keycal}

In this section, we provide the full calibration data of the hybrid key verifier
(\S\ref{sec:families}), the most complex verifier in our suite, which involves both header
matching and content-based key analysis.

\paragraph{Calibration dataset.} We sampled $400$ tunes from the IrishMAN corpus
at random (seeded), each with a key annotation in the
\texttt{K:} header. We then ran four key-verification strategies and measured their
agreement with that annotation (Table~\ref{tab:keycal}):

\begin{table}[t]
\centering
\caption{Key verifier calibration on $400$ IrishMAN tunes. Agreement is the
fraction of tunes where the verifier's answer matches the corpus key annotation.}
\label{tab:keycal}
\small
\begin{adjustbox}{max width=\columnwidth}
\begin{tabular}{lc}
\toprule
Strategy & Agreement \\
\midrule
Header only (trivially gameable) & 1.000 \\
Content analysis only (Aarden--Essen profiles) & 0.750 \\
Content analysis with relative-key tolerance & 0.890 \\
Hybrid: header AND content$\,\in\,\{$key, relative$\}$ & 0.890 \\
\bottomrule
\end{tabular}
\end{adjustbox}
\end{table}

\noindent The reference here is the corpus \texttt{K:} annotation, which the header-only policy
reproduces by construction. The hybrid policy matches the content-with-tolerance policy on this
set; its advantage appears when the model might game the header, since a policy that prints
\texttt{K:G} above music in C major passes the header-only check but fails the hybrid one. The
remaining $0.110$ is disagreement between the analysis routine and that annotation, which modal
tunes, extensive chromaticism and very short tunes can each move. The same policy applies to every
model we score.

\paragraph{Effect on reported numbers.} The key family's per-constraint satisfaction
rate (Table~\ref{tab:perfamily_all}) is $0.862$ on the single split and $0.671$ on
the high-$k$ split. If we used a header-only verifier, these would be higher
(the model rarely writes the wrong \texttt{K:} header), but header-only acceptance does not
inspect the notes. Content analysis with relative-key tolerance carries the $0.110$ disagreement
above; exact content-only matching disagrees on $0.250$ of the calibration tunes. We adopt the
hybrid policy and log all sub-signals, so future work can re-score under any policy.

\subsection{Held-out Pair Rationale}
\label{app:heldout}

In this section, we explain how the held-out pairs are chosen. The four held-out pairs
(\texttt{key}, \texttt{rhythm}), (\texttt{meter}, \texttt{leap}), (\texttt{final},
\texttt{structure}) and (\texttt{length}, \texttt{range}) were chosen to form a \emph{perfect matching} on the eight
families, ensuring that (1) every family participates in exactly one held-out pair,
(2) no family appears in multiple exclusions, and (3) the
matching constrains the maximum testable $k$ to $4$ without touching any held-out
pair, which naturally defines the high-$k$ split.

The specific pairing was selected to maximise the diversity of inter-family
interaction types:
\begin{itemize}
\item \texttt{key}--\texttt{rhythm}: one harmonic, one temporal, with no obvious
interaction (a tune can be in any key with any rhythm vocabulary).
\item \texttt{meter}--\texttt{leap}: one temporal, one melodic; the meter
determines bar duration but should not affect interval size.
\item \texttt{final}--\texttt{structure}: one local (last note), one global (16-bar
form); the most interesting test of whether a model can coordinate endpoint
and large-scale structure without having seen them together.
\item \texttt{length}--\texttt{range}: both are ``bounding'' constraints (number of
bars, pitch boundaries) that interact primarily through the amount of melodic
material available.
\end{itemize}

The supervised dataset contains zero instances of any held-out pair, verified by an exhaustive
check over all $12{,}890$ pairs before training.

\subsection{ABC Notation Primer}
\label{app:abc}

In this section, we introduce ABC notation, a plain-text format for monophonic music widely used
in folk-music archives. A minimal valid tune contains a header block and a body:

\begin{Verbatim}[breaklines,breakanywhere,fontsize=\scriptsize]
X:1               % reference number
T:My Tune          % title
M:4/4              % meter (time signature)
L:1/8              % default note length
K:G                % key signature
G2 BG dGBG | G2 Bd cAFA | G2 BG dGBG | cABG AGFG |
\end{Verbatim}

\noindent Key conventions: note names \texttt{C D E F G A B} denote the octave
starting at middle C; lowercase \texttt{c d e f g a b} denote one octave higher;
trailing apostrophes (e.g.\ \texttt{c'}) raise further. A digit after a note name
multiplies its duration by that factor relative to the default length \texttt{L:}.
Thus \texttt{G2} in a tune with \texttt{L:1/8} is a quarter note on G. The pipe
character \texttt{|} separates bars.

Because ABC is plain text, a language model generates it token by token in exactly
the way it generates code or prose. The \texttt{music21} library
\citep{cuthbert2010music21} parses ABC into a rich internal representation
(pitch, duration, key analysis, interval computation) that our verifiers operate on.
If parsing fails, for instance because the model outputs prose instead of notation
or uses an invalid header like \texttt{L:8} instead of \texttt{L:1/8}, the
gate returns $\rho(y) = \bot$, so $A(y, C) = 0$ (Equation~\ref{eq:allsat}).

\subsection{ABC Notation: A Worked Example}
\label{app:abcworked}

In this section, we provide a token-by-token walkthrough of a model-generated output that passes
all four constraints of a high-$k$ problem, for readers unfamiliar with ABC notation.

\paragraph{Problem.} Meter 3/4; exactly 8 bars; last note E; eighth and quarter notes
only.

\paragraph{Model output:}
{\small\begin{verbatim}
X:17011
T:Tune
L:1/8
M:3/4
K:Edor
B2 B2 AG | E2 E2 DE | F2 F2 ED | B2 B2 AG |
E2 E2 DE | F2 F2 ED | B2 B2 AG | E2 E2 DE |
\end{verbatim}}

\paragraph{Token-by-token analysis:}
\begin{itemize}
\item \texttt{X:17011}: Reference number (arbitrary, used by databases).
\item \texttt{T:Tune}: Title (arbitrary, not constrained).
\item \texttt{L:1/8}: Default note length is an eighth note. This means
\texttt{B} alone is an eighth note, \texttt{B2} is two eighth notes = one quarter
note.
\item \texttt{M:3/4}: Time signature: three quarter notes per bar. Since
\texttt{L:1/8}, each bar must contain six eighth-note equivalents. \checkmark\ meter.
\item \texttt{K:Edor}: Key of E Dorian (E minor with raised sixth). The key
constraint is not active in this problem, so any key is acceptable.
\item \texttt{B2 B2 AG}: Bar 1. \texttt{B2} = quarter note B (two eighths);
\texttt{B2} = quarter note B; \texttt{A} = eighth note A; \texttt{G} = eighth note
G. Total: $2 + 2 + 1 + 1 = 6$ eighth-note equivalents. \checkmark\ meter.
Durations used: quarter, eighth. \checkmark\ rhythm.
\item \texttt{|}: Bar line separator.
\item \texttt{E2 E2 DE}: Bar 2. Same structure. Total 6. \checkmark
\item \ldots (bars 3--7 follow the same pattern) \ldots
\item \texttt{E2 E2 DE |}: Bar 8 (last bar). The final token is \texttt{E},
an eighth note on pitch E. \checkmark\ final.
\end{itemize}

\noindent\textbf{Verification summary:} meter \checkmark\ (all 8 bars have 6
eighth-note equivalents); length \checkmark\ (exactly 8 bars); final \checkmark\
(last note is E); rhythm \checkmark\ (only eighth and quarter notes appear). All
four constraints satisfied; $A = 1$.

\subsection{Constraint Rendering Examples}
\label{app:rendering}

% <<< tables/families_full
% Hand-maintained from src/constraints.py (KEYS, METERS, BAR_COUNTS, RANGES,
% RHYTHM_VOCABS, LEAP_LIMITS, STRUCTURES); the prompt column is the sampler's text
% field verbatim, with one sampled value shown.
\begin{table*}[t]
\centering
\caption{Table~\ref{tab:families} with the prompt column restored. Under each family name is the span it
is checked over; the second column lists the default sampler's values (the verifier domain
$\Theta_f$ is wider), the third the sentence the prompt template emits for one such value,
verbatim from the renderer. \textcolor{famform}{\texttt{\textbf{structure}}} is set apart as the
only cross-section form family.}
\label{tab:families_full}
\small
\setlength{\tabcolsep}{5pt}
\renewcommand{\arraystretch}{1.15}
\rowcolors{2}{famshade}{white}
% No @{} at the outer edges: colortbl paints each cell's background with a
% \tabcolsep overhang on both sides, so with @{} the shading spills 5pt past
% the rules on each side. Keeping the outer separators makes the band land
% exactly on the rules.
\begin{tabularx}{\textwidth}{@{\hspace{\tabcolsep}}>{\raggedright\arraybackslash}p{1.72cm} >{\raggedright\arraybackslash}p{4.6cm} >{\raggedright\arraybackslash}X}
\toprule
\rowcolor{white}
\textbf{Family} & \textbf{Default sampler values} & \textbf{Prompt line (one sampled value)} \\
\midrule
\famcell{key}{global} & 13 keys: 8 major, 5 minor & \emph{the tune must be in D minor} \\
\famcell{meter}{every bar} & $4/4$, $3/4$, $6/8$, $2/4$; each bar must sum to the signature & \emph{the meter must be 6/8, and every bar must contain exactly the right total duration} \\
\famcell{length}{global} & $8$, $12$ or $16$ bars, no pickup & \emph{the tune must be exactly 16 bars long (no pickup bar)} \\
\famcell{range}{every note} & 4 inclusive pitch spans, e.g.\ G3--A5 & \emph{every note must lie between G3 and A5 inclusive} \\
\famcell{final}{last note} & a pitch class, any octave (a scale degree 1/3/5 when a key is present) & \emph{the last note of the tune must be E (any octave)} \\
\famcell{rhythm}{every note} & 4 admissible duration sets; no dotted values, no rests & \emph{use only eighth and quarter notes (no dotted notes, no other durations, no rests)} \\
\famcell{leap}{adjacent notes} & largest melodic interval: $7$, $9$ or $12$ semitones & \emph{melodic motion must be smooth: no melodic leap larger than a major sixth} \\
\famcellform{structure}{4-bar units} & \textsc{aaba}, \textsc{aabb} or \textsc{abab} over $16$ bars; equal letters identical, different letters distinct & \emph{the tune must be exactly 16 bars with an ABAB form in 4-bar units: labeled units with the same letter must be note-for-note identical, and units with different letters must differ} \\
\bottomrule
\end{tabularx}
\end{table*}

% >>> tables/families_full

In Table~\ref{tab:rendering}, we show the complete set of constraint-text renderers: for
each family, the table lists the parameter types and one representative rendering. These renderers are deterministic functions, so every benchmark item with
the same family and parameters produces the same constraint text.

\begin{table*}[t]
\centering
\caption{Constraint rendering for all eight families. The ``Parameters'' column
shows the parameter space; the ``Rendered text'' column shows the template output
for one example value. Paraphrase variants (Appendix~\ref{app:paraexamples}) replace
these templates with alternative wordings.}
\label{tab:rendering}
\small
\begin{tabular}{lp{0.22\textwidth}p{0.40\textwidth}}
\toprule
Family & Parameters & Rendered text (example) \\
\midrule
\texttt{key} & key name $\times$ mode & ``The tune must be in G major.'' \\
\texttt{meter} & time signature & ``The meter must be 3/4, and every bar must
contain exactly the right total duration.'' \\
\texttt{length} & bar count & ``The tune must be exactly 16 bars long (no pickup
bar).'' \\
\texttt{range} & low pitch, high pitch & ``Every note must lie between C4 and C6
inclusive.'' \\
\texttt{final} & pitch class; any octave & ``The last note of the tune must be D
(any octave).'' \\
\texttt{rhythm} & admissible duration set & ``Use only eighth and quarter notes (no
dotted notes, no other durations, no rests).'' \\
\texttt{leap} & max interval name & ``Melodic motion must be smooth: no melodic leap
larger than a perfect fifth.'' \\
\texttt{structure} & form label, unit size & ``The tune must be exactly 16 bars with
an AABA form in 4-bar units: labeled units with the same letter must be note-for-note
identical, and units with different letters must differ.'' \\
\bottomrule
\end{tabular}
\end{table*}

\noindent The constraint texts are designed to be unambiguous to a musically literate
reader. The \texttt{meter} renderer explicitly states ``every bar must contain exactly
the right total duration'' because simply saying ``the meter is 3/4'' could be
interpreted as a header-only requirement. Similarly, the \texttt{structure} renderer
spells out the matching/differing requirement rather than assuming the model knows
what AABA form means.

\subsection{Paraphrase Examples}
\label{app:paraexamples}

In Table~\ref{tab:paraex}, we show three examples of the paraphrase suite, which re-renders the
same constraint values with alternative wordings; they illustrate the range of surface variation,
and the constraint parameters are unchanged.

\begin{table*}[t]
\centering
\caption{Paraphrase examples. Each row shows the original constraint text and one
paraphrased variant; the constraint parameters are identical.}
\label{tab:paraex}
\small
\begin{tabular}{p{0.42\textwidth}p{0.42\textwidth}}
\toprule
Original & Paraphrased \\
\midrule
The tune must be in A minor. &
Compose the melody in the key of A minor. \\[4pt]
The meter must be 4/4, and every bar must contain exactly the right total
duration. &
Write it in 4/4 time, and make every measure add up to exactly the right
duration. \\[4pt]
Melodic motion must be smooth: no melodic leap larger than an octave. &
Keep the melodic motion conjunct: no interval between consecutive notes may
exceed 12 semitones. \\
\bottomrule
\end{tabular}
\end{table*}

\FloatBarrier
\section{Additional Experiments}
\label{app:addexp}

In this section, we present additional experiments. Appendix~\ref{app:matched} contains the
matched comparison; all other RL analyses in this section preserve the diagnostic study, with SFT
initialisation unless a Direct row or another configuration is explicitly named, and
Table~\ref{tab:historicalmain} collects those historical reference scores.

Unless explicitly labelled otherwise, all-satisfied rates use $A$ (Equation~\ref{eq:allsat}),
as in the main comparison; per-family statistics report the individual verifier verdicts.
Appendix~\ref{app:strictaudit} separately reports the additional format-gate audit.

\subsection{Per-Family Satisfaction Across Models}
\label{app:familybreak}

\begin{table*}[t]
\centering
\caption{Per-family satisfaction over all $2{,}180$ items for every model of Table~\ref{tab:main}.
Each column marks its {\fboxsep=1.2pt\colorbox{Best}{best}} and
{\fboxsep=1.2pt\colorbox{Secondbest}{runner-up}}; the {\fboxsep=1.2pt\colorbox{Ours}{shaded
row}} is our main configuration.}
\label{tab:familybreak}
\footnotesize
\setlength{\tabcolsep}{4pt}
\begin{adjustbox}{max width=\textwidth}
% <<< tables/family_breakdown (make_family_breakdown_table.py)
\begin{tabular}{lccccccccc}
\toprule
Model & \texttt{key} & \texttt{meter} & \texttt{length} & \texttt{range} & \texttt{final} & \texttt{rhythm} & \texttt{leap} & \texttt{structure} & All \\
Requests & 786 & 773 & 609 & 744 & 756 & 761 & 756 & 733 & 2180 \\
\midrule
\multicolumn{10}{l}{\emph{Zero-shot}} \\
Qwen3-4B & 0.242 & 0.109 & 0.016 & 0.293 & 0.278 & 0.332 & 0.336 & 0.057 & 0.068 \\
Qwen3-8B & 0.125 & 0.079 & 0.043 & 0.179 & 0.168 & 0.152 & 0.246 & 0.010 & 0.064 \\
Qwen3-32B & \cellcolor{Secondbest}\underline{0.767} & 0.219 & 0.148 & 0.628 & \cellcolor{Secondbest}\underline{0.868} & 0.585 & 0.574 & 0.150 & 0.204 \\
Llama-3.1-70B & 0.719 & 0.234 & 0.097 & 0.671 & 0.837 & 0.608 & 0.640 & 0.117 & 0.197 \\
MiniMax-M3 & 0.468 & 0.248 & 0.281 & 0.456 & 0.460 & 0.250 & 0.452 & 0.030 & 0.215 \\
ChatMusician-7B & 0.159 & 0.378 & 0.049 & 0.384 & 0.194 & 0.322 & 0.565 & 0.000 & 0.084 \\
\midrule
\multicolumn{10}{l}{\emph{Trained (Qwen3-4B-Instruct-2507)}} \\
SFT & 0.660 & 0.724 & \cellcolor{Secondbest}\underline{0.819} & 0.777 & 0.333 & 0.603 & 0.683 & 0.610 & 0.346 \\
RFT & 0.617 & 0.464 & 0.256 & 0.761 & 0.660 & 0.812 & 0.737 & 0.538 & 0.342 \\
Direct GRPO ($R=A$) & 0.719 & 0.820 & 0.565 & \cellcolor{Best}\textbf{0.907} & 0.788 & \cellcolor{Best}\textbf{0.937} & \cellcolor{Secondbest}\underline{0.819} & \cellcolor{Secondbest}\underline{0.644} & \cellcolor{Secondbest}\underline{0.545} \\
\method{} (SFT init.) & 0.670 & \cellcolor{Best}\textbf{0.872} & \cellcolor{Best}\textbf{0.844} & \cellcolor{Secondbest}\underline{0.867} & 0.487 & 0.629 & 0.798 & 0.506 & 0.424 \\
\rowcolor{Ours}
\method{} (Direct) & 0.735 & 0.820 & 0.657 & \cellcolor{Secondbest}\underline{0.867} & 0.796 & \cellcolor{Secondbest}\underline{0.928} & \cellcolor{Best}\textbf{0.853} & \cellcolor{Best}\textbf{0.658} & \cellcolor{Best}\textbf{0.580} \\
\midrule
\multicolumn{10}{l}{\emph{Trained (Qwen3-8B)}} \\
\method{} (Direct) & \cellcolor{Best}\textbf{0.805} & \cellcolor{Secondbest}\underline{0.841} & 0.731 & 0.817 & \cellcolor{Best}\textbf{0.905} & 0.913 & 0.792 & 0.307 & 0.530 \\
\bottomrule
\end{tabular}
% >>> tables/family_breakdown
\end{adjustbox}
\end{table*}

In Table~\ref{tab:familybreak}, we break the all-satisfied rate of every model of
Table~\ref{tab:main} down by constraint family over all $2{,}180$ items. Each cell is the fraction
of items requesting the family whose output satisfies it; an output rejected by the validation gate
fails every family it requests. The first row counts the items requesting each family.

\subsection{Instruction Wording}
\label{app:wording}

In this section, we test robustness to instruction wording. Rendering the same constraint values
in different words costs the matched Direct arm
$\ParaDropDirect$ joint satisfaction, against $\ParaDropBase$ for the untrained base model.
Drawing each training prompt from three renderings removes the effect ($\ParaDropAug$,
$p=\ParaPAug$) and costs $\AugMixedCost$ on Mixed,
so the wording sensitivity comes from the training distribution.

Every row pairs item for item with its own original, so the exact McNemar test applies; the
constraint values are identical in both members of a pair and only the surface form changes.
Paraphrase re-renders the $400$ single and seen-combination problems in a third English wording,
never used in training. Chinese re-renders the $300$ Mixed items, and no arm is trained on the
wording used there.

\begin{table*}[t]
\centering
\caption{Instruction-wording robustness on the matched runs. ``Lost'' and ``gained'' count items
that flip in each direction. The wording-augmented arm draws each training prompt from three
English renderings; the multilingual arm adds a fourth, Chinese one. Qwen3-8B's Chinese figure is
set by the parse gate: $0.74$ of its completions there omit the \texttt{X:} reference number,
which no other arm does.}
\label{tab:wording}
\small
\begin{tabular}{lcccrrc}
\toprule
Model & Variant & Original & Variant & Lost & Gained & $p$ \\
\midrule
Qwen3-4B zero-shot        & Paraphrase & 0.163 & 0.113 &  49 & 29 & $0.031$ \\
SFT (4B)                  & Paraphrase & 0.580 & 0.540 &  40 & 24 & $0.060$ \\
Direct GRPO ($R{=}A$)     & Paraphrase & 0.853 & 0.805 &  29 & 10 & $0.0034$ \\
\method{} (Direct)        & Paraphrase & 0.900 & 0.777 &  54 &  5 & $1.9{\times}10^{-11}$ \\
\method{} (Qwen3-8B)      & Paraphrase & 0.810 & 0.787 &  27 & 18 & $0.233$ \\
\method{} wording-aug.    & Paraphrase & 0.815 & 0.797 &  21 & 14 & $0.311$ \\
\method{} multiling.-aug. & Paraphrase & 0.797 & 0.838 &  13 & 29 & $0.020$ \\
\midrule
Qwen3-4B zero-shot        & Chinese & 0.160 & 0.013 &  46 &  2 & $8.4{\times}10^{-12}$ \\
SFT (4B)                  & Chinese & 0.477 & 0.377 &  52 & 22 & $0.00064$ \\
Direct GRPO ($R{=}A$)     & Chinese & 0.797 & 0.447 & 111 &  6 & $4.0{\times}10^{-26}$ \\
\method{} (Direct)        & Chinese & 0.807 & 0.553 &  78 &  2 & $5.4{\times}10^{-21}$ \\
\method{} (Qwen3-8B)      & Chinese & 0.657 & 0.187 & 143 &  2 & $4.7{\times}10^{-40}$ \\
\method{} wording-aug.    & Chinese & 0.740 & 0.587 &  50 &  4 & $3.8{\times}10^{-11}$ \\
\method{} multiling.-aug. & Chinese & 0.737 & 0.543 &  74 & 16 & $4.4{\times}10^{-10}$ \\
\bottomrule
\end{tabular}
\end{table*}

Three observations. First, the paraphrase effect scales with how much a run was trained on one
wording: the matched Direct arm, the highest on the original wording, loses $0.123$ under
paraphrase, and the untrained model loses $0.050$.
Second, wording variation in training buys invariance at a price (Table~\ref{tab:wordingcost}). The
English-augmented arm loses $0.05$ to $0.11$ on Single, Seen, High-$k$ and Structure, whose items
contain no held-out pair, and stays within $0.005$ of the matched Direct arm on Unseen and Dense,
where every item contains one; the multilingual arm loses on every suite.
Third, adding a Chinese rendering to training does not improve Chinese compliance: the
multilingual arm reaches $0.543$ against the English-only augmented arm's $0.587$ and the matched
Direct arm's $0.553$, with every completion clearing the parse gate. What transfers across
languages is wording variety rather than the language itself, and the bottleneck is
wording generalisation within a language.

\begin{table}[t]
\centering
\caption{What wording augmentation costs, against the matched Direct run under the same recipe,
seed, budget and group size. Both augmented arms spend the same $300$ updates over more
renderings.}
\label{tab:wordingcost}
\small
\begin{adjustbox}{max width=\columnwidth}
\begin{tabular}{lccc}
\toprule
Suite & \method{} (Direct) & wording-aug. & multiling.-aug. \\
\midrule
Mixed            & 0.807 & 0.740 & 0.737 \\
Single           & 0.985 & 0.915 & 0.915 \\
Seen             & 0.815 & 0.715 & 0.680 \\
Unseen           & 0.595 & 0.600 & 0.500 \\
High-$k$         & 0.575 & 0.463 & 0.506 \\
Structure        & 0.967 & 0.917 & 0.667 \\
Dense ($k{=}4$)  & 0.362 & 0.360 & 0.294 \\
Edge             & 0.310 & 0.292 & 0.270 \\
\midrule
Paraphrase       & 0.777 & 0.797 & 0.838 \\
Chinese          & 0.553 & 0.587 & 0.543 \\
\bottomrule
\end{tabular}
\end{adjustbox}
\end{table}

\subsection{Free-Form Requests}
\label{app:freeform}

In this section, we test requests worded as a user would write them, whereas every \bench{}
prompt lists its constraints in template sentences. We construct $60$ free-form requests ($12$, $16$, $16$ and $16$ at $k=1$ to $4$)
that state each constraint in free-form English, with varied order, synonyms (measures, time
signature, lowest and highest note) and indirect phrasing (``bring it home on C'', ``no
anacrusis''), for example ``I'd like something in waltz time, so 3/4, and please make sure every
measure adds up exactly.'' Parameter values stay inside the default sampler domain and no request
contains a held-out pair, so the set changes only the wording. Each request keeps the format rules of the template and has a twin that renders the same constraints with the template. Both sets are scored by the unchanged verifiers
under the protocol of Table~\ref{tab:main} (\texttt{scripts/make\_bench\_freeform.py};
\texttt{paper/make\_robustness\_tables.py}).

Table~\ref{tab:freeform} reports every model we ran. Direct \method{} has the highest free-form
rate among the 4B models, $0.583$, and item by item solves more free-form requests than each of
them: $11$ to $3$ against Direct GRPO ($R=A$), $21$ to $9$ against the SFT-initialised run,
$18$ to $8$ against RFT, $21$ to $5$ against SFT and $32$ to $1$ against Qwen3-4B. Every trained
model clears the validation gate on at least $58$ of the $60$ free-form requests, against $24$
and $26$ for the untrained Qwen3-4B and Qwen3-8B. The Qwen3-8B policy reaches $0.717$, the
highest rate overall.

\begin{table*}[t]
\centering
\caption{Free-form requests ($60$ items) against their template-worded twins, which request the
same constraints. ``Free-form gate'' counts completions that pass the validation gate; ``lost''
and ``gained'' count items that flip from the template twin to the free-form request; $p$ is the
exact McNemar test on these flips.}
\label{tab:freeform}
\small
% <<< tables/freeform (make_robustness_tables.py)
\begin{tabular}{lcccrrc}
\toprule
Model & Template & Free-form & Free-form gate & Lost & Gained & $p$ \\
\midrule
Qwen3-4B (zero-shot) & 0.117 & 0.067 & 24/60 & 6 & 3 & $0.51$ \\
SFT & 0.483 & 0.317 & 60/60 & 12 & 2 & $0.013$ \\
RFT & 0.500 & 0.417 & 58/60 & 7 & 2 & $0.18$ \\
Direct GRPO ($R=A$) & 0.700 & 0.450 & 60/60 & 16 & 1 & $<0.001$ \\
\method{} (SFT init.) & 0.617 & 0.383 & 60/60 & 15 & 1 & $<0.001$ \\
\method{} (Direct) & 0.717 & 0.583 & 60/60 & 10 & 2 & $0.039$ \\
\midrule
Qwen3-8B (zero-shot) & 0.150 & 0.133 & 26/60 & 4 & 3 & $1.00$ \\
\method{} (Direct, Qwen3-8B) & 0.767 & 0.717 & 60/60 & 8 & 5 & $0.58$ \\
\bottomrule
\end{tabular}
% >>> tables/freeform
\end{table*}

\subsection{General Capability}
\label{app:general}

In this section, we evaluate both Direct policies and their base models on three general
benchmarks: IFBench
\citep{pyatkin2026generalizing}, all $300$ test prompts scored by the official strict and loose
checkers at the prompt level; MMLU-Redux \citep{gema2025we}, a fixed subject-proportional sample
of $1{,}000$ questions answered zero-shot; and GSM8K \citep{cobbe2021training}, the full
$1{,}319$-question test split with zero-shot
chain of thought. Decoding is greedy under each model's own chat template with at most $4{,}096$
new tokens, and Qwen3-8B runs with thinking disabled, as in Table~\ref{tab:main}.
Table~\ref{tab:general} gives the item-level comparison behind Table~\ref{tab:generalmain}; no
difference is significant at either scale (exact McNemar, $p \ge 0.096$).

\begin{table*}[t]
\centering
\caption{General benchmarks before and after training (accuracy, \%). \method{} is the Direct
policy at each scale; ``only ours / base'' counts items that only one of the two models answers
correctly, and $p$ is the exact McNemar test on them.}
\label{tab:general}
\small
\setlength{\tabcolsep}{4pt}
\begin{adjustbox}{max width=\textwidth}
% <<< tables/general (make_robustness_tables.py)
\begin{tabular}{lcccccccc}
\toprule
& \multicolumn{4}{c}{Qwen3-4B} & \multicolumn{4}{c}{Qwen3-8B} \\
\cmidrule(lr){2-5}\cmidrule(lr){6-9}
Benchmark & Base & \method{} & Only ours / base & $p$ & Base & \method{} & Only ours / base & $p$ \\
\midrule
IFBench, strict ($n=300$) & 32.0 & 29.3 & 5 / 13 & $0.096$ & 25.7 & 25.3 & 4 / 5 & $1.00$ \\
IFBench, loose ($n=300$) & 33.3 & 31.7 & 9 / 14 & $0.40$ & 30.3 & 30.0 & 5 / 6 & $1.00$ \\
MMLU-Redux ($n=1{,}000$) & 83.9 & 84.4 & 14 / 9 & $0.40$ & 81.5 & 82.6 & 25 / 14 & $0.11$ \\
GSM8K ($n=1{,}319$) & 93.9 & 94.1 & 15 / 13 & $0.85$ & 93.0 & 93.3 & 16 / 12 & $0.57$ \\
\bottomrule
\end{tabular}
% >>> tables/general
\end{adjustbox}
\end{table*}

\subsection{Paraphrase and Training Dynamics}
\label{app:extra}

In Table~\ref{tab:paraphrase}, we re-render the single and seen-combination problems with
alternative wordings of identical constraint values, and in Table~\ref{tab:dynamics}, we give the
numeric values behind Figure~\ref{fig:dyn6}(d).

\begin{table}[t]
\centering
\caption{Diagnostic study: instruction paraphrase, scored by $A$ (the format audit leaves these verdicts
unchanged). The original column averages the single and seen-combination splits; the paraphrased
column re-renders the same $400$ problems with alternative wordings of identical constraint
values. Rows are the diagnostic lineage; Table~\ref{tab:wording} reports the matched runs.}
\label{tab:paraphrase}
\small
% <<< tables/paraphrase
\begin{adjustbox}{max width=\columnwidth}
\begin{tabular}{lccc}
\toprule
Model & Original wording & Paraphrased & $\Delta$ \\
\midrule
SFT & 0.580 & 0.535 & -0.045 \\
\method{} (diag.) & 0.735 & 0.688 & -0.047 \\
\bottomrule
\end{tabular}
\end{adjustbox}

% >>> tables/paraphrase
\end{table}

\begin{table*}[t]
\centering
\caption{Diagnostic study: numeric values behind Figure~\ref{fig:dyn6}(d). Curve rows come from a
full-retention re-run of the SFT-initialised reference configuration; ``original run'' gives the
three checkpoints the original run retained. Its step-$300$ value $0.660$ is a separate
evaluation of the weights Table~\ref{tab:ablationdiag} scores at $0.663$, within the re-evaluation
shift of Appendix~\ref{app:evalnoise}; the two are reported separately. The last row
uses the symmetric structure reward of Appendix~\ref{app:structvariants}; it and the
symmetric row of Table~\ref{tab:ablationdiag} ($0.590$ on Mixed) are that reward's two separate
runs.}
\label{tab:dynamics}
\small
% <<< tables/dynamics
\begin{tabular}{lcccccc}
\toprule
Step & 50 & 100 & 150 & 200 & 250 & 300 \\
\midrule
All satisfied & 0.570 & 0.593 & 0.633 & 0.687 & 0.693 & 0.703 \\
\quad\texttt{meter} & 0.716 & 0.852 & 0.901 & 0.926 & 0.914 & 0.926 \\
\quad\texttt{final} & 0.505 & 0.505 & 0.558 & 0.642 & 0.632 & 0.632 \\
\quad\texttt{key} & 0.838 & 0.838 & 0.787 & 0.838 & 0.800 & 0.812 \\
\quad\texttt{structure} & 0.767 & 0.836 & 0.808 & 0.740 & 0.822 & 0.781 \\
\midrule
All satisfied (original run, retained ckpts) & -- & -- & -- & 0.630 & 0.640 & 0.660 \\
All satisfied (symmetric structure reward) & 0.477 & 0.463 & 0.463 & 0.530 & 0.537 & 0.553 \\
\bottomrule
\end{tabular}

% >>> tables/dynamics
\end{table*}

\subsection{Rejection Sampling and a Music-Specialised Baseline}
\label{app:bon}

In Table~\ref{tab:qualitymatched}(b), we spend the verifier at inference time instead: for each
item we draw eight completions, matching the GRPO group size $G = 8$ and the rollout
temperature ($1.0$), from both trained models, so selection sees the same per-prompt candidate
count as one GRPO group,
and report pass@$k$ under the unbiased estimator ($1 - \binom{n-c}{k}/\binom{n}{k}$
for $c$ of $n$ samples satisfying all constraints, averaged over items). The results show that selection
helps both models and \method{} stays ahead at every matched candidate budget, on both suites, so
verifier-guided selection stacks with training rather than substituting for it; note
that, unlike the trained policy alone, best-of-$N$ requires the verifier and
$N$ generations and $N$ verifications at deployment.

ChatMusician-7B \citep{yuan2024chatmusician} is evaluated zero-shot under the same
protocol as every other untrained model, through its released
\texttt{Human:}/\texttt{Assistant:} template and greedy decoding
(context $2{,}048$ tokens). Under $A$, it reaches $\Chatmusiciankilltestsuite$ on the mixed
suite and $\Chatmusicianbenchstructure$ on the structure split, where its parse rate
is $0.158$. Under the additional format gate its score is zero on all eight suites
(Tables~\ref{tab:strictscores} and~\ref{tab:strictextra}), mainly because required headers are missing.
Trained on corpus ABC, it writes repeat signs on most form prompts,
which the instruction forbids and the parse gate rejects. Its training objective is
stylistic generation rather than stated requirements.

\subsection{Rejection-Sampling Fine-Tuning}
\label{app:rft}

In this section, we use rejection-sampling fine-tuning (RFT) to ask whether the verifier's signal
is better spent filtering supervised targets than as a reward. For each of the first $1{,}800$ prompts of the reinforcement pool we draw $8$ completions
from Qwen3-4B-Instruct-2507 at temperature $1.0$, $14{,}400$ in all, the number GRPO samples over $300$
updates of $48$. For each prompt we keep the first completion that passes the gate and satisfies
every requested constraint, which leaves $773$ pairs, and fine-tune the full model on them for three
epochs (learning rate $10^{-5}$, effective batch $16$, $147$ steps, seed $43$). Evaluation follows
the protocol of Table~\ref{tab:main}.

\subsection{Training versus Inference-Time Selection}
\label{app:passk}

In this section, we compare training with inference-time selection. For each Mixed item we draw
$64$ completions at temperature $1.0$ and report pass@$k$ under the unbiased estimator of
Appendix~\ref{app:bon}; with the verifier as the selector, pass@$k$ is the
accuracy of best-of-$k$ reranking. The results in Table~\ref{tab:passk} show that, at every
budget, the trained policy outscores its base model reranked from the same number of samples. To match the trained policy's single
greedy decode, Qwen3-4B needs more than $64$ reranked samples and Qwen3-8B between $32$ and $64$.

Table~\ref{tab:rerank8} extends the comparison to all eight suites at a budget of
eight samples per item. Mixed reuses the $64$ samples above under the same estimator; the other
seven suites draw eight completions per item at temperature $1.0$
(\texttt{scripts/run\_rerank8\_card0.sh} and \texttt{run\_rerank8\_card4.sh}, summarised by
\texttt{src/analyse\_rerank8.py}). Reranking raises the trained policy on every suite, by
$0.008$ to $0.262$, and on every suite the untrained model's best of eight stays
below the trained policy's single greedy decode.

\begin{table}[t]
\centering
\caption{Mixed suite: greedy all-satisfied rate and pass@$k$ from $64$ samples at temperature $1.0$.}
\label{tab:passk}
\small
\begin{adjustbox}{max width=\columnwidth}
\begin{tabular}{lcccc}
\toprule
Model & Greedy & pass@1 & pass@8 & pass@64 \\
\midrule
Qwen3-4B zero-shot        & 0.160 & 0.132 & 0.393 & 0.637 \\
SFT (4B)                  & 0.477 & 0.417 & 0.847 & 0.973 \\
RFT (4B)                  & 0.543 & 0.499 & 0.740 & 0.867 \\
Direct GRPO ($R=A$)       & 0.797 & 0.765 & 0.878 & 0.920 \\
\method{} (SFT init.)     & 0.577 & 0.653 & 0.946 & 1.000 \\
\method{} (Direct)        & 0.807 & 0.778 & 0.897 & 0.937 \\
\midrule
Qwen3-8B zero-shot        & 0.123 & 0.135 & 0.398 & 0.713 \\
\method{} (Direct, 8B)    & 0.657 & 0.667 & 0.767 & 0.823 \\
\bottomrule
\end{tabular}
\end{adjustbox}
\end{table}

\subsection{Constraint-Count Ladder}
\label{app:kladder}

In this section, we extend the number of constraints beyond the training range, as training
prompts use $k\in\{1,2,3\}$. \texttt{src/gen\_k\_ladder.py} draws suites of $300$ items at
$k=2$, $3$, $5$, $6$ and $7$ with the sampler and satisfiability filter of Dense, which is the
$k=4$ rung. The four held-out pairs form a perfect matching over the eight families, so any set
of five or more families contains one; the ladder therefore keeps a held-out pair in every item
at every rung, as Dense does, and varies only $k$. Because \texttt{length} and \texttt{structure}
cannot both be requested, $k=7$ is the largest admissible value. Table~\ref{tab:kladderfull}
lists every arm evaluated on the ladder.

\begin{table}[!ht]
\centering
\caption{All-satisfied rate on the constraint-count ladder ($k=4$ is Dense). Every item contains
a held-out pair; training prompts hold at most three constraints.}
\label{tab:kladderfull}
\footnotesize
\setlength{\tabcolsep}{3pt}
\begin{adjustbox}{max width=\columnwidth}
% <<< tables/kladder_full (make_kladder_tables.py)
\begin{tabular}{l*{6}{c}}
\toprule
Model & $k=2$ & $k=3$ & $k=4$ & $k=5$ & $k=6$ & $k=7$ \\
\midrule
Qwen3-4B (zero-shot) & 0.073 & 0.037 & 0.008 & 0.003 & 0.000 & 0.000 \\
SFT & 0.467 & 0.357 & 0.290 & 0.140 & 0.117 & 0.073 \\
RFT & 0.530 & 0.207 & 0.098 & 0.053 & 0.007 & 0.003 \\
Direct GRPO ($R=A$) & 0.710 & 0.487 & 0.322 & 0.170 & 0.060 & 0.063 \\
\method{} (SFT init.) & 0.630 & 0.403 & 0.366 & 0.187 & 0.200 & 0.117 \\
\method{} (Direct) & 0.647 & 0.530 & 0.362 & 0.230 & 0.160 & 0.087 \\
\method{} (Direct, Qwen3-8B) & 0.707 & 0.493 & 0.404 & 0.273 & 0.250 & 0.153 \\
\method{} (Direct, wording-aug.) & 0.680 & 0.477 & 0.360 & 0.167 & 0.117 & 0.087 \\
\bottomrule
\end{tabular}
% >>> tables/kladder_full
\end{adjustbox}
\end{table}

\subsection{Direct Reward Ablations}
\label{app:directabl}

In Table~\ref{tab:ablation}, we vary one part of Equation~\ref{eq:reward} at a time from the Direct
configuration of Table~\ref{tab:main}. Every arm uses seed $43$, $300$ updates of $48$
completions, the same rollout server and the evaluation of Table~\ref{tab:main}; at the first
update every arm except the $21\%$ pool samples exactly the default's rollouts. The arms are:
\begin{itemize}[leftmargin=1.2em,itemsep=1pt,topsep=2pt,parsep=0pt]
\item \textbf{Gate on $A$ only.} An output that the gate rejects for repeat signs
or for a degenerate or polyphonic score, but that music21 still parses, receives
$\lambda\frac{1}{k}\sum_{(f,\theta)\in C} s_f$ instead of $0$. $A$ keeps the gate, so the joint
term and the evaluation criterion are unchanged; outputs with no ABC block or no parse still
score $0$.
\item \textbf{$\lambda=0$, $0.5$ and $1$.} The outer mixture of Equation~\ref{eq:reward};
$\lambda=0$ is Direct GRPO ($R=A$) of Table~\ref{tab:main}.
\item \textbf{Binary form credit.} $s_{\texttt{structure}}=v_{\texttt{structure}}$.
\item \textbf{Symmetric form credit} and \textbf{equal form weights.} $0.5\bar e+0.5\bar d$, with
credit $1$ for a passing form and $0$ for four identical units, and
$0.5\bar e+0.5v_{\text{contrast}}$ (Appendix~\ref{app:structvariants}).
\item \textbf{Form share $21\%$.} The prompt pool before upsampling: $1{,}282$ of $6{,}000$
prompts contain \texttt{structure}, and $k=1/2/3$ is split $1{,}863/3{,}076/1{,}061$ against
$2{,}046/3{,}060/894$.
\end{itemize}
Table~\ref{tab:ablationdirectfull} gives the counts behind the All column of
Table~\ref{tab:ablation}, with exact McNemar tests against the default, Holm-corrected over the
eight arms. Without the all-satisfied bonus ($\lambda=1$) the policy fails to parse $25$ outputs: $19$
repeat notes without bar lines until the token limit and $6$ end before four bars.

\begin{table*}[t]
\centering
\caption{Direct reward ablations over all $2{,}180$ items: parsed outputs, solved items, and the
items solved only by the variant or only by the default, with exact McNemar $p$ and its Holm
adjustment over the eight arms.}
\label{tab:ablationdirectfull}
\small
% <<< tables/ablation_direct_full (make_ablation_direct_tables.py)
\begin{tabular}{lccccccc}
\toprule
Variant & Parsed & Solved & Rate & Only variant & Only default & $p$ & Holm $p$ \\
\midrule
Default & 2176 & 1265 & 0.580 & -- & -- & -- & -- \\
\midrule
Gate on $A$ only & 2179 & 1164 & 0.534 & 125 & 226 & $7.7{\times}10^{-8}$ & $3.9{\times}10^{-7}$ \\
\midrule
$\lambda=0$ ($R=A$) & 2180 & 1188 & 0.545 & 152 & 229 & $0.0001$ & $0.0002$ \\
$\lambda=0.5$ & 2168 & 1095 & 0.502 & 150 & 320 & $3.2{\times}10^{-15}$ & $2.6{\times}10^{-14}$ \\
$\lambda=1$ (no bonus) & 2155 & 1206 & 0.553 & 146 & 205 & $0.0019$ & $0.0019$ \\
\midrule
Binary form credit & 2180 & 1190 & 0.546 & 129 & 204 & $4.7{\times}10^{-5}$ & $0.0002$ \\
Symmetric form credit & 2180 & 1122 & 0.515 & 134 & 277 & $1.5{\times}10^{-12}$ & $1.0{\times}10^{-11}$ \\
Equal form weights & 2178 & 1186 & 0.544 & 142 & 221 & $4.0{\times}10^{-5}$ & $1.6{\times}10^{-4}$ \\
\midrule
Form share $21\%$ & 2179 & 1130 & 0.518 & 188 & 323 & $2.5{\times}10^{-9}$ & $1.5{\times}10^{-8}$ \\
\bottomrule
\end{tabular}
% >>> tables/ablation_direct_full
\end{table*}

\subsection{Edge Subsets and a Stricter Key Criterion}
\label{app:rescore}

In this section, we rescore Edge subsets and apply a stricter key criterion; both analyses use
saved per-item verdicts, without regeneration or retraining.
\texttt{src/rescore\_subsets.py} computes the key-sensitivity audit under the historical gate;
the final \texttt{make\_checkpoint\_assets.py} pass uses the same $A$ verdicts for Edge subsets.

Edge mixes two factors. Table~\ref{tab:edgesubsets} splits it by whether an item requests one of
the four held-out pairs. On the $\EdgeFreeN$ items without a held-out family pair,
the diagnostic reference reaches $\EdgeFreeOurs$ against $\EdgeFreeSft$ for the warm start
(exact McNemar $p=\PEdgeFreeSft$) and $\EdgeFreeMinimax$ for MiniMax-M3
($p=\PEdgeFreeMinimax$), a wider margin over the warm start than on Edge as a whole.

Table~\ref{tab:edgefactors} instead splits Edge by the factor it extrapolates: items that request
a meter outside $\{4/4, 3/4, 6/8, 2/4\}$ and items that request a length outside $\{8, 12, 16\}$
bars, for every model of Table~\ref{tab:main}.

\begin{table}[!ht]
\centering
\caption{Edge all-satisfied rate on the items whose requested meter or length lies outside the
RL prompt domain.}
\label{tab:edgefactors}
\footnotesize
\setlength{\tabcolsep}{3pt}
\begin{adjustbox}{max width=\columnwidth}
% <<< tables/edge_factors (make_edge_factor_table.py)
\begin{tabular}{lccc}
\toprule
Model & Unseen meter ($n=160$) & Unseen length ($n=127$) & All Edge \\
\midrule
Qwen3-4B & 0.000 & 0.000 & 0.010 \\
Qwen3-8B & 0.019 & 0.024 & 0.032 \\
Qwen3-32B & 0.037 & 0.063 & 0.112 \\
Llama-3.1-70B & 0.019 & 0.055 & 0.058 \\
MiniMax-M3 & 0.138 & 0.142 & 0.160 \\
ChatMusician-7B & 0.044 & 0.008 & 0.046 \\
SFT & 0.144 & 0.055 & 0.076 \\
RFT & 0.131 & 0.102 & 0.128 \\
Direct GRPO ($R=A$) & 0.269 & 0.189 & 0.300 \\
\method{} (SFT init.) & 0.188 & 0.094 & 0.134 \\
\method{} (Direct) & 0.287 & 0.181 & 0.310 \\
\method{} (Direct, Qwen3-8B) & 0.300 & 0.260 & 0.252 \\
\bottomrule
\end{tabular}
% >>> tables/edge_factors
\end{adjustbox}
\end{table}

\begin{table*}[t]
\centering
\caption{Diagnostic study: Edge split by held-out-pair membership, using the main-table success criterion.
Pair-free items contain none of the four held-out family pairs.}
\label{tab:edgesubsets}
\small
% <<< tables/edge_subsets
\begin{tabular}{lccc}
\toprule
Model & All ($n=500$) & Pair-free ($n=329$) & With held-out pair ($n=171$) \\
\midrule
Qwen3-4B & 0.010 & 0.012 & 0.006 \\
Qwen3-8B & 0.032 & 0.049 & 0.000 \\
Qwen3-32B & 0.112 & 0.125 & 0.088 \\
Llama-3.1-70B & 0.058 & 0.067 & 0.041 \\
MiniMax-M3 & 0.160 & 0.195 & 0.094 \\
ChatMusician-7B & 0.046 & 0.055 & 0.029 \\
SFT & 0.076 & 0.100 & 0.029 \\
\method{} (re-evaluated) & 0.162 & 0.210 & 0.070 \\
\bottomrule
\end{tabular}

% >>> tables/edge_subsets
\end{table*}

The deployed key check requires the header to match and the analyzed key to be the requested key
or its relative (\S\ref{app:keycal}). Tables~\ref{tab:strictkeyscores}
and~\ref{tab:strictkeyextra} hold the historical format gate fixed and report every rate with
that tolerance and without it, that is with
\texttt{key} satisfied only when header and analysis both give the requested key. Rates stay
unchanged or decrease. The largest baseline decrease is $0.060$ (Qwen3-32B on Unseen);
the largest decrease for the fixed \method{} checkpoint is $0.030$ on Seen.
The diagnostic \method{} checkpoint remains the highest-scoring model on seven of the eight splits.

\begin{table*}[t]
\centering
\caption{Diagnostic study: All-satisfied rate with the relative-key tolerance and without it
(hybrid/exact key), six original suites, holding the historical format gate fixed.
Baselines show the paired rates; the two
\method{} rows report the fixed-checkpoint rate under each key rule.}
\label{tab:strictkeyscores}
\scriptsize
\setlength{\tabcolsep}{3pt}
% <<< tables/strictkey_scores
\begin{tabular}{lcccccc}
\toprule
Model / key rule & Mixed & Single & Seen & Unseen & High-$k$ & Structure \\
\midrule
Qwen3-4B & 0.160/0.157 & 0.225/0.210 & 0.100/0.090 & 0.040/0.030 & 0.037/0.031 & 0.100/0.100 \\
Qwen3-8B & 0.123/0.117 & 0.210/0.200 & 0.115/0.110 & 0.050/0.050 & 0.006/0.006 & 0.000/0.000 \\
Qwen3-32B & 0.363/0.340 & 0.535/0.485 & 0.250/0.240 & 0.190/0.130 & 0.094/0.062 & 0.267/0.267 \\
Llama-3.1-70B & 0.380/0.370 & 0.695/0.675 & 0.300/0.295 & 0.195/0.185 & 0.056/0.056 & 0.167/0.167 \\
MiniMax-M3 & 0.367/0.363 & 0.580/0.550 & 0.315/0.310 & 0.190/0.185 & 0.100/0.100 & 0.008/0.008 \\
ChatMusician-7B & 0.160/0.147 & 0.290/0.280 & 0.130/0.120 & 0.090/0.085 & 0.006/0.000 & 0.000/0.000 \\
SFT & 0.477/0.453 & 0.745/0.715 & 0.415/0.385 & 0.440/0.415 & 0.231/0.225 & 0.600/0.592 \\
\method{} (hybrid) & $0.663$ & $0.835$ & $0.635$ & $0.595$ & $0.356$ & $0.808$ \\
\method{} (strict key) & $0.637$ & $0.815$ & $0.605$ & $0.580$ & $0.344$ & $0.783$ \\
\bottomrule
\end{tabular}

% >>> tables/strictkey_scores
\end{table*}

\begin{table}[t]
\centering
\caption{Diagnostic study: Hybrid/exact key on Dense and Edge, holding the historical format gate fixed;
\method{} uses the single re-evaluation.}
\label{tab:strictkeyextra}
\small
% <<< tables/strictkey_extra
\begin{adjustbox}{max width=\columnwidth}
\begin{tabular}{lcc}
\toprule
Model & Dense & Edge \\
\midrule
Qwen3-4B & 0.008/0.006 & 0.010/0.008 \\
Qwen3-8B & 0.020/0.018 & 0.032/0.032 \\
Qwen3-32B & 0.074/0.056 & 0.112/0.102 \\
Llama-3.1-70B & 0.040/0.032 & 0.058/0.048 \\
MiniMax-M3 & 0.088/0.084 & 0.160/0.152 \\
ChatMusician-7B & 0.020/0.014 & 0.046/0.042 \\
SFT & 0.290/0.266 & 0.076/0.070 \\
\method{} (re-evaluated) & 0.370/0.342 & 0.162/0.150 \\
\bottomrule
\end{tabular}
\end{adjustbox}

% >>> tables/strictkey_extra
\end{table}

\subsection{Significance Tests}
\label{app:significance}

In this section, we report significance tests for the diagnostic reward-design comparisons; the
matched Direct-versus-SFT-initialised tests are in Appendix~\ref{app:matched}.

Table~\ref{tab:sigstruct} reports exact McNemar tests for the single-run diagnostic
reward-design experiment under the main-table success criterion, pairing the same item IDs;
Appendix~\ref{app:checkpoint} lists the success counts
underlying the historical reference rates. Table~\ref{tab:pairedaudit}
reports the Edge comparison.

\begin{table*}[t]
\centering
\caption{Diagnostic study: Single-run reward-design comparisons (nominal exact McNemar $p$) over per-item outcomes.}
\label{tab:sigstruct}
\small
% <<< tables/sig_only
\setlength{\tabcolsep}{4.5pt}
\begin{tabular}{lcccccc}
\toprule
Comparison & Mixed & Single & Seen & Unseen & High-$k$ & Struct. \\
\midrule
Reference run vs.\ SFT & $<10^{-4}$ & $0.002$ & $<10^{-4}$ & $<10^{-4}$ & $0.007$ & $<10^{-4}$ \\
Reference run vs.\ binary credit, $21\%$ share & $0.076$ & $0.597$ & $1.000$ & $0.312$ & $1.000$ & $<10^{-4}$ \\
Reference run vs.\ binary credit, $33\%$ share & $0.822$ & $0.583$ & $0.625$ & $0.771$ & $0.678$ & $<10^{-4}$ \\
\bottomrule
\end{tabular}

% >>> tables/sig_only
\end{table*}

\begin{table}[t]
\centering
\caption{Diagnostic study: Paired comparison on Edge. ``Only RL'' and ``Only baseline'' count discordant
items in the designated single evaluation of each model.}
\label{tab:pairedaudit}
\scriptsize
% <<< tables/paired_audit
\begin{adjustbox}{max width=\columnwidth}
\begin{tabular}{lcccc}
\toprule
Comparison & Only RL & Only baseline & Exact $p$ & Direction \\
\midrule
Edge: RL vs.\ MiniMax-M3 & 62 & 61 & 1.000 & No detected difference \\
\bottomrule
\end{tabular}
\end{adjustbox}

% >>> tables/paired_audit
\end{table}

\subsection{Matched Comparison}
\label{app:matched}

In this section, we detail the matched comparison, which supplies the three 4B RL rows in
Table~\ref{tab:main}. All runs use seed $43$,
checkpoint $300$, the same $6{,}000$-prompt training pool and $48$ completions per update.
Two policy accelerators use micro-batch $2$ and accumulation $12$; a third serves rollouts
with memory fraction $0.92$. The remaining GRPO settings match Table~\ref{tab:hyperparams}.
Direct binary uses $\lambda=0$ ($R=A$), and both \method{} routes use $\lambda=0.7$.
The reward weight is the only changed training setting between the two Direct arms.
The SFT-initialised arm changes both the initial policy and its fixed KL reference.
All evaluations use the same item IDs and the all-satisfied criterion $A$.

\begin{table*}[t]
\centering
\caption{Success counts and rates for the matched final checkpoints.}
\label{tab:matchedcounts}
\small
% <<< inline/tables/matched_s43_counts.tex
\begin{tabular}{lccc}
\toprule
Split & Direct, binary & Direct, MusicRLVR & SFT init., MusicRLVR \\
\midrule
Mixed & 239/300 (0.797) & 242/300 (0.807) & 173/300 (0.577) \\
Single & 192/200 (0.960) & 197/200 (0.985) & 176/200 (0.880) \\
Seen & 149/200 (0.745) & 163/200 (0.815) & 117/200 (0.585) \\
Unseen & 113/200 (0.565) & 119/200 (0.595) & 106/200 (0.530) \\
High-$k$ & 75/160 (0.469) & 92/160 (0.575) & 56/160 (0.350) \\
Structure & 109/120 (0.908) & 116/120 (0.967) & 46/120 (0.383) \\
Dense & 161/500 (0.322) & 181/500 (0.362) & 183/500 (0.366) \\
Edge & 150/500 (0.300) & 155/500 (0.310) & 67/500 (0.134) \\
\bottomrule
\end{tabular}
% >>> inline/tables/matched_s43_counts.tex
\end{table*}

\begin{table*}[t]
\centering
\caption{Parse-gate pass rates on the six original suites, for every model of
Table~\ref{tab:main}. SFT and every GRPO-trained checkpoint clear the gate on every item.}
\label{tab:gate6}
\footnotesize
\renewcommand{\arraystretch}{0.85}
% <<< inline/tables/matched_s43_gate6.tex
\begin{tabular*}{\textwidth}{@{\extracolsep{\fill}}l*{6}{c}}
\toprule
Model & Mixed & Single & Seen & Unseen & High-$k$ & Structure \\
\midrule
Qwen3-4B & 0.513 & 0.450 & 0.510 & 0.545 & 0.362 & 0.925 \\
Qwen3-8B & 0.367 & 0.465 & 0.345 & 0.305 & 0.219 & 0.158 \\
Qwen3-32B & 0.973 & 1.000 & 0.960 & 0.945 & 0.938 & 0.983 \\
Llama-3.1-70B & 0.970 & 1.000 & 0.980 & 0.980 & 0.944 & 1.000 \\
MiniMax-M3 & 0.577 & 0.670 & 0.565 & 0.465 & 0.531 & 0.083 \\
ChatMusician-7B & 0.693 & 0.800 & 0.680 & 0.670 & 0.575 & 0.158 \\
\midrule
SFT (4B) & 1.000 & 1.000 & 1.000 & 1.000 & 1.000 & 1.000 \\
\rowcolor{black!6}
Direct GRPO ($R=A$) & 1.000 & 1.000 & 1.000 & 1.000 & 1.000 & 1.000 \\
\rowcolor{black!6}
\method{} (SFT init.) & 1.000 & 1.000 & 1.000 & 1.000 & 1.000 & 1.000 \\
\rowcolor{black!6}
\method{} (Direct) & 1.000 & 1.000 & 1.000 & 1.000 & 1.000 & 1.000 \\
\midrule
RFT (4B) & 1.000 & 0.980 & 0.995 & 1.000 & 0.988 & 1.000 \\
\method{} (Direct, Qwen3-8B) & 1.000 & 1.000 & 1.000 & 1.000 & 1.000 & 1.000 \\
\bottomrule
\end{tabular*}
% RFT and Qwen3-8B rows added by hand from outputs/eval_<tag>_<suite>.verified.jsonl ("parsed").
% >>> inline/tables/matched_s43_gate6.tex
\end{table*}

\begin{table}[t]
\centering
\caption{Parse-gate pass rates for the matched evaluations.}
\label{tab:matchedparse}
\scriptsize
\setlength{\tabcolsep}{3pt}
% <<< inline/tables/matched_s43_parse.tex
\begin{adjustbox}{max width=\columnwidth}
\begin{tabular}{lrrrrrrrr}
\toprule
Model & Mixed & Single & Seen & Unseen & High-$k$ & Structure & Dense & Edge \\
\midrule
Direct GRPO ($R=A$) & 1.000 & 1.000 & 1.000 & 1.000 & 1.000 & 1.000 & 1.000 & 1.000 \\
\method{} (SFT init.) & 1.000 & 1.000 & 1.000 & 1.000 & 1.000 & 1.000 & 1.000 & 1.000 \\
\method{} (Direct) & 1.000 & 1.000 & 1.000 & 1.000 & 1.000 & 1.000 & 1.000 & 0.992 \\
\bottomrule
\end{tabular}
\end{adjustbox}
% >>> inline/tables/matched_s43_parse.tex
\end{table}

\begin{table*}[t]
\centering
\caption{Exact two-sided McNemar tests for the matched Direct \method{} checkpoint against each
control; discordant counts are successes by only one checkpoint, and Holm correction is applied
over the eight suites within each comparison.}
\label{tab:matchedtests}
\small
% <<< inline/tables/matched_s43_tests.tex
\begin{tabular}{llrrrr}
\toprule
Comparison & Split & Only Direct & Only control & Exact $p$ & Holm $p$ \\
\midrule
Binary reward & Mixed & 19 & 16 & $0.7359$ & $1.0000$ \\
 & Single & 5 & 0 & $0.0625$ & $0.3125$ \\
 & Seen & 25 & 11 & $0.0288$ & $0.1729$ \\
 & Unseen & 21 & 15 & $0.4050$ & $1.0000$ \\
 & High-$k$ & 32 & 15 & $0.0186$ & $0.1304$ \\
 & Structure & 7 & 0 & $0.0156$ & $0.1250$ \\
 & Dense & 78 & 58 & $0.1029$ & $0.4117$ \\
 & Edge & 42 & 37 & $0.6530$ & $1.0000$ \\
\midrule
SFT init. & Mixed & 90 & 21 & $<10^{-4}$ & $<10^{-4}$ \\
 & Single & 22 & 1 & $<10^{-4}$ & $<10^{-4}$ \\
 & Seen & 65 & 19 & $<10^{-4}$ & $<10^{-4}$ \\
 & Unseen & 36 & 23 & $0.1175$ & $0.2350$ \\
 & High-$k$ & 58 & 22 & $<10^{-4}$ & $0.0002$ \\
 & Structure & 71 & 1 & $<10^{-4}$ & $<10^{-4}$ \\
 & Dense & 90 & 92 & $0.9409$ & $0.9409$ \\
 & Edge & 122 & 34 & $<10^{-4}$ & $<10^{-4}$ \\
\bottomrule
\end{tabular}
% >>> inline/tables/matched_s43_tests.tex
\end{table*}

\paragraph{Pooled comparison.} The eight suites are disjoint item sets whose union is the
$\PoolN$ items of \bench{}, so the same paired test applies to the union. Against binary
reward, Direct \method{} satisfies $\PoolGradedCount$ items against $\PoolBinaryCount$
($\PoolGraded$ against $\PoolBinary$), winning $\PoolGradedWins$ and losing $\PoolBinaryWins$
of the $\the\numexpr\PoolGradedWins+\PoolBinaryWins\relax$ items on which the two disagree
(exact two-sided $p = \PoolGradedBinaryP$).
Against the SFT-initialised route it satisfies $\PoolGradedCount$ against $\PoolWarmCount$,
winning $\PoolGradedWarmWins$ and losing $\PoolWarmWins$ ($p\PoolGradedWarmP$); the
binary-reward arm also leads that route, $\PoolBinaryCount$ against $\PoolWarmCount$
($p\PoolBinaryWarmP$).

\paragraph{Per-suite breakdown.} Graded credit has the higher rate on all eight of the suites,
and the pooled comparison above is significant ($p = \PoolGradedBinaryP$). Against the
SFT-initialised route, Direct is ahead on seven suites, six of them significant after the
eight-suite Holm correction of Table~\ref{tab:matchedtests}. All three models pass the parse gate on every original-suite item; on Edge
the Direct \method{} model passes $496/500$, while the binary-reward and SFT-initialised models pass
$500/500$ (Tables~\ref{tab:gate6} and~\ref{tab:matchedparse}).

\paragraph{What the residual failures look like.}
Table~\ref{tab:matchedfail} decomposes the $58$ Mixed items the Direct \method{} checkpoint
does not satisfy, from the same saved per-item verdicts as Table~\ref{tab:matchedcounts}.
The failures are shallow rather than pervasive: $48$ of the $58$ miss a single family and only
$2$ miss three, and none fails the parse gate. \texttt{structure} has the largest per-family
deficit ($22$ of the $73$ items requesting it) and the top sole miss ($18$ items);
\texttt{length} is next in missed count ($14$) and \texttt{key} and \texttt{final} each miss $11$. The split-level patterns of Appendix~\ref{app:failpatterns} come from the diagnostic checkpoint.

\begin{table*}[t]
\centering
\caption{Matched study: the $58$ unsatisfied Mixed items of the Direct \method{} checkpoint.
``Missed'' counts items requesting a family whose verifier fails; ``sole miss'' restricts to the
$48$ items that miss exactly one family. \texttt{rhythm} is omitted, with no miss among the items
requesting it.}
\label{tab:matchedfail}
\small
\begin{tabular}{lccc}
\toprule
Family & Requested & Missed & Sole miss \\
\midrule
\texttt{final} & 95 & 11 & 7 \\
\texttt{structure} & 73 & 22 & 18 \\
\texttt{key} & 80 & 11 & 9 \\
\texttt{length} & 65 & 14 & 10 \\
\texttt{leap} & 89 & 7 & 4 \\
\texttt{meter} & 81 & 3 & 0 \\
\texttt{range} & 80 & 2 & 0 \\
\midrule
\multicolumn{4}{l}{\emph{Items by number of families missed:} 1: $48$, 2: $8$, 3: $2$; parse-gate
failures: $0$.} \\
\bottomrule
\end{tabular}
\end{table*}

\subsection{Diagnostic Reference Results}
\label{app:historical}

In Table~\ref{tab:historicalmain}, we give the diagnostic evaluations. Unless a diagnostic
explicitly states another setting, its RL results refer to the SFT-initialised checkpoint in
this table. The Direct route is reported only from the matched runs of
Table~\ref{tab:main}, so no Direct row appears here.

\begin{table*}[t]
\centering
\caption{Diagnostic reference results (SFT-initialised route): all-satisfied rates above,
parse-gate pass rates below. A different RL checkpoint from the matched comparison of Table~\ref{tab:main}. Colours rank rows within this table; missing entries were not evaluated.}
\label{tab:historicalmain}
\footnotesize
\renewcommand{\arraystretch}{0.85}
% Historical results preserved from the reference study.
% Models in rows; suites in columns. Best and runner-up are ranked per suite.
\setlength{\tabcolsep}{3.0pt}
\begin{tabular*}{\textwidth}{@{\hspace{\tabcolsep}\extracolsep{\fill}}l*{8}{c}@{\hspace{\tabcolsep}}}
\toprule
& \multicolumn{6}{c}{Six original splits} & \multicolumn{2}{c}{Later splits} \\
\cmidrule(lr){2-7}\cmidrule(lr){8-9}
Model & Mixed & Single & Seen & Unseen & High-$k$ & Structure & Dense & Edge \\
\midrule
\rowcolor{famname!8}
\multicolumn{9}{l}{\textcolor{famname}{\textit{Zero-shot}}} \\
Qwen3-4B & 0.160 & 0.225 & 0.100 & 0.040 & 0.037 & 0.100 & 0.008 & 0.010 \\
Qwen3-8B & 0.123 & 0.210 & 0.115 & 0.050 & 0.006 & 0.000 & 0.020 & 0.032 \\
Qwen3-32B & 0.363 & 0.535 & 0.250 & 0.190 & 0.094 & 0.267 & 0.074 & 0.112 \\
Llama-3.1-70B & 0.380 & 0.695 & 0.300 & 0.195 & 0.056 & 0.167 & 0.040 & 0.058 \\
MiniMax-M3 & 0.367 & 0.580 & 0.315 & 0.190 & 0.100 & 0.008 & 0.088 & \cellcolor{Secondbest}\underline{0.160} \\
ChatMusician-7B & 0.160 & 0.290 & 0.130 & 0.090 & 0.006 & 0.000 & 0.020 & 0.046 \\
\midrule
\rowcolor{famname!8}
\multicolumn{9}{l}{\textcolor{famname}{\textit{Trained (4B)}}} \\
SFT (4B) & \cellcolor{Secondbest}\underline{0.477} & \cellcolor{Secondbest}\underline{0.745} & \cellcolor{Secondbest}\underline{0.415} & \cellcolor{Secondbest}\underline{0.440} & \cellcolor{Secondbest}\underline{0.231} & \cellcolor{Secondbest}\underline{0.600} & \cellcolor{Secondbest}\underline{0.290} & 0.076 \\
\rowcolor{black!6}
\method{} (diag.) & \cellcolor{Best}\textbf{0.663} & \cellcolor{Best}\textbf{0.835} & \cellcolor{Best}\textbf{0.635} & \cellcolor{Best}\textbf{0.595} & \cellcolor{Best}\textbf{0.356} & \cellcolor{Best}\textbf{0.808} & \cellcolor{Best}\textbf{0.370} & \cellcolor{Best}\textbf{0.162} \\
\rowcolor{black!6}
\bottomrule
\end{tabular*}
\par\vspace{4pt}
{\footnotesize
\renewcommand{\arraystretch}{0.85}
\begin{tabular*}{\textwidth}{@{\extracolsep{\fill}}l*{6}{c}}
\toprule
\rowcolor{famname!8}
\multicolumn{7}{l}{\textcolor{famname}{\textit{Parse-gate pass rate (six original splits)}}} \\
Model & Mixed & Single & Seen & Unseen & High-$k$ & Structure \\
\midrule
Qwen3-4B & 0.513 & 0.450 & 0.510 & 0.545 & 0.362 & 0.925 \\
Qwen3-8B & 0.367 & 0.465 & 0.345 & 0.305 & 0.219 & 0.158 \\
Qwen3-32B & 0.973 & 1.000 & 0.960 & 0.945 & 0.938 & 0.983 \\
Llama-3.1-70B & 0.970 & 1.000 & 0.980 & 0.980 & 0.944 & 1.000 \\
MiniMax-M3 & 0.577 & 0.670 & 0.565 & 0.465 & 0.531 & 0.083 \\
ChatMusician-7B & 0.693 & 0.800 & 0.680 & 0.670 & 0.575 & 0.158 \\
\midrule
SFT (4B) & 1.000 & 1.000 & 1.000 & 1.000 & 1.000 & 1.000 \\
\rowcolor{black!6}
\method{} (diag.) & 1.000 & 1.000 & 1.000 & 1.000 & 1.000 & 1.000 \\
\bottomrule
\end{tabular*}
}

\end{table*}

\subsection{Diagnostic Reward-Design Ablations}
\label{app:diagablation}

\begin{figure*}[t]
\centering
\includegraphics[width=0.85\textwidth]{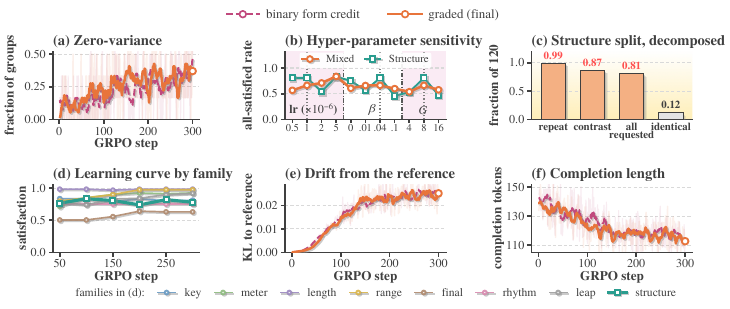}
\caption{\textbf{Reward-design ablations (diagnostic study).} \textbf{(a)} Zero-variance groups.
\textbf{(b)} Sensitivity to learning rate, $\beta$ and $G$ (dotted: reference). \textbf{(c)} Form
verifier decomposed. \textbf{(d)} Per-family satisfaction. \textbf{(e--f)} KL and length.}
\label{fig:dyn6}
\end{figure*}

In Table~\ref{tab:ablationdiag}, we run the reward-design comparison from the SFT checkpoint of the
diagnostic study. The results show that raising form exposure from $21\%$ to $33\%$ takes Structure from $0.250$ to
$0.525$, and graded form credit then takes it to $0.808$ while the other five suites move by at
most $0.025$; grading contrast as well (symmetric) drops it to $0.200$. Figure~\ref{fig:dyn6}
traces these runs: under graded credit Structure outputs pass repetition on $0.99$ and contrast on
$0.87$ of items (c), and every learning rate, KL coefficient and group size tested keeps Mixed
above the SFT start, $0.540$--$0.837$ against $0.477$ (b). In the Direct configuration
(Table~\ref{tab:ablation}) Structure stays between $0.867$ and $0.967$ under every variant, and
every variant loses most on High-$k$.

\begin{table}[t]
\centering
\caption{Reward-design ablations (diagnostic study): one $300$-step run per RL row from the same
SFT checkpoint. Rows give the credit for the form constraint and the share of training problems
containing it (symmetric credit: Appendix~\ref{app:structvariants}). Colours as in
Table~\ref{tab:main}; the {\fboxsep=1.2pt\colorbox{Ours}{shaded row}} is the default reward.}
\label{tab:ablationdiag}
\footnotesize
\setlength{\tabcolsep}{3pt}
\begin{adjustbox}{max width=\columnwidth}
\begin{tabular}{lcccccc}
\toprule
Form credit & Mixed & Single & Seen & Unseen & High-$k$ & Struct. \\
\midrule
SFT, no GRPO & 0.477 & 0.745 & 0.415 & 0.440 & 0.231 & \cellcolor{Secondbest}\underline{0.600} \\
Binary, $21\%$ & 0.610 & \cellcolor{Secondbest}\underline{0.810} & \cellcolor{Secondbest}\underline{0.630} & \cellcolor{Best}\textbf{0.635} & 0.350 & 0.250 \\
Binary, $33\%$ & \cellcolor{Secondbest}\underline{0.653} & \cellcolor{Secondbest}\underline{0.810} & 0.610 & 0.610 & 0.331 & 0.525 \\
\rowcolor{Ours}
Graded, $33\%$ & \cellcolor{Best}\textbf{0.663} & \cellcolor{Best}\textbf{0.835} & \cellcolor{Best}\textbf{0.635} & 0.595 & \cellcolor{Secondbest}\underline{0.356} & \cellcolor{Best}\textbf{0.808} \\
Symmetric, $33\%$ & 0.590 & 0.745 & 0.590 & \cellcolor{Secondbest}\underline{0.630} & \cellcolor{Best}\textbf{0.388} & 0.200 \\
\bottomrule
\end{tabular}
\end{adjustbox}
\end{table}

\subsection{Fixed-Checkpoint Evaluation}
\label{app:checkpoint}
In this section, we report success counts for the diagnostic reference checkpoint,
\texttt{outputs/grpo\_4b\_v2/checkpoint-300}. Table~\ref{tab:checkpoint} gives its counts on the
six original suites; on Dense and
Edge it reaches $185/500$ and $81/500$. Its first evaluation supplies the original-suite rates;
the separate re-evaluation is documented in Appendix~\ref{app:evalnoise}. Success counts for
the three matched checkpoints are in Table~\ref{tab:matchedcounts}.

\begin{table}[ht]
\centering
\caption{Diagnostic study: success counts and all-satisfied rates for the
SFT-initialised reference checkpoint on the six original suites, matching Table~\ref{tab:historicalmain}.}
\label{tab:checkpoint}
\small
% <<< tables/checkpoint_scores
\begin{tabular}{lcc}
\toprule
Split & SFT init. count & Rate \\
\midrule
Mixed & 199/300 & 0.663 \\
Single & 167/200 & 0.835 \\
Seen & 127/200 & 0.635 \\
Unseen & 119/200 & 0.595 \\
High-$k$ & 57/160 & 0.356 \\
Structure & 97/120 & 0.808 \\
\bottomrule
\end{tabular}

% >>> tables/checkpoint_scores
\end{table}

\subsection{Hyper-parameter Sweeps}
\label{app:sweeps}

Table~\ref{tab:sensdirect} gives the seven seed-$43$ Direct sensitivity runs
in Figure~\ref{fig:matched}(g), each changing one setting from the default,
with $300$ updates and $48$ completions per update. The repeated Mixed reference
is taken from the matched Direct result in Table~\ref{tab:main}
($242/300=0.807$); the Structure reference likewise uses the measured
$116/120=0.967$. The seven new runs use their measured scores.

\begin{table}[ht]
\centering
\caption{Direct sensitivity at step $300$: greedy all-satisfied rates for
seven runs and the displayed default reference ($^{\star}$), repeated in
each parameter group. The reference provenance is explained in the text.}
\label{tab:sensdirect}
\small
\begin{tabular}{llcc}
\toprule
Hyper-parameter & Value & Mixed & Structure \\
\midrule
Learning rate & $5{\times}10^{-7}$ & 0.573 & 0.550 \\
 & $1{\times}10^{-6}$$^{\star}$ & 0.807 & 0.967 \\
 & $2{\times}10^{-6}$ & 0.767 & 0.933 \\
\midrule
KL coefficient $\beta$ & $0$ & 0.793 & 0.967 \\
 & $0.01$ & 0.837 & 0.942 \\
 & $0.04$$^{\star}$ & 0.807 & 0.967 \\
 & $0.1$ & 0.680 & 0.908 \\
\midrule
Group size $G$ & $4$ & 0.713 & 0.842 \\
 & $8$$^{\star}$ & 0.807 & 0.967 \\
 & $16$ & 0.787 & 0.958 \\
\bottomrule
\end{tabular}
\end{table}

\label{app:sensdetail}
In Table~\ref{tab:sensdetail}, we list the values plotted in Figure~\ref{fig:dyn6}(b). Each cell is
one training run per setting with a single parameter moved from the SFT-initialised
reference configuration, scored on Mixed and Structure. The reward mixing weight is swept the same way
in Appendix~\ref{app:lambda}.

\begin{table}[ht]
\centering
\caption{Diagnostic study: All-satisfied rate on the mixed and structure suites for each
hyper-parameter setting, one training run each. $^{\star}$ marks the SFT-initialised reference configuration.}
\label{tab:sensdetail}
\small
% <<< tables/sweeps
\begin{tabular}{llcc}
\toprule
Hyper-parameter & Value & Mixed & Structure \\
\midrule
Learning rate & $5{\times}10^{-7}$ & 0.567 & 0.808 \\
 & $1{\times}10^{-6}$$^{\star}$ & 0.663 & 0.808 \\
 & $2{\times}10^{-6}$ & 0.710 & 0.542 \\
 & $5{\times}10^{-6}$ & 0.837 & 0.833 \\
\midrule
KL coefficient $\beta$ & $0$ & 0.607 & 0.750 \\
 & $0.01$ & 0.660 & 0.567 \\
 & $0.04$$^{\star}$ & 0.663 & 0.808 \\
 & $0.1$ & 0.600 & 0.458 \\
\midrule
Group size $G$ & $4$ & 0.540 & 0.533 \\
 & $8$$^{\star}$ & 0.663 & 0.808 \\
 & $16$ & 0.577 & 0.475 \\
\bottomrule
\end{tabular}

% >>> tables/sweeps
\end{table}

\paragraph{Stability across settings.}
Every setting exceeds the SFT Mixed baseline of $0.477$. Learning rate moves Mixed most
($0.270$), against $0.063$ for the KL coefficient and $0.123$ for group size.

The $G$ sweep fixes $48$ completions per optimiser step. Accordingly $G=4,8,16$
uses $12,6,3$ prompt groups per step, for $3{,}600,1{,}800,900$ prompt draws in
$300$ steps. The corresponding logged epochs are $0.60,0.30,0.15$. The sweep
therefore compares allocations of a fixed completion budget, with prompt exposure falling as $G$
grows.

\subsection{Reward Mixing Weight}
\label{app:lambda}

In Table~\ref{tab:lambdasweep}, we move $\lambda$ of Equation~\ref{eq:reward}, which weights the
per-family average against the binary all-satisfied term, while every other setting stays at the
SFT-initialised reference configuration. It is reported separately from Table~\ref{tab:sensdetail} because that
table lists the values plotted in Figure~\ref{fig:dyn6}(b), and because the form decomposition
below is what makes this sweep interpretable.

The default scores highest on both suites among the three weights tested. Against $\lambda = 0.7$, the low arm's Mixed comparison gives exact McNemar
$p = \PLamzerofiveMixed$, while its Structure score is lower ($p = \PLamzerofiveStructure$);
the high arm scores lower on both ($p = \PLamonezeroMixed$ and $p = \PLamonezeroStructure$).
Table~\ref{tab:ablation} varies $\lambda$ over $0$, $0.5$, $0.7$ and $1$ in the Direct
configuration, where $0.7$ is likewise highest over all $2{,}180$ items.

\begin{table}[ht]
\centering
\caption{Diagnostic study: Reward mixing weight, one training run per setting, scored by the binary criterion of
Equation~\ref{eq:allsat}. Repetition and contrast count the structure-split items passing all
required-equal and all required-distinct unit-pair checks, respectively; $^{\star}$ marks the default.}
\label{tab:lambdasweep}
\small
% <<< tables/lambda_sweep
\begin{tabular}{lcccc}
\toprule
$\lambda$ & Mixed & Structure & Repetition & Contrast \\
\midrule
$0.5$ & 0.633 & 0.425 & 120/120 & 55/120 \\
$0.7$$^{\star}$ & 0.663 & 0.808 & 119/120 & 104/120 \\
$1.0$ & 0.570 & 0.325 & 119/120 & 45/120 \\
\bottomrule
\end{tabular}

% >>> tables/lambda_sweep
\end{table}

\paragraph{Contrast distinguishes the tested weights.}
The last two columns separate the halves of the form constraint. Repetition is satisfied on
$119$--$120$ of the $120$ items at every weight, while contrast passes $\LamContrastzerofive$
items at $\lambda = 0.5$ and $\LamContrastonezero$ at $\lambda = 1.0$ against
$\LamContrastzeroseven$ at the default: the decrease on Structure accompanies a drop in contrast,
not in repetition. The all-satisfied counts behind those rates are $51$, $97$ and $39$;
each deficit relative to the contrast count consists of items passing the form check and missing
a co-requested family. Outputs whose four
units are all identical -- which satisfy every required-equal pair and no required-distinct one --
rise from $\LamIdentzeroseven$ items at the default to $\LamIdentzerofive$ and
$\LamIdentonezero$ at the two ends. The same pattern holds across the rows of
Table~\ref{tab:ablationdiag}: repetition passes $119$--$120$ items, and the binary form pass count
equals the contrast pass count.

\paragraph{Where the weight can act.}
Write $m_i$ for the per-family average in Equation~\ref{eq:reward}, setting it to zero for a
gate-rejected rollout. For $\lambda > 0$, a group in which every rollout has $A(y_i,C)=0$
has $R(y_i,C)=\lambda m_i$. Equation~\ref{eq:grpoadv} then gives
$\hat{A}_i=(m_i-\bar{m})/(\sigma_m+\epsilon/\lambda)$, where $\sigma_m$ is the group sample
standard deviation of $m$: the scaling cancels except through the stabiliser.
The mixture also controls the reward gap between full and partial success. Consider two
gate-passing rollouts for the same prompt: one satisfies all $k$ families; the other fails
exactly one family, retaining graded credit $s\in[0,1)$ for that family. Their reward gap is
\[
\Delta R=(1-\lambda)+\frac{\lambda(1-s)}{k},
\]
which is $(1-s)/k$ at $\lambda=1$ and $0.3+0.7(1-s)/k$ at the default. The binary term thus
adds an explicit bonus for joint satisfaction; its effect on normalised advantages also
depends on the other rewards in the group. The default mixture gives the highest
compliance and the fewest all-identical outputs.

\subsection{Per-Family Satisfaction on the Original Suites}
\label{app:perfamily}

\begin{figure*}[t]
\centering
\includegraphics[width=0.62\textwidth]{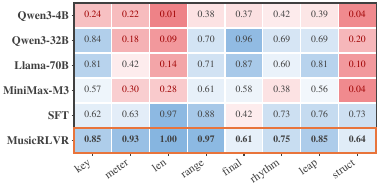}
\caption{Diagnostic study: Per-family satisfaction on the mixed suite; \method{} cells use the fixed checkpoint (Table~\ref{tab:perfam}); \texttt{structure} stays red for
every untrained model.}
\label{fig:heat}
\end{figure*}

In this section, we break the all-satisfied rate down by family. Table~\ref{tab:historicalmain}
reports the all-satisfied rate, which is zero whenever any single constraint fails;
Table~\ref{tab:perfamily_all} reports each family's
satisfaction rate on each split for the same checkpoint, and Figure~\ref{fig:heat} gives the
same per-family view on Mixed across models.

\begin{table}[t]
\centering
\caption{Diagnostic study: Per-family satisfaction on Mixed: the fraction of items requesting a family
whose verifier passes. $\Delta$ is the fixed-checkpoint rate minus the SFT rate.}
\label{tab:perfam}
\small
\renewcommand{\arraystretch}{0.88}
% <<< tables/perfamily_mixed
\begin{tabular}{lcccc}
\toprule
Family & Base & SFT & \method{} & $\Delta$ \\
\midrule
\texttt{key} & 0.237 & 0.625 & $0.850$ & +0.225 \\
\texttt{meter} & 0.222 & 0.630 & $0.926$ & +0.296 \\
\texttt{length} & 0.015 & 0.969 & $1.000$ & +0.031 \\
\texttt{range} & 0.375 & 0.875 & $0.975$ & +0.100 \\
\texttt{final} & 0.368 & 0.421 & $0.611$ & +0.189 \\
\texttt{rhythm} & 0.416 & 0.727 & $0.753$ & +0.026 \\
\texttt{leap} & 0.393 & 0.764 & $0.854$ & +0.090 \\
\texttt{structure} & 0.041 & 0.726 & $0.644$ & -0.082 \\
\bottomrule
\end{tabular}

% >>> tables/perfamily_mixed
\vspace{-8pt}
\end{table}

The mixed suite includes $73$ structure-family items. Their satisfaction rate is
$\AggMixedStruct$ for \method{} against $0.726$ for SFT. On the dedicated
structure split, the all-satisfied rate is
$\Oursbenchstructure$ against $\Sftbenchstructure$;
equal prompt weighting gives $\AggUniqueStruct$ against $0.522$
(Appendix~\ref{app:coverage}). The splits differ in their requested constraints and prompt weights.

\begin{table}[t]
\centering
\caption{Diagnostic study: per-family satisfaction rate for \method{} across the splits. Each cell is the
fraction of problems mentioning that family whose own verifier passes at the fixed checkpoint
(other constraints may fail); families mentioned by fewer than five problems in a split are
marked ``--''.}
\label{tab:perfamily_all}
\small
% <<< tables/perfamily_all
\begin{adjustbox}{max width=\columnwidth}
\begin{tabular}{lccccc}
\toprule
Family & Single & Seen & Unseen & High-$k$ & Structure \\
\midrule
\texttt{final} & $0.783$ & $0.636$ & $0.407$ & $0.635$ & -- \\
\texttt{key} & $0.862$ & $0.887$ & $0.565$ & $0.671$ & $0.643$ \\
\texttt{leap} & $0.292$ & $0.935$ & $0.985$ & $0.872$ & $1.000$ \\
\texttt{length} & $1.000$ & $1.000$ & $1.000$ & $1.000$ & -- \\
\texttt{meter} & $0.812$ & $0.903$ & $0.897$ & $0.890$ & $1.000$ \\
\texttt{range} & $1.000$ & $0.982$ & $0.955$ & $0.980$ & $1.000$ \\
\texttt{rhythm} & $0.833$ & $0.790$ & $0.843$ & $0.655$ & $0.692$ \\
\texttt{structure} & $1.000$ & $0.510$ & $0.667$ & $0.554$ & $0.867$ \\
\bottomrule
\end{tabular}
\end{adjustbox}

% >>> tables/perfamily_all
\end{table}

\noindent \texttt{length} and \texttt{range} maintain high satisfaction across the
splits.

\subsection{Performance by Number of Constraints (\texorpdfstring{$k$}{k})}
\label{app:byk}
\label{app:crossbyk}

In Table~\ref{tab:byk}, we break the mixed suite ($n=300$; $100/100/60/40$ items at
$k=1/2/3/4$) down by the number of constraints, using the same saved diagnostic reference evaluations as Table~\ref{tab:historicalmain}. Figure~\ref{fig:degrade} instead pools all $2{,}180$
items for the matched runs; Table~\ref{tab:bykpooled} lists its values.

\begin{table}[t]
\centering
\caption{All-satisfied rate by number of constraints $k$ over all eight suites ($2{,}180$
items), the values plotted in Figure~\ref{fig:degrade}. Each slice pools the suites containing
that $k$; the trained rows are the matched runs of Table~\ref{tab:main}.}
\label{tab:bykpooled}
\footnotesize
\setlength{\tabcolsep}{3pt}
\begin{adjustbox}{max width=\columnwidth}
% <<< tables/byk_pooled (make_fig_degrade_pooled.py)
\begin{tabular}{lccccc}
\toprule
Model & $k=1$ & $k=2$ & $k=3$ & $k=4$ & $k{=}1/k{=}4$ \\
Items & $354$ & $614$ & $512$ & $700$ & \\
\midrule
Qwen3-4B & 0.243 & 0.068 & 0.018 & 0.016 & $15.5\times$ \\
Qwen3-32B & 0.559 & 0.225 & 0.098 & 0.083 & $6.8\times$ \\
Llama-3.1-70B & 0.630 & 0.212 & 0.090 & 0.044 & $14.2\times$ \\
MiniMax-M3 & 0.486 & 0.267 & 0.137 & 0.089 & $5.5\times$ \\
SFT & 0.723 & 0.349 & 0.178 & 0.277 & $2.6\times$ \\
Direct GRPO ($R=A$) & 0.977 & 0.661 & 0.357 & 0.361 & $2.7\times$ \\
\method{} (SFT init.) & 0.757 & 0.430 & 0.283 & 0.353 & $2.1\times$ \\
\method{} (Direct) & 0.992 & 0.673 & 0.406 & 0.419 & $2.4\times$ \\
\bottomrule
\end{tabular}
% >>> tables/byk_pooled
\end{adjustbox}
\end{table}

\begin{table*}[t]
\centering
\caption{Diagnostic study: all-satisfied rate on the mixed suite by number of constraints $k$. The last
column is the $k{=}1$ rate over the $k{=}4$ rate; a dagger marks curves not monotone in $k$ (the
$k{=}3$ and $k{=}4$ cells hold only $60$ and $40$ items, so one item moves them by $0.017$ and
$0.025$).}
\label{tab:byk}
\small
% <<< tables/byk_audit
\begin{tabular}{lccccc}
\toprule
Model & $k=1$ & $k=2$ & $k=3$ & $k=4$ & $k{=}1/k{=}4$ \\
\midrule
\method{} (diag.) & $0.850$ & $0.640$ & $0.517$ & $0.475$ & $1.8\times$ \\
SFT (4B) & 0.720 & 0.430 & 0.267 & 0.300 & $2.4\times^\dagger$ \\
Qwen3-32B & 0.650 & 0.300 & 0.133 & 0.150 & $4.3\times^\dagger$ \\
Llama-3.1-70B & 0.720 & 0.340 & 0.100 & 0.050 & $14.4\times$ \\
MiniMax-M3 & 0.560 & 0.360 & 0.267 & 0.050 & $11.2\times$ \\
Qwen3-4B & 0.340 & 0.110 & 0.033 & 0.025 & $13.6\times$ \\
Qwen3-8B & 0.240 & 0.120 & 0.017 & 0.000 & -- \\
\bottomrule
\end{tabular}

% >>> tables/byk_audit
\end{table*}

\noindent The endpoint ratios vary across models. Llama-3.1-70B falls from $0.720$
at $k=1$ to $\MainLlamaMixedKFour$ at $k=4$, a $\MainLlamaKRatio{\times}$ drop; MiniMax-M3 falls
$11.2{\times}$ and the untrained 4B base $13.6{\times}$. The diagnostic reference rates are
$\AggKOne$ and $\AggKFour$, giving the smallest endpoint ratio in the table
($\AggKRatio{\times}$), against $2.4{\times}$ for SFT. The matched by-$k$ curves, pooled over all suites,
are in Figure~\ref{fig:degrade}.

\begin{table}[t]
\centering
\caption{Variety and degeneration over all $2{,}180$ items ($n$: completions that \texttt{music21} parses into at least four notes, a looser test
than the validation gate), with the
IrishMAN reference sample of Table~\ref{tab:qual}. Distinctness uses the first eight bars.}
\label{tab:qualitymain}
\scriptsize
\setlength{\tabcolsep}{2.5pt}
\begin{adjustbox}{max width=\columnwidth}
% <<< tables/quality_main (src/audit_quality_main.py)
\begin{tabular}{lrcccccc}
\toprule
Source & $n$ & PC ent. & Dist.\ bars & Dist.\ 4-gr. & $\le$3 pitches & Same bars & Rep.\ notes \\
\midrule
IrishMAN corpus & 299 & 2.63 & 0.828 & 0.769 & 0.0\% & 0.0\% & 0.7\% \\
Qwen3-4B & 1053 & 2.06 & 0.449 & 0.503 & 28.9\% & 5.4\% & 8.5\% \\
Qwen3-8B & 891 & 1.97 & 0.457 & 0.393 & 24.2\% & 18.0\% & 40.6\% \\
Qwen3-32B & 2076 & 2.64 & 0.775 & 0.752 & 0.8\% & 0.1\% & 5.0\% \\
Llama-3.1-70B & 2093 & 2.42 & 0.610 & 0.607 & 3.0\% & 0.2\% & 1.1\% \\
MiniMax-M3 & 2143 & 2.29 & 0.725 & 0.794 & 9.2\% & 4.4\% & 12.7\% \\
ChatMusician-7B & 1710 & 1.83 & 0.599 & 0.551 & 31.2\% & 12.1\% & 32.8\% \\
SFT & 2180 & 1.87 & 0.389 & 0.390 & 23.0\% & 7.7\% & 32.0\% \\
RFT & 2172 & 2.07 & 0.472 & 0.521 & 31.7\% & 6.6\% & 11.2\% \\
Direct GRPO ($R=A$) & 2180 & 2.14 & 0.389 & 0.444 & 22.9\% & 7.8\% & 1.4\% \\
\method{} (SFT init.) & 2180 & 1.62 & 0.359 & 0.342 & 35.1\% & 11.4\% & 40.7\% \\
\method{} (Direct) & 2178 & 2.15 & 0.381 & 0.434 & 22.9\% & 6.5\% & 1.2\% \\
\method{} (Direct, Qwen3-8B) & 2180 & 2.02 & 0.508 & 0.494 & 28.1\% & 7.0\% & 52.0\% \\
\bottomrule
\end{tabular}
% >>> tables/quality_main
\end{adjustbox}
\end{table}

\subsection{Variety and Degeneration of the Matched Checkpoints}
\label{app:qualitymain}

In Tables~\ref{tab:qualitymain} and~\ref{tab:qualitycontrolled}, we repeat the within-tune
statistics of Table~\ref{tab:qual} for every model of Table~\ref{tab:main} over all $2{,}180$ items
(\texttt{src/audit\_quality\_main.py}). Pitch-class entropy uses the whole tune; the distinct-bar
and distinct pitch 4-gram ratios use the first eight bars, so tunes of different lengths are
compared on equal spans. Three flags mark degenerate tunes: at most three distinct pitches, every
bar identical, and more than half of consecutive notes repeating the previous pitch.

Against its base model, Direct \method{} raises pitch-class entropy from $2.06$ to $2.15$, lowers
the share of tunes with at most three pitches from $28.9\%$ to $22.9\%$ and of tunes dominated by
repeated notes from $8.5\%$ to $1.2\%$. Table~\ref{tab:qualitycontrolled} keeps only the $332$ items without a form constraint,
which requests repetition by design, on which both models render at least eight bars. The same
pattern holds there: entropy $2.02$ to $2.17$ and repeated-note tunes $12.3\%$ to $2.1\%$.

\begin{table}[t]
\centering
\caption{The same statistics on the $332$ items without a form constraint that both Qwen3-4B and
Direct \method{} render to at least eight bars; every model is scored on those items.}
\label{tab:qualitycontrolled}
\scriptsize
\setlength{\tabcolsep}{2.5pt}
\begin{adjustbox}{max width=\columnwidth}
% <<< tables/quality_controlled (src/audit_quality_main.py)
\begin{tabular}{lrcccccc}
\toprule
Model & $n$ & PC ent. & Dist.\ bars & Dist.\ 4-gr. & $\le$3 pitches & Same bars & Rep.\ notes \\
\midrule
Qwen3-4B & 332 & 2.02 & 0.530 & 0.562 & 30.1\% & 4.2\% & 12.3\% \\
SFT & 332 & 1.82 & 0.372 & 0.366 & 24.4\% & 9.3\% & 30.1\% \\
Direct GRPO ($R=A$) & 332 & 2.10 & 0.406 & 0.430 & 31.6\% & 16.9\% & 0.6\% \\
\method{} (Direct) & 332 & 2.17 & 0.410 & 0.442 & 25.0\% & 7.5\% & 2.1\% \\
\method{} (Direct, Qwen3-8B) & 332 & 2.23 & 0.627 & 0.634 & 16.3\% & 3.9\% & 61.1\% \\
Llama-3.1-70B & 322 & 2.40 & 0.665 & 0.631 & 4.7\% & 0.3\% & 2.2\% \\
\bottomrule
\end{tabular}
% >>> tables/quality_controlled
\end{adjustbox}
\end{table}

\subsection{Length-Controlled Output Statistics}
\label{app:qualitymatched}

\begin{table}[H]
\centering
\scriptsize
\setlength{\tabcolsep}{3pt}
\caption{\textbf{Diagnostic output analysis.}
\textbf{(a)}~Surface statistics (first 8 bars, $\QualityMatchedItems$ matched items).
\textbf{(b)}~Best-of-$N$ ($T{=}1.0$, $n{=}8$, binary).}
\label{tab:qualitymatched}
\vspace{2pt}
\textbf{(a)} \\[2pt]
% <<< tables/quality_matched
\begin{tabular*}{\linewidth}{@{\extracolsep{\fill}}lcccc@{}}
\toprule
Model & Notes & PC ent. & Dist.\ bars & Dist.\ 4-gr. \\
\midrule
SFT & 42.7 & 1.930 & 0.417 & 0.419 \\
\method{} (diag.) & 40.5 & 1.826 & 0.426 & 0.403 \\
\bottomrule
\end{tabular*}
% >>> tables/quality_matched
\\[4pt]
\textbf{(b)} \\[2pt]
% <<< tables/bon
\begin{tabular*}{\linewidth}{@{\extracolsep{\fill}}llcccc@{}}
\toprule
Suite & Model & p@1 & p@2 & p@4 & p@8 \\
\midrule
Mixed & SFT & 0.415 & 0.584 & 0.728 & 0.827 \\
 & \method{} (diag.) & 0.676 & 0.817 & 0.900 & 0.950 \\
\addlinespace
Struct. & SFT & 0.692 & 0.861 & 0.938 & 0.967 \\
 & \method{} (diag.) & 0.844 & 0.953 & 0.978 & 0.983 \\
\bottomrule
\end{tabular*}
% >>> tables/bon
\end{table}

In Table~\ref{tab:qualitymatched}, we compare the first eight parsed bars of a common item set of
the diagnostic study, because the full-tune ratios of Table~\ref{tab:qual} have length-dependent
denominators. In this section, we give the uncontrolled statistics those ratios come from, and the cross-prompt
duplication counts of Table~\ref{tab:crossprompt}. Both read historical-gate outputs, but they measure different things: the
first are within-tune proxies, the second counts how often two different prompts produce the same
score.

\begin{table}[H]
\centering
\caption{Diagnostic study: distributional statistics of the single-run mixed-suite completions and $299$
corpus tunes. No constraint names entropy or either distinctness measure; bar count is
constrained on the $138/300$ items requesting \texttt{length} or \texttt{structure}. Mean length
differs between rows, so distinctness denominators are unmatched; Table~\ref{tab:qualitymatched}
controls for that.}
\label{tab:qual}
\small
\renewcommand{\arraystretch}{0.88}
% <<< tables/quality_stats
\setlength{\tabcolsep}{3pt}
\begin{tabular}{lcccc}
\toprule
Model & Bars & PC ent. & Dist.\ bars & Dist.\ 4-gr. \\
\midrule
Corpus & 19.9 & 2.63 & 0.758 & 0.676 \\
Base & 29.3 & 2.06 & 0.366 & 0.391 \\
SFT & 39.2 & 2.00 & 0.276 & 0.294 \\
RL, binary credit & 32.2 & 1.85 & 0.284 & 0.289 \\
\midrule
\textsc{MusicRLVR} & 28.7 & 1.89 & 0.313 & 0.292 \\
\bottomrule
\end{tabular}

% >>> tables/quality_stats
\end{table}

\begin{table}[H]
\centering
\caption{Diagnostic study: full-score duplication for the first eligible occurrence of each of
$\QualityUniquePrompts$ distinct prompts in that common set, identified by parsed event
signatures and ignoring metadata, which measures duplication across prompts.}
\label{tab:crossprompt}
\small
% <<< tables/quality_crossprompt
\begin{adjustbox}{max width=\columnwidth}
\begin{tabular}{lcc}
\toprule
Model & Unique full scores & Most frequent score \\
\midrule
SFT & 167/220 & 7/220 \\
\method{} (diagnostic) & 182/220 & 5/220 \\
\bottomrule
\end{tabular}
\end{adjustbox}

% >>> tables/quality_crossprompt
\end{table}

On the full completions, \method{} raises distinct-bar fraction from $0.276$ to $0.313$
while mean length falls $39.2\rightarrow28.7$ bars. Under the length control of
Table~\ref{tab:qualitymatched} the distinct-bar comparison is $\QualitySftDistinctBars \rightarrow \QualityRlDistinctBars$ on matched eight-bar
prefixes. The matched checkpoints of Table~\ref{tab:main} are analysed in
Appendix~\ref{app:qualitymain}.

\subsection{Output Length Statistics}
\label{app:outlength}

\begin{table}[!t]
\centering
\caption{Diagnostic study: Output length in the single-run mixed-suite diagnostic ($n = 300$ problems). Only completions
that pass the historical parse gate are included, so the parsed count differs by model; the
IrishMAN reference ($299$ tunes, $19.9$ bars) is in Table~\ref{tab:qual}.}
\label{tab:outlength}
\small
% <<< tables/outlength
\begin{adjustbox}{max width=\columnwidth}
\begin{tabular}{lccc}
\toprule
Source & Parsed $n$ & Mean bars & Mean notes \\
\midrule
Qwen3-4B (zero-shot) & 154 & 29.3 & 78.4 \\
Qwen3-32B (zero-shot) & 292 & 18.4 & 58.1 \\
Llama-3.1-70B (zero-shot) & 291 & 21.2 & 66.1 \\
MiniMax-M3 (zero-shot) & 173 & 17.9 & 37.0 \\
SFT & 300 & 39.2 & 167.5 \\
\method{} & 300 & 28.7 & 130.3 \\
\bottomrule
\end{tabular}
\end{adjustbox}

% >>> tables/outlength
\end{table}

In Table~\ref{tab:outlength}, we report the average number of bars and notes per parsed
completion on the mixed suite, supplementing the quality metrics in
Table~\ref{tab:qual}.

\noindent The supervised model over-generates ($39.2$ bars against $19.9$ for the
corpus), and the reinforcement stage pulls this back to $28.7$ bars, consistent with a
reward that favours concision without an unconditional length term, since a longer tune has more places
to violate a constraint. The three larger zero-shot models sit near the corpus length
($17.9$--$21.2$ bars; the 4B base averages $29.3$), and MiniMax-M3 parses on only $173$ of
$300$ problems.

\subsection{Zero-Variance Groups During Training}
\label{app:zerostd}
In this section, we track groups without a reward gradient during training.
Equation~\ref{eq:grpoadv} implies that a group whose $G$ rollouts all receive the
same reward contributes no reward gradient. The trainer logs the fraction of such groups
per step (\texttt{frac\_reward\_zero\_std}); Table~\ref{tab:zerostd} summarises it
for the two reward variants of the main ablation.

\begin{table*}[t]
\centering
\caption{Diagnostic study: Fraction of GRPO groups with zero reward variance (duplicate step records from restarts
removed), averaged over the first
and last quarter of training and over all $300$ steps, with the mean within-group
reward standard deviation alongside.}
\label{tab:zerostd}
\small
\begin{tabular}{lccccc}
\toprule
& \multicolumn{3}{c}{Zero-variance groups} & \multicolumn{2}{c}{Reward std} \\
\cmidrule(lr){2-4}\cmidrule(lr){5-6}
Run & first 25\% & last 25\% & all & first 25\% & last 25\% \\
\midrule
Binary form credit, original form share ($21\%$) & 0.116 & 0.307 & 0.221 & 0.301 & 0.214 \\
Graded form credit, $33\%$ form share (final) & 0.133 & 0.322 & 0.254 & 0.296 & 0.210 \\
\bottomrule
\end{tabular}
\end{table*}

The two diagnostic rows differ in both the reward and the share of structure prompts; the
matched runs below change the reward alone.

The matched runs measure the same quantity on two arms that differ only in
$\lambda$, holding the prompt mix fixed. Averaged over the first $50$ updates, binary reward
leaves $\DeadBinaryEarly$ of groups with zero reward variance against $\DeadGradedEarly$ for
the graded reward; over the last $50$ the two reach $\DeadBinaryLate$ and $\DeadGradedLate$,
once most groups succeed outright, and over all $300$ updates they average $\DeadBinaryAll$ and
$\DeadGradedAll$ (Figure~\ref{fig:matched}(b)). The SFT-initialised arm runs lower throughout
($\DeadWarmEarly$, $\DeadWarmLate$, $\DeadWarmAll$), so the fraction is compared between arms
that share an initialisation.

\subsection{Additional Training Curves}
\label{app:curves}

Figure~\ref{fig:matched}(h) uses the $14{,}400$ logged rollouts of the seed-$43$ Direct
\method{} run, re-scored with the training-time verifiers and pooled as passing/requested
counts in centred $25$-step windows (truncated at the training boundaries),
with gate-rejected outputs counted as failures.

\label{app:dyncurves}

\begin{figure*}[ht]
\centering
\includegraphics[width=0.95\textwidth]{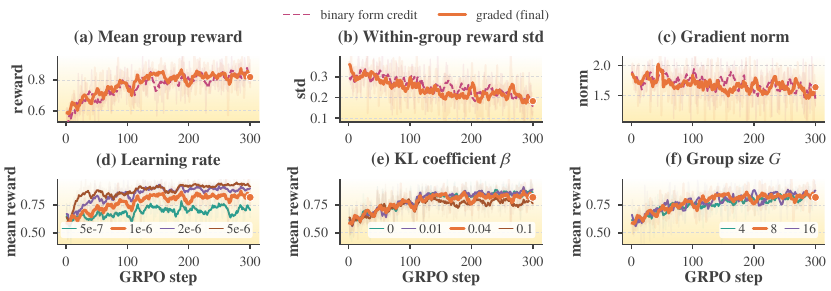}
\caption{Diagnostic study: \textbf{trainer-log quantities behind Figure~\ref{fig:dyn6}.} \textbf{(a--c)}
Mean group reward, within-group reward SD and gradient norm, binary form credit against the
final graded run. \textbf{(d--f)} Mean group reward for every hyper-parameter setting of
Figure~\ref{fig:dyn6}(b) on one shared scale; the thick orange curve is the default run
($\text{lr}=1\mathrm{e}{-}6$, $\beta=0.04$, $G=8$). Curves are smoothed over $9$ steps; each
axis spans the $2$nd--$98$th percentile of its raw trace, so a few steps clip.}
\label{fig:logs_app}
\end{figure*}

In Figure~\ref{fig:logs_app}, we give the trainer-log quantities behind the diagnostics of
Figure~\ref{fig:dyn6}; the numbers behind it are as follows. Final KL to the reference falls
$\KlBetaLo \rightarrow \KlBetaMid \rightarrow \KlBetaHi$ as $\beta$ rises
$0.01 \rightarrow 0.04 \rightarrow 0.1$ (the $\beta = 0$ run logs no KL term), and yet the observed
mixed rates lie in the narrow range $0.600$--$0.663$ (Table~\ref{tab:sensdetail}): the KL term therefore tracks $\beta$ as designed, while
Mixed compliance is flat across the range. Along
learning rate, final KL $\KlLrA$, $\KlLrB$, $\KlLrC$, $\KlLrD$ and mean group reward $\RwLrA$,
$\RwLrB$, $\RwLrC$, $\RwLrD$ over $5{\times}10^{-7}$ to $5{\times}10^{-6}$ run in the same order as
the mixed suite.

\subsection{Additional Diagnostics}
\label{app:comprehensive}
\label{app:analysis}

\begin{figure*}[ht]
\centering
\includegraphics[width=\textwidth]{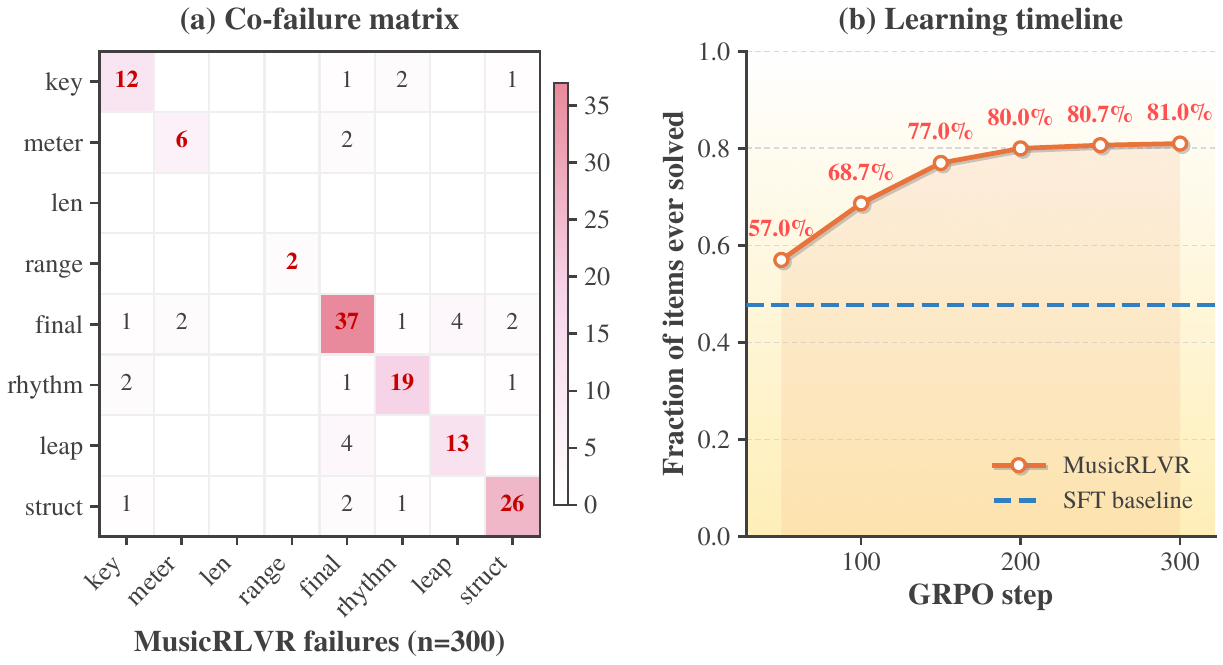}
\caption{Diagnostic study: per-item diagnostics on the mixed suite ($n = 300$). \textbf{(a)} Co-failure
matrix of \method{}: diagonal entries are per-family failure counts, off-diagonal entries
co-failures, which are rare: \texttt{key} shares at most two with any family. \textbf{(b)}
CDF of the step at which each item is first solved: $57\%$ by step $50$, $81\%$ at least once by
step $300$.}
\label{fig:analysis}
\end{figure*}

In Figure~\ref{fig:analysis}, we show two further per-item diagnostics: which families fail
together, and when each item is first solved during training.

\section{Error and Structure Analysis}
\label{app:erroranalysis}

In this section, we analyse which constraints bind on each split and how form credit behaves across
reward variants and form types.

\subsection{Which Constraint Binds, by Split}
\label{app:failpatterns}

\begin{figure*}[t]
\centering
\includegraphics[width=\textwidth]{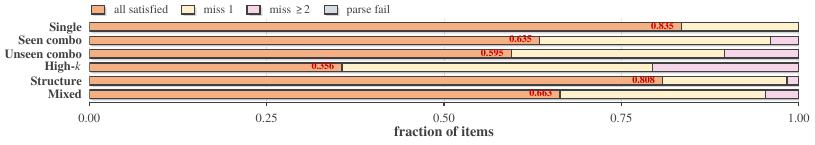}
\caption{Diagnostic study: \textbf{What an unsatisfied item is missing.} Outcome composition for the
fixed checkpoint; printed all-satisfied rates match Table~\ref{tab:historicalmain}.
Parse failures are zero on all six original splits.}
\label{fig:failmodes}
\end{figure*}

\begin{figure*}[t]
\centering
\includegraphics[width=\textwidth]{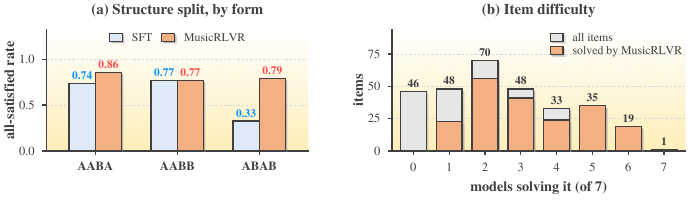}
\caption{Diagnostic study: \textbf{Structure and item-level diagnostics.} \textbf{(a)} Satisfaction by
requested form for the fixed checkpoint.
\textbf{(b)} Item counts by the number of seven evaluated models solving them under the historical $A$.}
\label{fig:fail3}
\end{figure*}

In Figure~\ref{fig:failmodes}, we give the outcome composition of each split, and in
Table~\ref{tab:failpatterns}, we add the most frequent failing families, using the diagnostic
reference evaluation. Figure~\ref{fig:fail3} breaks the same evaluation down by requested form
and by how many models solve each item.

\begin{table*}[t]
\centering
\caption{Diagnostic study: which constraints fail, by split. ``Not all satisfied'' is one minus the
all-satisfied rate; the next column counts items failing two or more families at once; the last
lists the three most frequent failing families as fractions of the split's items (an item
failing two is counted under both). No completion fails the parse gate on any split.}
\label{tab:failpatterns}
\small
\resizebox{\textwidth}{!}{
% <<< tables/failtop
\begin{tabular}{lccc}
\toprule
Split & Not all satisfied & Failing $\geq 2$ families & Most frequent failing families \\
\midrule
Single & $0.165$ & $0.000$ & \texttt{leap} ($0.085$), \texttt{final} ($0.025$), \texttt{key} ($0.020$) \\
Seen & $0.365$ & $0.040$ & \texttt{structure} ($0.125$), \texttt{final} ($0.120$), \texttt{rhythm} ($0.065$) \\
Unseen & $0.405$ & $0.105$ & \texttt{final} ($0.160$), \texttt{key} ($0.150$), \texttt{structure} ($0.090$) \\
High-$k$ & $0.644$ & $0.206$ & \texttt{final} ($0.237$), \texttt{rhythm} ($0.181$), \texttt{key} ($0.156$) \\
Structure & $0.192$ & $0.017$ & \texttt{structure} ($0.133$), \texttt{key} ($0.042$), \texttt{rhythm} ($0.033$) \\
\bottomrule
\end{tabular}

% >>> tables/failtop
}
\end{table*}

\noindent Most unsatisfied items miss one family. The fraction missing multiple
families is $0.040$ on Seen and $0.105$ on Unseen, rising to
$0.206$ on High-$k$. The most frequent failing family is \texttt{final}
on Unseen and High-$k$. \texttt{structure} is among the three most frequent
failures on Seen and Unseen; \texttt{leap} appears among
the top three only on Single.

\subsection{Structure Reward Variants}
\label{app:structvariants}

In Table~\ref{tab:structdecomp}, we compare the fixed default checkpoint with the
equal-weight control and the symmetric-credit study. The default and equal-weight
rows each use one run; symmetric credit reports the mean and sample SD of its two runs. The default form term weights repetition
$0.7$ and contrast $0.3$; the following controls examine that choice.

\paragraph{Symmetric form credit.} The contrast counterpart of Equation~\ref{eq:gradedstruct}
is
\begin{equation*}
\bar{d} = \frac{1}{|D| M}\sum_{(a,b) \in D} \sum_{m=1}^{M}\mathbb{I}\bigl[u_a[m] \neq u_b[m]\bigr],
\end{equation*}
and the symmetric term is
\begin{equation*}
s^{\text{sym}} =
\begin{cases}
1, & v_{\texttt{structure}} = 1,\\
0, & u_1 = u_2 = u_3 = u_4,\\
0.5\,\bar e + 0.5\,\bar d, & \text{else}.
\end{cases}
\end{equation*}
The outer reward mixture is unchanged.
Across two runs, contrast satisfaction is $\AggSymContrast$ and the all-identical
fraction is $\AggSymIdent$. The control jointly changes the weights, the contrast
factor and the credit for all-identical outputs. Under the default term, an all-identical
output earns $s_{\texttt{structure}}=0.7$ but has $A=0$, because it fails contrast.
Appendix~\ref{app:lambda} separately varies the outer mixing weight.

\paragraph{Reweighting repetition and contrast.} The single-variable version
moves only the weights, to $0.5/0.5$: an all-identical output still banks the repetition half,
the binary contrast factor is kept, and the rest of the reward and the training setup are
unchanged. Two things follow.
\begin{enumerate}[label=(\arabic*)]
\item In the single-run diagnostic comparison, the structure split moves $0.808 \rightarrow \StructW$ and all-identical outputs
$15 \rightarrow \StructWident$ of $120$, the same direction as the joint control
(Table~\ref{tab:structdecomp}).
\item The equal-weight control scores $\StructWmix$ on Mixed, compared with
$\Ourskilltestsuite$ for the default.
\end{enumerate}
Appendix~\ref{app:lambda} extends this comparison to the outer reward mixture. In the single-run diagnostic study,
the two tested $\lambda$ endpoints yield larger increases in all-identical outputs than the
equal-weight form control. The default achieves the highest Structure score and fewest
all-identical outputs among these tested weight settings. In the Direct configuration
(Table~\ref{tab:ablation}), the default again gives the highest Structure, $0.967$ against
$0.942$ for both variants, and the highest rate over all items, $0.580$ against $0.515$
(symmetric) and $0.544$ (equal weights).

\begin{table*}[t]
\centering
\caption{Diagnostic study: structure decomposition. ``Runs'' is the number of training runs: default and
equal-weight use one each, symmetric credit reports mean $\pm$ sample SD over two. All-identical
is a fraction; ``all requested'' additionally requires the companion constraint, present in $66$
of $120$ items.}
\label{tab:structdecomp}
\small
\setlength{\tabcolsep}{5.5pt}
% <<< tables/structdecomp
\begin{tabular}{lccccc}
\toprule
Reward & Runs & Repetition & Contrast & All requested & All-identical \\
\midrule
\method{} & 1 & $0.992$ & $0.867$ & $0.808$ & $0.125$ \\
Equal weights & 1 & $1.000$ & $0.733$ & $0.675$ & $0.267$ \\
Symmetric credit & 2 & $0.996\pm0.006$ & $0.233\pm0.024$ & $0.208\pm0.012$ & $0.762\pm0.018$ \\
\bottomrule
\end{tabular}

% >>> tables/structdecomp
\end{table*}

\subsection{Structure Results by Form Type}
\label{app:formtype}

In Table~\ref{tab:formtype}, we decompose performance on the three form types of the structure
split, which come in unequal numbers: \textsc{AABA} ($n = 42$), \textsc{AABB} ($n = 35$) and
\textsc{ABAB} ($n = 43$).

\begin{table}[H]
\centering
\caption{Diagnostic study: Structure results by form for the fixed \method{} and SFT checkpoints.
Contrast and all-identical are fractions
of items of that form.}
\label{tab:formtype}
\small
% <<< tables/formtype
\begin{adjustbox}{max width=\columnwidth}
\begin{tabular}{lcccc}
\toprule
Model & Form & All-sat. & Contrast & All-identical \\
\midrule
SFT & \textsc{aaba} & $0.738$ & $0.810$ & $0.190$ \\
SFT & \textsc{aabb} & $0.771$ & $0.857$ & $0.143$ \\
SFT & \textsc{abab} & $0.326$ & $0.395$ & $0.605$ \\
\method{} & \textsc{aaba} & $0.857$ & $0.905$ & $0.071$ \\
\method{} & \textsc{aabb} & $0.771$ & $0.829$ & $0.171$ \\
\method{} & \textsc{abab} & $0.791$ & $0.860$ & $0.140$ \\
\bottomrule
\end{tabular}
\end{adjustbox}

% >>> tables/formtype
\end{table}

\noindent \textsc{ABAB} is the form supervision handles worst, and by a wide margin:
SFT reaches $0.771$ on \textsc{AABB} but $0.326$ on \textsc{ABAB}, with $26/43$ of
its \textsc{ABAB} outputs written as four identical units. All three forms require distinct A and B units, so the gap tracks the
arrangement rather than the distinctness requirement: \textsc{ABAB} is the only one that
interleaves the two units instead of repeating each in place.

At the fixed checkpoint, \textsc{ABAB} has the largest improvement over SFT among the
three forms: its all-satisfied rate rises from $0.326$ to $\AggAbab$, while the
all-identical fraction changes from $0.605$ to $\AggAbabIdent$.
Table~\ref{tab:formtype} reports the rate for every form.

\section{Qualitative Examples}
\label{app:qualitative}
\label{app:examples}

\begin{figure}[t]
\centering
\includegraphics[width=\columnwidth]{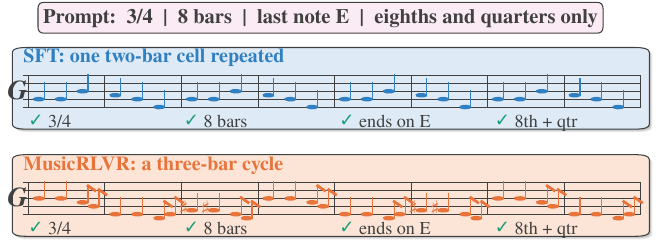}
\caption{One High-$k$ item, diagnostic study (Appendix~\ref{app:historical}): both outputs
satisfy all four constraints.}
\label{fig:staffdiag}
\end{figure}

% Example 1 is placed here so it can open the page where this introduction ends.
\begin{figure*}[t]
\small\raggedright
\textbf{Example 1: unseen-combination split ($k = 2$, held-out pair \texttt{key}--\texttt{rhythm}).}\par
\noindent\textbf{Item} \texttt{uc-0022}. \textbf{Constraints:} (1) the tune must be in G major; (2) use only sixteenth, eighth, and quarter notes (no dotted notes, no other durations, no rests)\par\smallskip
{\centering\includegraphics[width=0.8\textwidth]{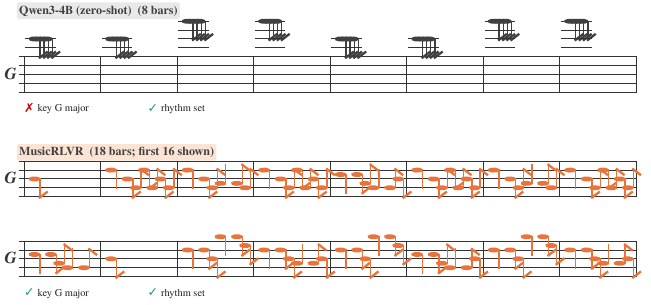}\par}\smallskip
\noindent Qwen3-4B (zero-shot) fails \texttt{key}; \method{} satisfies every constraint.
\end{figure*}

\label{app:crossmodel}

In this section, we present qualitative examples. Figure~\ref{fig:staffdiag} shows two outputs
that satisfy the same four constraints with different repeating patterns. Every example below is a benchmark item chosen
by a fixed rule (the first item in file order with the stated constraint families and pass/fail
pattern), its
completions are read verbatim from the frozen evaluation files, and the notation is
drawn from \texttt{music21}'s parse of the output: pitch, duration and bar lines. The key and
meter checks additionally read the header-derived objects, and the key check analyses the whole
score. Marks under each staff are the verifiers' decisions on that output. Outputs rejected at
the gate are shown as a short excerpt together with the gate's reason. Examples~1--6 compare the
zero-shot base model with \method{} (diagnostic run), and Example~2 shows SFT as well; Examples~7--8
compare \method{} with the larger zero-shot models on the same items.

Examples~9--14 use the matched Direct \method{} of Table~\ref{tab:main} and one rule: the first
item in file order, among those with the stated families, that Direct \method{} satisfies while
Qwen3-32B, Llama-3.1-70B and MiniMax-M3 all fail. Such items are common. Over the $2{,}180$
benchmark items, Direct \method{} satisfies $609$ that all three of these models fail, whereas all
three satisfy only $5$ items that Direct \method{} fails. The six examples cover four constraints
on Mixed, two held-out pairs on Dense (with the Qwen3-8B policy alongside), a held-out pair with
a form on Unseen, the interleaved \textsc{abab} form, and two Edge items: one with out-of-range
meter ($5/4$), and one combining a rhythm restriction with \textsc{abab} structure.

% <<< tables/examples
% Two-column adaptation of make_fig_examples.py output: each example is one full-width
% float, since the staff images are unreadable at column width. Regenerate paper.tex,
% not this copy.
\begin{figure*}[t]
\small\raggedright
\textbf{Example 2: high-$k$ split ($k = 4$).}\par
\noindent\textbf{Item} \texttt{hk-0004}. \textbf{Constraints:} (1) the meter must be 3/4, and every bar must contain exactly the right total duration; (2) the tune must be exactly 8 bars long (no pickup bar); (3) the last note of the tune must be E (any octave); (4) use only eighth and quarter notes (no dotted notes, no other durations, no rests)\par\smallskip
{\centering\includegraphics[width=0.8\textwidth]{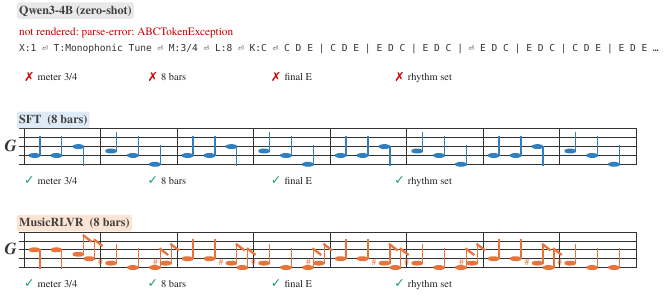}\par}\smallskip
\noindent Qwen3-4B (zero-shot) fails at the parse gate (parse-error: ABCTokenException); SFT satisfies every constraint; \method{} satisfies every constraint.
\end{figure*}

\begin{figure*}[t]
\small\raggedright
\textbf{Example 3: single-constraint split ($k = 1$).}\par
\noindent\textbf{Item} \texttt{sg-0000}. \textbf{Constraints:} (1) melodic motion must be smooth: no melodic leap larger than an octave\par\smallskip
{\centering\includegraphics[width=0.8\textwidth]{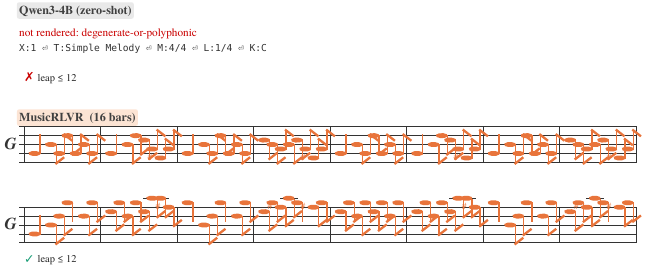}\par}\smallskip
\noindent Qwen3-4B (zero-shot) fails at the parse gate (degenerate-or-polyphonic); \method{} satisfies every constraint.
\end{figure*}

\begin{figure*}[t]
\small\raggedright
\textbf{Example 4: seen-combination split ($k = 3$).}\par
\noindent\textbf{Item} \texttt{sc-0005}. \textbf{Constraints:} (1) the last note of the tune must be D (any octave); (2) use only sixteenth, eighth, and quarter notes (no dotted notes, no other durations, no rests); (3) melodic motion must be smooth: no melodic leap larger than a perfect fifth\par\smallskip
{\centering\includegraphics[width=0.8\textwidth]{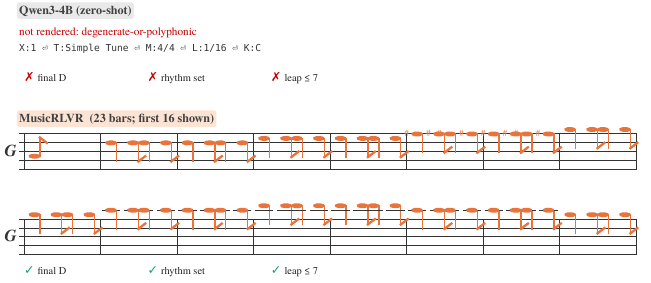}\par}\smallskip
\noindent Qwen3-4B (zero-shot) fails at the parse gate (degenerate-or-polyphonic); \method{} satisfies every constraint.
\end{figure*}

\begin{figure*}[t]
\small\raggedright
\textbf{Example 5: seen-combination split ($k = 2$).}\par
\noindent\textbf{Item} \texttt{sc-0001}. \textbf{Constraints:} (1) use only sixteenth, eighth, and quarter notes (no dotted notes, no other durations, no rests); (2) the tune must be exactly 16 bars with an AABB form in 4-bar units: labeled units with the same letter must be note-for-note identical, and units with different letters must differ\par\smallskip
{\centering\includegraphics[width=0.8\textwidth]{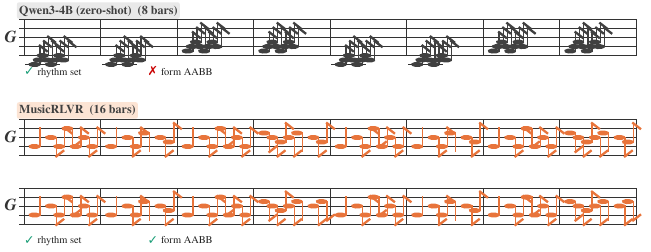}\par}\smallskip
\noindent Qwen3-4B (zero-shot) fails \texttt{structure}; \method{} satisfies every constraint.
\end{figure*}

\begin{figure*}[t]
\small\raggedright
\textbf{Example 6: unseen-combination split ($k = 2$, held-out pair \texttt{length}--\texttt{range}).}\par
\noindent\textbf{Item} \texttt{uc-0001}. \textbf{Constraints:} (1) the tune must be exactly 16 bars long (no pickup bar); (2) every note must lie between G3 and A5 inclusive\par\smallskip
{\centering\includegraphics[width=0.8\textwidth]{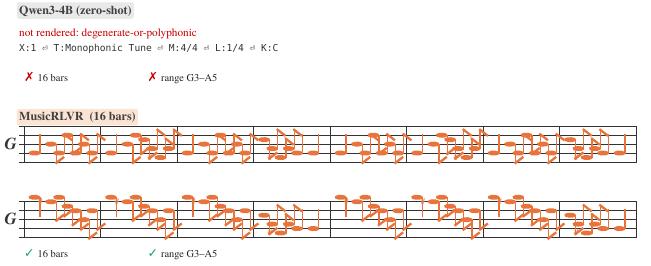}\par}\smallskip
\noindent Qwen3-4B (zero-shot) fails at the parse gate (degenerate-or-polyphonic); \method{} satisfies every constraint.
\end{figure*}

\begin{figure*}[t]
\small\raggedright
\textbf{Example 7: structure split ($k = 1$, zero-shot models against \method{}).}\par
\noindent\textbf{Item} \texttt{st-0023}. \textbf{Constraints:} (1) the tune must be exactly 16 bars with an AABA form in 4-bar units: labeled units with the same letter must be note-for-note identical, and units with different letters must differ\par\smallskip
{\centering\includegraphics[width=0.55\textwidth]{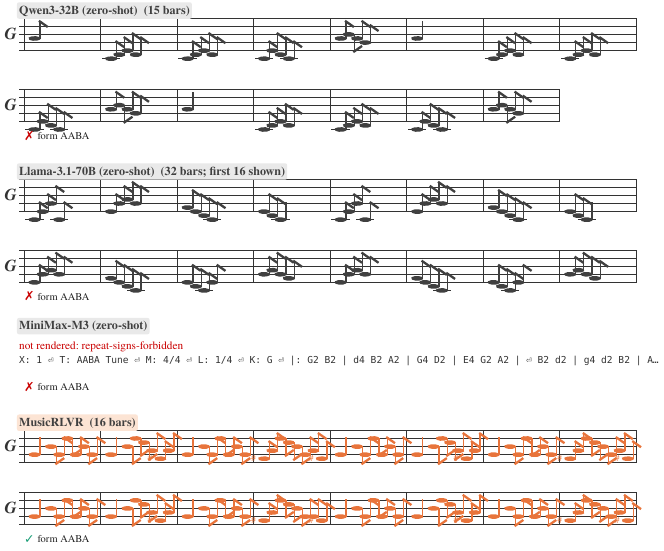}\par}\smallskip
\noindent Qwen3-32B (zero-shot) fails \texttt{structure}; Llama-3.1-70B (zero-shot) fails \texttt{structure}; MiniMax-M3 (zero-shot) fails at the parse gate (repeat-signs-forbidden); \method{} satisfies every constraint.
\end{figure*}

\begin{figure*}[t]
\small\raggedright
\textbf{Example 8: high-$k$ split ($k = 4$, zero-shot models against \method{}).}\par
\noindent\textbf{Item} \texttt{hk-0278}. \textbf{Constraints:} (1) the tune must be in C major; (2) every note must lie between D4 and D6 inclusive; (3) the last note of the tune must be G (any octave); (4) melodic motion must be smooth: no melodic leap larger than a perfect fifth\par\smallskip
{\centering\includegraphics[width=0.55\textwidth]{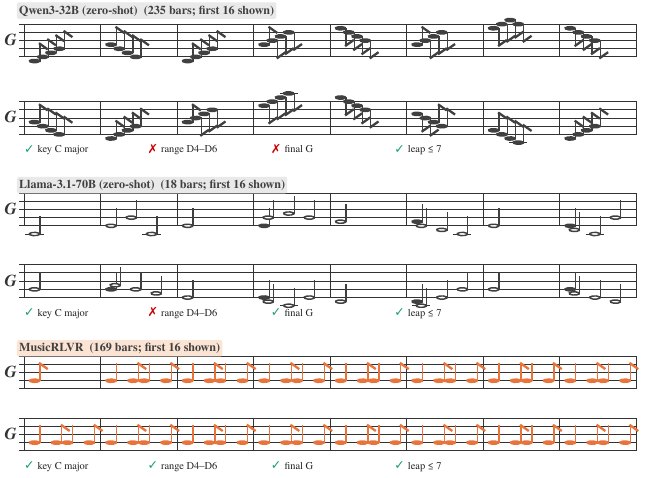}\par}\smallskip
\noindent Qwen3-32B (zero-shot) fails \texttt{range}, \texttt{final}; Llama-3.1-70B (zero-shot) fails \texttt{range}; \method{} satisfies every constraint.
\end{figure*}

% >>> tables/examples

% <<< tables/examples_direct (make_fig_examples_direct.py)
\begin{figure*}[t]
\small\raggedright
\textbf{Example 9: Mixed split ($k = 4$).}\par
\noindent\textbf{Item} \texttt{k4-0002}. \textbf{Constraints:} (1) the tune must be in A minor; (2) the meter must be 3/4, and every bar must contain exactly the right total duration; (3) the last note of the tune must be C (any octave); (4) melodic motion must be smooth: no melodic leap larger than a perfect fifth\par\smallskip
{\centering\includegraphics[width=0.75\textwidth]{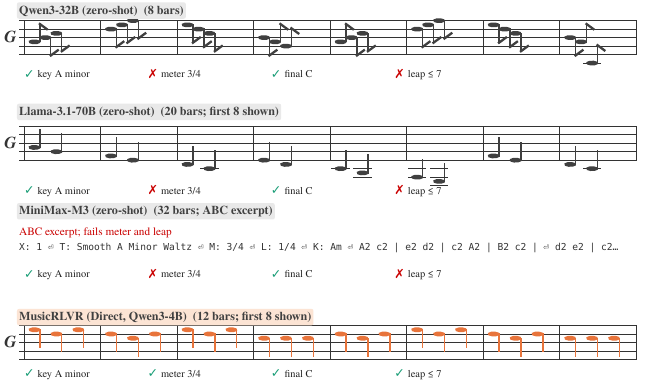}\par}\smallskip
\noindent Qwen3-32B (zero-shot) fails \texttt{meter}, \texttt{leap}; Llama-3.1-70B (zero-shot) fails \texttt{meter}, \texttt{leap}; MiniMax-M3 (zero-shot) fails \texttt{meter}, \texttt{leap}; \method{} (Direct, 4B) satisfies every constraint.
\end{figure*}

\begin{figure*}[t]
\small\raggedright
\textbf{Example 10: Dense split ($k = 4$, two held-out pairs, \method{} on both base models).}\par
\noindent\textbf{Item} \texttt{dn-0003}. \textbf{Constraints:} (1) the tune must be in D minor; (2) the tune must be exactly 16 bars long (no pickup bar); (3) every note must lie between G3 and A5 inclusive; (4) use only eighth, quarter, and half notes (no dotted notes, no other durations, no rests)\par\smallskip
{\centering\includegraphics[width=0.75\textwidth]{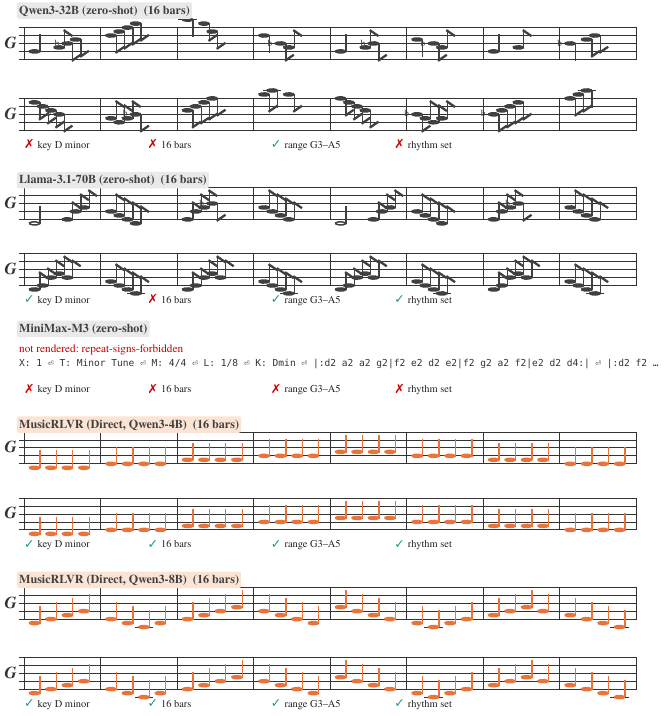}\par}\smallskip
\noindent Qwen3-32B (zero-shot) fails \texttt{key}, \texttt{length}, \texttt{rhythm}; Llama-3.1-70B (zero-shot) fails \texttt{length}; MiniMax-M3 (zero-shot) fails at the parse gate (repeat-signs-forbidden); \method{} (Direct, 4B) satisfies every constraint; \method{} (Direct, 8B) satisfies every constraint.
\end{figure*}

\begin{figure*}[t]
\small\raggedright
\textbf{Example 11: unseen-combination split ($k = 3$, held-out pair \texttt{meter}--\texttt{leap} plus a form).}\par
\noindent\textbf{Item} \texttt{uc-0006}. \textbf{Constraints:} (1) the meter must be 6/8, and every bar must contain exactly the right total duration; (2) melodic motion must be smooth: no melodic leap larger than an octave; (3) the tune must be exactly 16 bars with an ABAB form in 4-bar units: labeled units with the same letter must be note-for-note identical, and units with different letters must differ\par\smallskip
{\centering\includegraphics[width=0.75\textwidth]{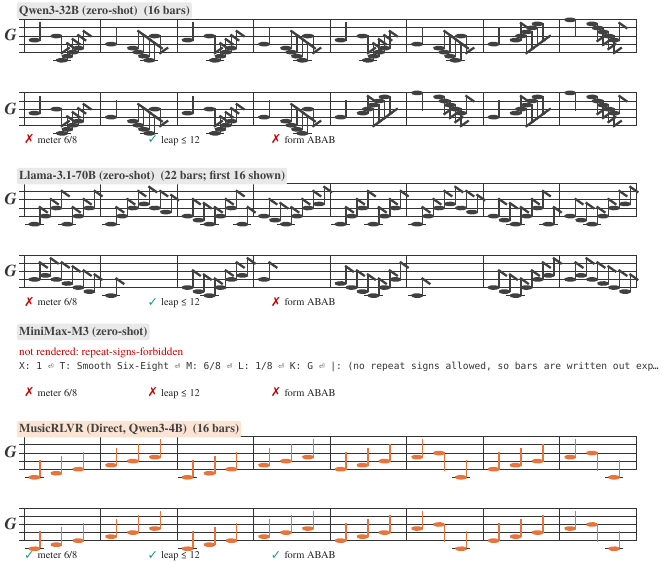}\par}\smallskip
\noindent Qwen3-32B (zero-shot) fails \texttt{meter}, \texttt{structure}; Llama-3.1-70B (zero-shot) fails \texttt{meter}, \texttt{structure}; MiniMax-M3 (zero-shot) fails at the parse gate (repeat-signs-forbidden); \method{} (Direct, 4B) satisfies every constraint.
\end{figure*}

\begin{figure*}[t]
\small\raggedright
\textbf{Example 12: structure split ($k = 2$, the interleaved form \textsc{abab}).}\par
\noindent\textbf{Item} \texttt{st-0033}. \textbf{Constraints:} (1) the meter must be 6/8, and every bar must contain exactly the right total duration; (2) the tune must be exactly 16 bars with an ABAB form in 4-bar units: labeled units with the same letter must be note-for-note identical, and units with different letters must differ\par\smallskip
{\centering\includegraphics[width=0.75\textwidth]{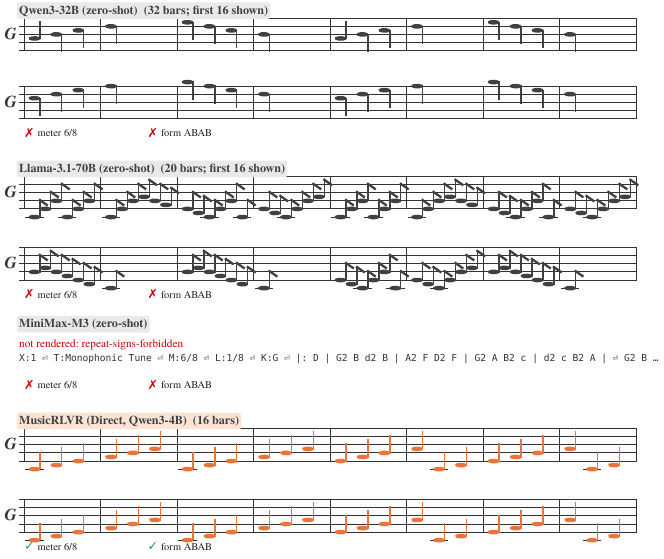}\par}\smallskip
\noindent Qwen3-32B (zero-shot) fails \texttt{meter}, \texttt{structure}; Llama-3.1-70B (zero-shot) fails \texttt{meter}, \texttt{structure}; MiniMax-M3 (zero-shot) fails at the parse gate (repeat-signs-forbidden); \method{} (Direct, 4B) satisfies every constraint.
\end{figure*}

\begin{figure*}[t]
\small\raggedright
\textbf{Example 13: Edge split ($k = 2$, a $5/4$ meter, outside the RL prompt range).}\par
\noindent\textbf{Item} \texttt{eg-0033}. \textbf{Constraints:} (1) the meter must be 5/4, and every bar must contain exactly the right total duration; (2) every note must lie between C4 and C5 inclusive\par\smallskip
{\centering\includegraphics[width=0.75\textwidth]{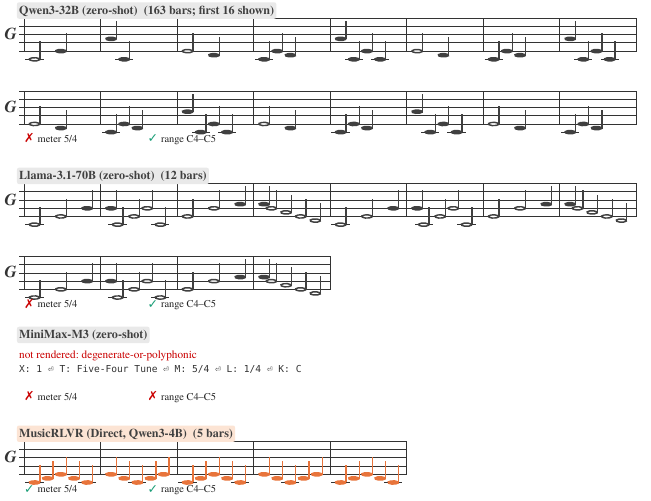}\par}\smallskip
\noindent Qwen3-32B (zero-shot) fails \texttt{meter}; Llama-3.1-70B (zero-shot) fails \texttt{meter}; MiniMax-M3 (zero-shot) fails at the parse gate (degenerate-or-polyphonic); \method{} (Direct, 4B) satisfies every constraint.
\end{figure*}

\begin{figure*}[t]
\small\raggedright
\textbf{Example 14: Edge split ($k = 2$, rhythm restriction + \textsc{abab} structure).}\par
\noindent\textbf{Item} \texttt{eg-0005}. \textbf{Constraints:} (1) use only quarter and half notes (no dotted notes, no other durations, no rests); (2) the tune must be exactly 16 bars with an ABAB form in 4-bar units: labeled units with the same letter must be note-for-note identical, and units with different letters must differ\par\smallskip
{\centering\includegraphics[width=0.75\textwidth]{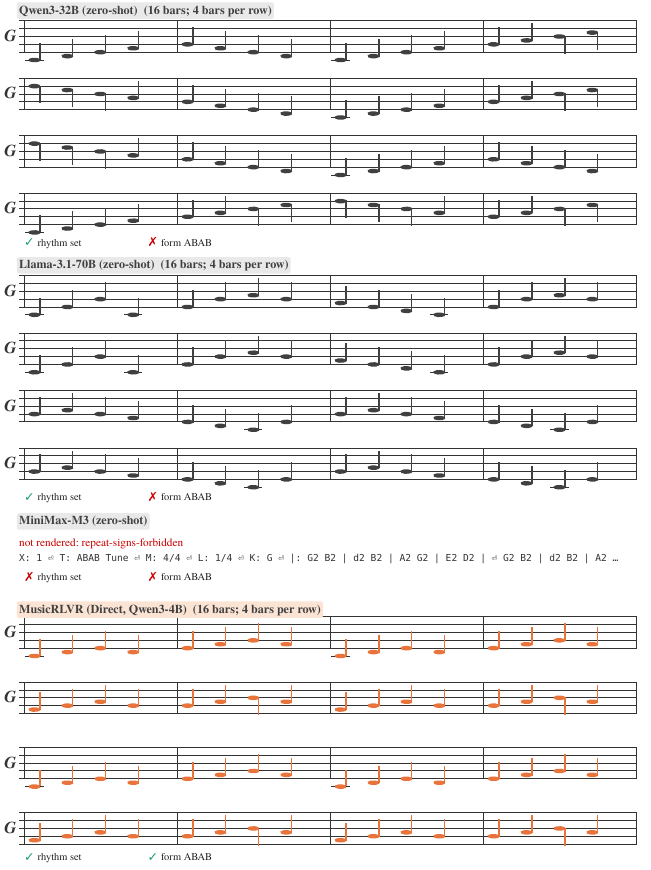}\par}\smallskip
\noindent Qwen3-32B (zero-shot) fails \texttt{structure}; Llama-3.1-70B (zero-shot) fails \texttt{structure}; MiniMax-M3 (zero-shot) fails at the parse gate (repeat-signs-forbidden); \method{} (Direct, 4B) satisfies every constraint.
\end{figure*}
% >>> tables/examples_direct

\FloatBarrier
\section{Broader Impact}
\label{app:impact}

In this section, we expand the ethics statement. The method makes music-generation models follow
explicit specifications, which helps wherever a score must meet a specification: music education,
game audio, accessibility tools. We do not foresee significant negative societal impacts specific
to this work, which concerns short monophonic melodies in ABC notation under explicit constraints.
The likeliest misreading is to take a high all-satisfied rate as a quality score; it certifies
that the requested properties hold. All supervised targets derive from the publicly
available IrishMAN corpus of traditional Irish music; the reinforcement prompts are generated
synthetically. The benchmark and verifiers will be released to support reproducibility and
further research.

\end{document}